\documentclass{jfm}
\usepackage{graphicx}
\usepackage{tikz}
\usetikzlibrary{shapes}
\usepackage{epstopdf, epsfig}
\usepackage{amsmath}
\usepackage{color}
\usepackage{float}
\usepackage{placeins}
\usepackage{subcaption}
\usepackage{natbib}
\usepackage{hyperref}
\usepackage{booktabs}% nicer horizontal lines
\usepackage{caption}% fix vertical spacing of table captions
\usepackage{siunitx}% align numbers at the decimal point
\definecolor{mycolor}{rgb}{0.3,0,0.6}

\newcommand{\uu}{\mbox{\boldmath $u$} {}}
\newcommand{\ee}{\mbox{\boldmath $e$} {}}
\newcommand{\vv}{\mbox{\boldmath $v$} {}}
\newcommand{\xx}{\mbox{\boldmath $x$} {}}
\newcommand{\ii}{\mbox{\boldmath $I$} {}}
\newcommand{\ttau}{\mbox{\boldmath $\tau$} {}}
\newcommand{\ppi}{\mbox{\boldmath $\pi$} {}}
\newcommand{\tuu}{\mbox{\boldmath $\tilde{u}$} {}}
\newcommand{\tvv}{\mbox{\boldmath $\tilde{v}$} {}}

\newcommand{\tp}{\mbox{$\tilde{P}$} {}}

\newcommand{\tn}{\mbox{$\tilde{n}$} {}}
\newcommand{\tS}{\mbox{$\tilde{S}$} {}}
\newcommand{\tu}{\mbox{$\tilde{u}$} {}}

\newcommand{\overbar}[1]{\mkern 1.5mu\overline{\mkern-1.5mu#1\mkern-1.5mu}\mkern 1.5mu}
\newcommand{\blackline}{\raisebox{2pt}{\tikz{\draw[-,black,solid,line width = 0.9pt](0,0) -- (5mm,0);}}}
\newcommand{\redline}{\raisebox{2pt}{\tikz{\draw[-,red,dashed,line width = 1.5pt](0,0) -- (5mm,0);}}}
\newcommand{\blueline}{\raisebox{2pt}{\tikz{\draw[-,blue,dash dot,line width = 1.5pt](0,0) -- (5mm,0);}}}
\newcommand{\bluesline}{\raisebox{2pt}{\tikz{\draw[-,blue,solid,line width = 1.5pt](0,0) -- (5mm,0);}}}
\newcommand{\greenline}{\raisebox{2pt}{\tikz{\draw[-,black!60!green,loosely dotted,line width = 1.5pt](0,0) -- (5mm,0);}}}
\newcommand{\diamondline}{\raisebox{0pt}{\tikz{\node[draw,scale=0.4,diamond,fill=black!60!green](){};}}}
\newcommand{\starline}{\raisebox{0.5pt}{\tikz{\node[draw,scale=0.25,star,star points=5,star point ratio=2.5,fill=black!60!green](){};}}}
\newcommand{\sq}{\raisebox{0.5pt}{\tikz{\node[draw,blue,scale=0.6,regular polygon, regular polygon sides=4,fill=none](){};}}}
\newcommand{\circleline}{\raisebox{0pt}{\tikz{\draw[black,solid,line width = 1.0pt,fill=black](2.8mm,0.8mm) circle (0.8mm);\draw[-,black,solid,line width = 1.0pt](0.,0.8mm) -- (5.5mm,0.8mm)}}}
\newcommand{\circlelinedot}{\raisebox{0pt}{\tikz{\draw[black,solid,line width = 1.0pt,fill=none](2.8mm,0.8mm) circle (0.8mm);\draw[-,black,dash dot,line width = 1.0pt](0.,0.8mm) -- (5.5mm,0.8mm)}}}
\newcommand{\circlelinered}{\raisebox{0pt}{\tikz{\draw[red,solid,line width = 1.0pt,fill=none](2.8mm,0.8mm) circle (0.8mm);\draw[-,red,solid,line width = 1.0pt](0.,0.8mm) -- (5.5mm,0.8mm)}}}
\newcommand{\circlelinegrn}{\raisebox{0pt}{\tikz{\draw[-,black!60!green,solid,line width = 1.0pt,fill=none](2.8mm,0.8mm) circle (0.8mm);\draw[-,black!60!green, dotted,line width = 1.0pt](0.,0.8mm) -- (5.5mm,0.8mm)}}}
\newcommand{\pcircleline}{\raisebox{0pt}{\tikz{\draw[mycolor,solid,line width = 1.0pt,fill=none](2.8mm,0.8mm) circle (0.8mm);\draw[-,mycolor,solid,line width = 1.0pt](0.,0.8mm) -- (5.5mm,0.8mm)}}}
\newcommand{\circlenline}{\raisebox{0pt}{\tikz{\draw[black,solid,line width = 1.0pt,fill=none](2.8mm,0.8mm) circle (0.8mm);\draw[-,black,dashed,line width = 1.0pt](0.,0.8mm) -- (5.5mm,0.8mm)}}}
\newcommand{\blackdashline}{\raisebox{2pt}{\tikz{\draw[-,black,dashed,line width = 1.0pt](0.,0.8mm) -- (5.5mm,0.8mm);}}}
\newcommand{\greendashline}{\raisebox{2pt}{\tikz{\draw[-,black!60!green,dashed,line width = 1.0pt](0.,0.8mm) -- (5.5mm,0.8mm);}}}

\newcommand{\figone}{\raisebox{0.5pt}{\tikz{\node[draw,black,scale=0.4,circle,line width = 1.0pt,fill=none](){};}}}
\newcommand{\figtwo}{\raisebox{0pt}{\tikz{\node[draw,blue,scale=0.4,diamond,line width = 1.0pt,fill=none](){};}}}
\newcommand{\figthree}{\raisebox{0.5pt}{\tikz{\node[draw,black!60!green,scale=0.4,regular polygon, regular polygon sides=4,line width = 1.0pt,fill=none](){};}}}
\shorttitle{Population Balance Modeling in a Large Eddy Simulation Framework}
\shortauthor{A.K. Aiyer, D. Yang, M. Chamecki, C. Meneveau}

\title{A Population Balance Model for Large Eddy Simulation of Polydisperse Droplet Evolution}

\author{A. K. Aiyer\aff{1}
  \corresp{\email{aaiyer1@jhu.edu}},
 D. Yang\aff{2},
 M. Chamecki\aff{3},
 and C. Meneveau\aff{1}
}
\affiliation{\aff{1}Department of Mechanical Engineering, Johns Hopkins University, MD, USA
\aff{2}Department of Mechanical Engineering, University of
Houston,  TX, USA
\aff{3}Department of Atmospheric and Oceanic Sciences, University of California, Los Angeles, CA, USA
}

\begin{document}

\maketitle

\begin{abstract}
 In the context of many applications of turbulent multi-phase flows, knowledge of the dispersed phase size distribution and its evolution is critical to predicting important macroscopic features. 
We develop a  large eddy simulation (LES) model that can predict the turbulent transport and evolution of size distributions, for a specific subset of applications in which the dispersed phase can be assumed to consist of spherical droplets, and occurring at low volume fraction.
 We use a population dynamics model for polydisperse droplet distributions specifically adapted to a LES framework including a model for droplet breakup due to turbulence, neglecting coalescence consistent with the assumed small dispersed phase volume fractions. We model the number density fields  using an Eulerian approach for each bin of the discretized droplet size distribution. Following earlier methods used in the Reynolds averaged Navier--Stokes framework, the droplet breakup due to turbulent fluctuations is modelled by treating droplet-eddy collisions as in kinetic theory of gases. Existing models assume the scale of droplet-eddy collision to be in the inertial range of turbulence. In order to also model smaller droplets comparable to or smaller than the Kolmogorov scale we extend the breakup kernels 
 using a structure function model that smoothly transitions from the inertial to the viscous range. 
 The model includes a dimensionless coefficient that is fitted by comparing predictions in a one-dimensional version of the model with a laboratory experiment of oil droplet breakup below breaking waves. After initial comparisons of the one-dimensional model to measurements of oil droplets in an axisymmetric jet, it is then applied in a three-dimensional LES of
 a jet in crossflow with large oil droplets of a single size being released at the source of the jet. We model the concentration fields using $N_d =15$ bins of discrete droplet sizes and solve scalar transport equations for each bin. The resulting droplet size distributions are compared with published experimental data, and good agreement for the relative size distribution is obtained. The LES results also enable us to quantify size distribution variability. We find that the probability distribution functions of key quantities such as the total surface area and the Sauter mean diameter of oil droplets are highly variable, some displaying strong non-Gaussian intermittent behavior. 
 \end{abstract}

\section{Introduction}
An understanding of liquid droplet size distributions in turbulent flows is important in 
the context of numerous natural and engineering processes. For instance, knowledge of 
the droplet size distribution is important for predicting fate and transport of oil from an
underwater spill \citep{North2015,Nissanka2016}. The size of 
oil droplets affects their rise velocity and can influence the entire plume's transport characteristics 
in the ocean \citep{chen2018}. A review of recent developments in oil spill modelling can be found in \citet{Nissanka2018}. In combustion of fuel sprays the droplet size distribution plays an 
important role, determining atomization quality and spray combustion performance 
\citep{Bossard1996}. The size distribution also plays a significant role in cloud dynamics, while also affecting the thermodynamic characteristics of the system \citep{Liu1995,Igel2016}.  Population balance equations (PBEs), initially formulated by  \citet{Smolu1916} to 
study aggregation processes, are commonly used in many problems involving particulate systems with particles of many sizes. The basic formulation of PBE's has been extended by 
\citet{Hulburt1964}, \citet{Randolph1964} and \citet{Ramkrishna1985} to include additional phenomena such as breakup, 
coalescence, nucleation, condensation, etc. 
PBE's can be classified into four 
main categories \citep{Jakobsen2014}: Method of moments, multi-class method, method of weighted residuals 
and stochastic methods. In this paper we use the multi-class formulation to study the 
evolution of polydisperse liquid droplets in a surrounding fluid. In this 
method the continuous size range is divided into a number of small contiguous 
subclasses, and the PBE is converted into a number of discretized transport equations for each class (bin). Methods such as the method of classes \citep{Marchal1988}, the fixed pivot method 
\citep{Kumar1996}, and the cell average technique \citep{Kumar2006}  have been 
developed to discretize the PBE.  

\citet{Coulaloglou1977} were among the first to 
introduce a simple macroscopic formulation to study breakup and coalescence in an 
agitated liquid-liquid dispersion (droplets in a liquid), typically in a turbulent flow. Over the years, a considerable number of studies have been performed for steady-state size distributions in stirred vessels   \citep[e.g.][among many others]{Coulaloglou1977,Tsouris1994,Prince1990,Wang2007}. These
models used the average dissipation rate of turbulent kinetic energy 
to characterize the breakup, an approach well suited  for 
use in Reynolds averaged Navier--Stokes (RANS) $k-\epsilon$ modeling, where the mean rate of dissipation $\langle \epsilon \rangle$ is a primary field variable available from the turbulence model. There have been several studies conducted in recent years that focus on predicting droplet size distribution either through models based on a RANS approach, or by using correlations for the mean size distribution \citep[e.g.][]{Bandara2011,Brandvik2013,Johansen2013,Zhao2014J,Zhao2014}.
 More recently \citet{Pedel2014} used large eddy simulation (LES) coupled with an Eulerian solver for the droplet phase to predict droplet concentrations using the direct quadrature methods of moment (DQMOM). \citet{Neuber2017} used an 
Eulerian LES coupled with a sparse Lagrangian droplet method to study aggregation and 
nucleation. Further work combining LES for the carrier flow with PBE for droplets can be found in \citet{Seubert2012}, \citet{Sewerin2017} and \citet{Salehi2017}. The latter of which  used LES for the carrier phase coupled with a statistical Lagrangian approach for the droplets.
However, Lagrangian methods can become expensive for large number of droplets. Even though limited to applications with relatively low volume fractions, Eulerian approaches can be advantageous since they are not limited by the number of droplets, as the distribution of droplets in each size range is described by a continuous concentration field. \citet{Yang2016} used an Eulerian-Eulerian LES model to study bubble-driven buoyant plumes in a stably stratified quiescent fluid. They showed that the plume structure depended on the diameter of the droplets, but their LES only used a single bubble size in each simulation. Thus, there is interest in modeling plumes of droplets of many sizes and there is a need to extend LES to systems with polydisperse size distributions. There is little existing body of literature of combining LES with PBE's using the method of classes in the context of Eulerian descriptions of droplet concentration fields. The present paper is devoted to developing and testing such a tool. 

The population balance equation is a transport equation that describes the evolution of the number density field representing the population of a type of particles due to advection, diffusion, breakup or coalescence. It can be written as
\begin{equation}
\label{eqn:pbe}
\frac{\partial n(d_i,\xx,t)}{\partial t} + \nabla \cdot \left[\vv(d_i,\xx,t)  n(d_i,\xx,t) \right] = S_{b,i} + S_{c,i} ,
\end{equation}
where $n(d_i,\xx,t)$ is the number density of droplets of diameter $d_i$, $\vv(d_i,\xx,t)$ is the droplet velocity, $S_{b,i}$ and $S_{c,i}$ are source terms for droplet breakup and coalescence affecting droplets of diameter $d_i$ (within bin $i$), respectively. The divergence is calculated w.r.t the spatial coordinate $\xx$. In this work we consider small dispersed phase volume fractions and neglect droplet coalescence ($S_{c,i}=0$). Effects of evaporation, growth and aggregation have also been neglected in this work. The droplet concentration transport velocity $\vv(d_i,\xx,t)$ can be specified using the approach of \citet{Ferry2001} and \citet{Yang2016} that includes effects of drag, buoyancy and relative acceleration (see \S \ref{sec:LES_model}).
 
In order to close the problem, one needs a model for the source term $S_{b,i}$.  A number of models have been developed for breakup due to turbulent fluctuations.
For a turbulent flow, breakup due to shearing-off processes and due to interfacial instability are often neglected. The turbulent fluctuations need to overcome the main resistive forces in the droplet, namely surface tension and viscosity. 
\citet{Prince1990} and \citet{Tsouris1994} treated the interaction between eddies and droplets similar to collisions between molecules in kinetic theory of gases. This allowed them to define a collision frequency based on the size of the eddy and droplet, and invoke a model for their typical relative velocity at that scale. The requirement for breakup in such a model is that the turbulent kinetic energy of the colliding eddy is greater than the potential energy associated with the resistive forces of the droplet. \citet{Skartlien2013} and \citet{Zhao2014}  extended the model of \citet{Prince1990} and \citet{Tsouris1994} to include the effect of viscosity in the resistive force. It should be noted that there are numerous other droplet breakup models in the literature based on criteria such as turbulent kinetic energy of the droplet being higher than a 
 critical value \citep{Coulaloglou1977,Chatzi1987}, or the inertial force of colliding eddy exceeding the interfacial force of the smallest droplet \citep{Lehr1999}, surface deformation and breakup due to turbulent stresses of surrounding fluid \citep{MARTINEZ-BAZAN1999}, to name a few. 
 
 In most of the previous formulations, the scale of the colliding droplets and eddies was assumed to be in the inertial range of turbulence. Hence Kolmogorov scaling for the inertial subrange was used to estimate the magnitude of eddy-velocity fluctuations at a particular scale. This precludes these models from being able to calculate the breakup frequency for droplets that fall in or near the viscous range, e.g. below $\eta$ up to about 15$\eta$, where $\eta$ is the Kolmogorov length scale. A unified treatment is needed to extend these models to the entire spectrum of turbulence. Recently, \citet{Solsvik2016a}  included the viscous range using a model energy spectrum  \citep{Pope2001} for the complete range of scales. Using a Fourier series transform, they derived the second-order structure function from this spectrum including the viscous range. The resulting expressions are quite complicated, owing to the particular functional form of the viscous cutoff in spectral space. However, a more direct approach and simpler expressions can already be found in the literature, based on the Batchelor blending function \citep{Batchelor1951} written directly for the structure function in physical space.  In this study, we adopt this particular physical space version for modeling the eddy velocity fluctuations, since it affords more intuitive understanding and more efficient numerical evaluations of the model terms.
 
 A significant number of experimental studies have been conducted in stirred tanks  \citep{Narsimhan1980,Calabrese1986,SATHYAGAL19961377} and have been used to measure the droplet breakup frequency. However, it is challenging to relate the measured transient droplet size distributions to models due to the difficulty in characterizing the turbulence in stirred tanks  which is highly anisotropic and spatially very heterogeneous. For instance, such flows contain highly localized  high shear regions near the surface of the impeller blades and strong tip vortices shed by the blades.  
 
 There have been efforts to design suitable experiments in which the turbulence is well characterized. A well-known reference is that of \citet{MARTINEZ-BAZAN1999} who designed and carried out a series of experiments where air bubbles were injected into a fully developed turbulent water jet. This ensured that the turbulence was well characterized and size distributions could be measured using non-intrusive optical techniques. \citet{eastwood2004} injected droplets of varying density, viscosity and interfacial tension into a fully developed water jet and tracked particle size distributions using digital image processing techniques. \citet{Brandvik2013} performed oil jet experiments 
 in a very large cylindrical tank.
 They measured droplet sized distributions using an in situ laser diffractometer. \citet{Murphy2016} used an oil jet in crossflow to study the droplet breakup and resulting size distributions. Their experiment consisted of a nozzle supported by a carriage that was moved at a constant speed thus setting up a crossflow. The size distribution was measured non-intrusively using in-line holographic techniques. \citet{Zhao2016} conducted a large scale experiment of underwater oil release through a $25.4\;\mbox{mm}$ pipe. They measured size distributions using two LISST's (Laser in-situ scattering and transmissometry) in the range of $2.5\mbox{--}500\;\mu\mbox{m}$. Recently, \citet{li2017} studied the time evolution of subsurface oil droplets generated by breaking waves. They performed experiments for varying wave energy and surface tension and measured the generated size distribution using digital inline holography at two magnifications. Such experiments, where the flow is well characterized and the size distributions can be measured non-intrusively are ideal to use as comparisons for models and we shall make use of the  two lab-scale oil-droplet experimental data sets in the present work. 
 
 Over the recent decade, progress has been made in simulating two-phase flows using sufficiently fine spatial resolution to describe detailed deforming interfaces and thus capture the formation of droplets from instabilities of liquid sheets \citep{Desjardins2008,gorokhovski2008,hermann2010,duret2012,hermann2013}. The aim is to, for example, simulate primary atomization and determine the resulting droplet size distributions. The LES tool described in the present work focuses on much coarser numerical meshes that cannot resolve such detailed dynamics and assume, for example, that the size distribution resulting from small-scale initial droplet formation processes is known.
 
The paper begins in \S \ref{sec:PBE} with a description of the breakup model that is based on prior approaches 
used in RANS or integral models \citep{Prince1990,Tsouris1994,Zhao2014}, extended here to include the correct scaling for droplets in the viscous range and adapted for applicability in LES. The modeling is focused on applications of LES where relatively coarse grids have to be employed and the small-scale details cannot be resolved.  The model contains an undetermined multiplicative dimensionless parameter that is chosen by matching predictions of a simplified version of the model to an experiment in breaking waves, and then verifying its robustness under very different flow conditions (a round jet), showcasing the robustness of the breakup model, as explained in \S \ref{sec:fit_K}.  An application to LES of a turbulent, droplet-laden jet in cross-flow is presented in \S \ref{sec:LES_model}, focusing on the 
droplet breakup and transport occurring in regions away from the nozzle. In the present LES application, the nozzle details will not be resolved and thus the initial breakup mechanisms of oil into large droplets near the nozzle will be replaced by an appropriately chosen initial inflow condition of droplets of a given diameter. 
The focus of the study will be on comparing the size distributions far away from the nozzle with available experimental data, and to showcase the advantage of LES in being able to predict variability and intermittency of the size distribution and characteristic scales of the droplets.  Conclusions are presented in \S \ref{sec:conclusions}.

\section{Droplet breakup model}\label{sec:PBE}
The population balance equation described in equation ($\ref{eqn:pbe}$), neglecting the effect of coalescence and written using the droplet size (diameter $d_i$) as the internal coordinate, is given by
 \begin{equation}\label{eqn:pbe_discrete}
 \frac{\partial n(d_i,\xx,t)}{\partial t} + \nabla \cdot [ \vv(d_i,\xx,t) n(d_i,\xx,t)  ] = S_b(d_i,\xx,t),
 \end{equation}
where $n(d_i,\xx,t)$ is the number density of droplets in the i-th bin representing droplets of diameter
around $d_i$, at location $\xx$ at time $t$. The divergence is calculated w.r.t the spatial coordinate $\xx$. The source term due to breakup is written according to \citet{Zhao2014},
\begin{equation}\label{eqn:break_source}
S_b(d_i,\xx,t) = \sum_{j=i+1}^n P(d_i,d_j)g(d_j,\xx,t)n(d_j,\xx,t)-g(d_i,\xx,t)n(d_i,\xx,t).
\end{equation}
The first term on the right-hand side of equation $(\ref{eqn:break_source})$ represents the birth of droplets of size $d_i$ due to the total contribution from breakups of larger droplets of 
diameter
$d_j$. The second term accounts for death of droplets of size $d_i$ due to breakup. The factor
$P(d_i,d_j)$ is the probability for the formation of a droplet 
of size $d_i$ due to the breakup of a parent droplet of size $d_j$, and $g(d_i,\xx,t)$ is the breakup frequency of a droplet of size $d_i$ (in bin $i$). The breakup probability can be related to the probability density function $\beta(d_i,d_j)$, i.e.\ $P(d_i,d_j) = \beta(d_i,d_j) \delta(d_i)$, where $\delta (d_i)$ is the width of the bin centered at $d_i$. Also note that the number density $n(d_i,\xx,t)$ for discrete bins can be related to the more general continuous number density distribution function $n^*$ according to
\begin{equation}
    n(d_i,\xx,t) = \int_{d_{i-1/2}}^{d_{i+1/2}} n^*(d,\xx,t)\mathrm{d} (d),
\end{equation}
where $[d_{i+1/2}-d_{i-1/2}]$ is the i-th bin width. In order to solve equation $(\ref{eqn:pbe_discrete})$,  models are needed for the probability $P(d_i,d_j)$ and the breakup frequency $g(d_i,\xx,t)$. 

\subsection{Model for the breakup probability.}

Models for the breakup probability  function $P(d_i,d_j)$ (or $\beta(d_i,d_j)$),  can broadly be classified as  statistical, phenomenological or empirical \citep[see][]{Lasheras2002,Liao2009}. In this study we use the phenomenological model proposed by \citet{Tsouris1994}  that leads to a ``U-shaped'' distribution. We keep in mind, however, that experiments for bubble breakup \citep{mbl_1999} have led to other possible shapes for $P(d_i,d_j)$ and that there remains considerable uncertainty about the best model to use. Here we proceed with the model of \citet{Tsouris1994} because it is based on a relatively simple physical reasoning as shown below.   

The breakup is considered to be binary, and $P(d_i,d_j)$ is formulated based on the formation energy required to form the daughter droplets of size $d_i$ and a complementary droplet to ensure volume conservation \citep{Tsouris1994}. The formation energy is proportional to the difference in initial and final surface areas according to  
\begin{equation}\label{eqn:form_energy}
E_f(d_i,d_j) = \pi\sigma\left[(d_j^3-d_i^3)^{2/3} + d_i^2 -d_j^2\right],
\end{equation}
where $\sigma$ is the interfacial tension between the dispersed and continuous phase.
It can be shown using equation $(\ref{eqn:form_energy})$ that the
breakup of a parent droplet into two equal size daughter droplets is a maximum energy process. Substituting $d_i = d_j/2^{1/3}$ in equation $(\ref{eqn:form_energy})$ we get a maximum formation energy equal to
\begin{equation}\label{eqn:max}
    E_{f,max} = \pi\sigma d_j^2(2^{1/3}-1).
\end{equation}
Equation $(\ref{eqn:form_energy})$ is minimized when $d_i=0$, that is no breakup of the parent droplet. To allow for breakup, a minimum diameter $d_{min}$ is specified and the corresponding surface formation energy is
\begin{equation}\label{eqn:min}
E_{f,min} = \pi\sigma\left[(d_j^3-d_{min}^3)^{2/3} + d_{min}^2 - d_j^2\right],    
\end{equation}
where $d_{min} = 1\;\mu\mbox{m}$ in this study.
Making the crucial assumption that the probability of breakup of a drop of size $d_j$ leading to a droplet of size such that it falls in a bin around $d_i$  decreases linearly with the required formation energy and remains within the bounds specified above, the discrete breakup probability $P(d_i,d_j)$ can be written as
\begin{equation}\label{eqn:br_prob}
    P(d_i,d_j) = \frac{[E_{f,min}+( E_{f,max} - E_f(d_i,d_j))]}{\sum_{k=1}^{j-1}\left[E_{f,min}+( E_{f,max} - E_f(d_k,d_j))\right]}.
\end{equation}
where $E_f(d_i,d_j)$ is the surface formation energy defined in equation ($\ref{eqn:form_energy}$). Also, we assume that the bin sizes are logarithmically distributed. Thus equation $(\ref{eqn:br_prob})$ is meant to model the discrete probability that a particle of size $d_j$ breaks up into a particle inside a bin centered at $d_i$ with a width of $\delta (\log(d_i))$, and its complement $d_c$ (to conserve volume). This distribution is U-shaped, with a minimum probability for the formation of two equally sized daughter droplets (when $E_f(d_i,d_j)=E_{f,max}$ which leads to a maximum of required energy), and probability maxima at the two ends (which have formation energy minima). 

\citet{Martinez2010} derived constraints that apply to the droplet size probability density function $\beta(d_i,d_j)$ for the breakup process to be volume conserving. The discrete probability of forming a droplet in bin $d_i$ must be equal to the probability of formation of the complement in bin $d_c$. The discrete breakup probability in equation (\ref{eqn:br_prob}) conserves volume, since $P(d_i,d_j) = P(d_c,d_j)$ (we note that expressing this probability in terms of a universal density $\beta(d_i,d_j)$ presents further challenges \citep{Martinez2010} that are left for future analysis, while here we use the discrete version).
 
\subsection{Model for breakup frequency}
Modeling breakup based on encounter rates of turbulent eddies and their characteristic fluctuations with droplets of a certain size has been a popular method in the literature. The phenomenological model   by 
\citet{Coulaloglou1977} postulates that a droplet in a liquid-liquid dispersion 
breaks up  when the kinetic energy transmitted from  droplet--eddy collisions exceeds the  surface energy. Many other papers have pursued this approach, mostly in the RANS context \citep[e.g][]{Narsimhan1979,Chatzi1987}. Here we follow the  approach 
of \citet{Prince1990} and \citet{Tsouris1994}, where the droplet--eddy collisions are treated akin to the of collisions between molecules in kinetic theory of gases.  The breakup frequency is computed as an integral over the product of a collision frequency and a breakup efficiency according to
\begin{equation}\label{eqn:br_freq}
g(d_i) = K \int_{0}^{d_i} \frac{\pi}{4}(d_i+d_e)^2 \, u_e(d_e) \, \Omega(d_i,d_e) \, \mathrm{d}n_e(d_e).
\end{equation}
Here $d_i$ is the diameter of droplet, $d_e$ is the eddy size, $n_e(d_e)$ is the number density of eddies of size $d_e$, $u_e(d_e)$ is the characteristic fluctuation velocity of  eddies of size $d_e$ (in a frame moving with the advection velocity caused by larger eddies), $\Omega(d_i,d_e)$ is a breakup efficiency and the integral is evaluated over all eddies, up to the size of the droplet (i.e. for $d_e$ up to $d_e=d_i$). A crucial assumption of the model is that eddies larger than the scale of the droplet are assumed to be only responsible for advection of the droplet, not contributing to collisions with the droplet that require relative velocity. One could develop a ``smoother'' model in which the lack of deformation due to eddies larger than $d_i$ is included as an additional cutoff behavior in the function $\Omega(d_i,d_e)$. Here we choose to include that cutoff behavior by following earlier work \citep{Tsouris1994,luo96} as a sharp cutoff, while lumping any possible dependencies on the exact cutoff scale into the unknown model parameter $K$, expected to be of order unity.

The number density of eddies, $n_e(d_e)$, can be estimated from the energy spectrum \citep{Azbel,Tsouris1994,Solsvik2016}, or more simply by assuming the eddies to be space-filling, i.e, $n_e(d_e) \propto d_e^{-3}$. The latter argument leads to $\mathrm{d}n_e(d_e) = C_1 d_e^{-4}\mathrm{d}(d_e)$, where $C_1$ is a constant of order $1$.  

 The eddy fluctuation velocity $u_e(r)$ written in terms of the two-point separation distance, $r$, is assumed to be expressed based on the second-order longitudinal structure function $S_2(r)$ as $u_e(r) \sim [S_2(r)]^{1/2}$. The structure function is defined according to \citep{Pope2001}:
 \begin{equation}
 S_2(r) = <\left[u_L({\bf x}+r {\bf e}_L)-u_L({\bf x}) \right]^2>,
 \end{equation}
 where $u_L$ is the fluid velocity component in the direction of unit vector ${\bf e}_L$
 and the angular brackets represent statistical averaging.
 In previous models \citep{Tsouris1994,Bandara2011,Zhao2014} a Kolmogorov scaling valid in the 
inertial range of turbulence was used for $S_2(r)$, leading to  $u_e(r) \sim (\epsilon r)^{1/3}$. However, this expression cannot be used if the size of the droplet is near the viscous range of turbulence. In order to capture both inertial and viscous ranges, as well as a smooth transition between the two ranges, we use   the approach of \citet{Batchelor1951}  with a blending function. In this approach, the structure function is given by  
\begin{equation}\label{eqn:structure_function}
S_2(r) = C_2\epsilon^{2/3}r^{2/3} \left[1+\left(\frac{r}{\gamma_2\eta}\right)^{-2}\right]^{-2/3},
\end{equation}
where $\eta$ is the Kolmogorov length scale. We choose the usual value for the Kolmogorov coefficient $C_2\approx2.1$ \citep{Pope2001}. 
The parameter $\gamma_2=(15C_2)^{3/4}\approx13$ sets the crossover scale between the inertial and viscous range.
We note that while most prior models are for use in a RANS framework using the average energy dissipation, $\epsilon$, in LES we can use a local value of the instantaneous rate of dissipation averaged over the grid scale, modeled as the subgrid-scale (SGS) dissipation rate. As a result, even though equation (\ref{eqn:structure_function}) is based on K41, in LES we only assume K41 scaling for the scales below the grid scale while intermittency in the resolved range of scales can be explicitly computed, and its effects on breakup rates taken into account in the LES model.

The breakup efficiency $\Omega(d_i,d_e)$ in equation $(\ref{eqn:br_freq})$ is the probability that a given eddy interacting with the droplet has sufficient energy to overcome the resistive forces in the system, namely surface tension and viscosity. It is assumed to be given by the usual formation potential in terms of an exponential \citep{Coulaloglou1977,Prince1990}
\begin{equation}
\Omega(d_i,d_e) = \exp\left(-\frac{E_{\sigma}(d_i)+E_{\nu}(d_i)}{E_e(d_e)}\right),
\label{eq:omega}
\end{equation}
where $E_{\sigma}(d_i)$ is resistive energy associated to a droplet of size $d_i$ due to surface tension, $E_{\nu}(d_i)$ is the viscous resistive energy and $E_e(d_e)$ is the kinetic energy of the turbulent eddy at scale $d_e$.
The resistive surface tension energy $E_{\sigma}$ is defined as the integral of the formation energy $E_f(d',d_i)$ multiplied by a measure of the breakup probability (see equation \ref{eqn:br_prob}):
\begin{equation}\label{eqn:e_sigma}
E_{\sigma}(d_i) = \int_{0}^{d_i}  c \, [E_{f,min}+( E_{f,max} - E_f(d',d_i))] \,  E_f(d',d_i) \mathrm{d}(d'),
\end{equation}
where $c$ is a normalization constant so that the integral of the probability between 0 and $d_i$ is unity. Using equation $(\ref{eqn:form_energy})$, changing the integration to $\xi' = d'/d_i$ and evaluating the integral 
numerically, we obtain 
\begin{equation}
    E_{\sigma}(d_i) = 0.0702\ \pi \sigma d_i^2,
\end{equation}
where $\sigma$ is the surface tension of the droplet. The viscous resistive energy of the droplet at steady state can be expressed as \citep{Calabrese1986,Skartlien2013,Zhao2014}
\begin{equation}
E_{\nu}(d_i) = \alpha \frac{\pi}{6}\epsilon^{1/3}d_i^{7/3}\mu_d\sqrt{\frac{\rho_c}{\rho_d}},
\end{equation}
where $\alpha\approx2$, $\rho_c$ and $\rho_d$ are the carrier and droplet phase density, and $\mu_d$ is the dynamic viscosity of the dispersed droplet phase.
For $E_e(d_e)$, the kinetic energy of the turbulent eddy, we use the longitudinal structure function $S_2$, defined in equation $(\ref{eqn:structure_function})$ applied to all three coordinate directions for the eddy fluctuation velocity, $u_e$. Assuming the volume of the eddy to be equal to that of a sphere, with density equal to the carrier phase density, the total energy contained in an eddy can be written as
\begin{equation}
E_e(d_e) = \frac{3}{2}\left(\frac{\pi}{6}\rho_c d_e^3\right) S_2(d_e).
\end{equation}

In order to formulate a parameterization for the breakup frequency, here we identify the important nondimensional numbers of the system. The breakup frequency can be re-written as a function of 
the Reynolds number ($Re_i$) based on droplet diameter and a velocity scale defined as $u_{d_i} = (\epsilon d_i)^{1/3}$, the Ohnesorge number ($Oh_i$) of the dispersed phase controlling the relative importance of viscosity to surface tension of the droplet, and the density and viscosity ratio of droplet to carrier flow fluid. These nondimensional numbers are defined below,
\begin{equation}\label{eqn:nond}
Re_i = \frac{\epsilon^{1/3} d_i^{4/3}}{\nu}\quad; \quad Oh_i = \frac{\mu_d}{\sqrt{\rho_d \sigma d_i}}\quad;\quad
      \Gamma = \frac{\mu_d}{\mu_c}\left(\frac{\rho_c}{\rho_d}\right)^{1/2}
\end{equation}
After some manipulation, equation $(\ref{eqn:br_freq})$ can be rewritten as
\begin{equation}\label{eqn:br_freq_nd}
g_i = K^{*} \frac{1}{\tau_{b,i}}\int_0^1 r_e^{-11/3} (r_e+1)^2 \left(1+\left(\frac{r_e Re_i^{3/4}}{\gamma_2}\right)^{-2}\right)^{-1/3} \Omega(Oh_i,Re_i,\Gamma;r_e) \, \mathrm{d}r_e,
\end{equation}
where, $r_e=d_e/d_i$ is the eddy size normalized by the droplet diameter, 
  $\tau_{b,i} =\epsilon^{-1/3}d_i^{2/3}$ is the breakup time scale for an eddy of size equal to that of the 
droplet as if it were in the inertial range (it does not have to be),  and $\Omega(Oh,Re,\Gamma;r_e)$ is the nondimensional breakup probability,
\begin{equation}
\Omega(Oh,Re,\Gamma;r_e) = \exp\left( -\frac{\Gamma f_2}{Re}\left[1+\left(\frac{r_e Re^{3/4}}{\gamma_2}\right)^{-2}\right]^{2/3} r_e^{-11/3}\right),
\end{equation}
where 
\begin{equation}
f_2 = 0.14\frac{\Gamma}{Re \, Oh^2} +0.583.
\end{equation}
All the nondimensional prefactors appearing when rewriting equation ($\ref{eqn:br_freq}$) have been absorbed into $K^{*}$.
Equation ($\ref{eqn:br_freq_nd}$) provides  a frequency for droplet breakup that depends on $Re_i$, $Oh_i$ and $\Gamma$. Note that if we had only considered an inertial range scaling for the eddy fluctuation velocity, we can combine $Re$ and $Oh$ into a Weber number, defined as $We = Re^2\ Oh^2$. The breakup frequency would then only depend on $We$ and $\Gamma$. The
integral represents a correction to the frequency calculated by solely considering an eddy 
equal to the size of the droplet, by evaluating the effect of collisions of eddies smaller 
than the droplet. If $d_i$ falls in the viscous range, the integral cancels the inertial range scaling assumed by the prefactor $\tau_{b,i} =\epsilon^{-1/3}d_i^{2/3}$ so that this situation is also accounted for. We note that the value of the integral $g_i \tau_{b,i}/K^*$ in equation (\ref{eqn:br_freq_nd}) will inevitably depend on the assumed maximum eddy size interacting with the droplet, which is currently taken to be exactly the droplet size $d_i$. However, if one were to take a different upper integration limit (still of order $d_i$ but not exactly $d_i$), the breakup frequency may not change much since the modified value of the integral will be largely (but not exactly) canceled when fitting the prefactor $K^∗$ to data. This behavior is demonstrated in the next section.

In LES, $g_i$ needs to be evaluated on every grid point and timestep, depending on the local Reynolds number $Re_i$ and the local rate of dissipation. Evaluating numerically the integral in equation (\ref{eqn:br_freq_nd}) at every timestep and grid point would be prohibitive in practice. Hence, we develop an empirical fit to prior numerical integrations. The speedup obtained from the fits is discussed in Appendix~$\ref{appA}$. We
develop the parameterization for a wide range of Reynolds and Ohnesorge numbers, for a fixed value of
$\Gamma$. We define $g_f(Re,Oh,\Gamma)$ as the integral in equation (\ref{eqn:br_freq_nd}), i.e. $ g_f(Re,Oh,\Gamma) = g_i(Re_i,Oh_i,\Gamma)\,\tau_{b,i}/K^*$, and evaluate it numerically for a range of $Re$ and $Oh$ values for a fixed $\Gamma$. Then, a fit can be developed in the following form,  
\begin{equation}
\log_{10}(g_{f}) = a x^b + c x^d - e,
\end{equation}
where $x=\log_{10}Re$, and $a,b,c,d,e$ are functions of $Oh$. Further details of the 
functional form of the coefficients are provided in appendix $\ref{appA}$. The final model for the breakup frequency (for a given value of $\Gamma$) thus has the form
\begin{equation}\label{eqn:br_final}
\begin{gathered}
g_i(Re_i,Oh_i;\Gamma) = \frac{ K^{*}} {\tau_{b,i}}\ 10^{G(Re_i,Oh_i)}, \\ G(Re_i,Oh_i) = a \, [\log_{10}(Re_i)]^b + c \, [\log_{10}(Re_i)]^d  - e ,
\end{gathered}
\end{equation}
where the fits for parameters $a$--$e$ as functions of $Oh$ are provided in appendix $\ref{appA}$ for a few representative values of $\Gamma$.

The breakup frequency model is thus complete except for the prefactor $K^*$ appearing in equation ($\ref{eqn:br_freq_nd}$). Its value is obtained by fitting results from an experiment (see next section), and the fitted value will then be used subsequently for comparisons with other data and for future applications.

\section{Determining the value of $K^*$}
\label{sec:fit_K}
\subsection {Wave breaking experiment of \cite{li2017}.}
\begin{figure}
\centering
\includegraphics[width=0.8\linewidth]{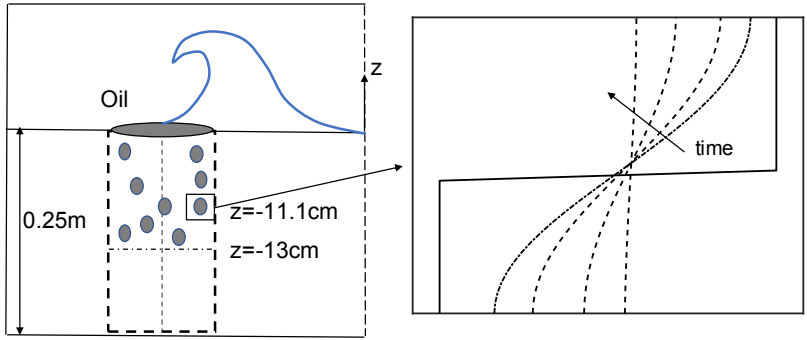}
\caption{Left: Sketch of the wave breaking experiment of \citet{li2017}. Right: schematic dependence of $\overline{n}_i(z,t)$ on time and height for a particular droplet size, starting from a step-function initial condition that is assumed to be well mixed initially down to a depth of $13\;\mbox{cm}$  below the surface and having zero concentration below. At increasing times, turbulent diffusion smooths the step and droplets rise towards the surface at different speeds depending on their size.}
\label{fig:sketch_wave}
\end{figure}

In order to fit a specific value for the parameter $K^*$  we use the data of a breaking wave experiment from \citet{li2017}.
The experiments were performed in an acrylic tank $6\;\mbox{m}$ long, $0.6\;\mbox{m}$  deep and $0.3\;\mbox{m}$ wide.
Breaking waves were generated mechanically using a piston-type wave maker consisting of a vertical plate that extends over the entire tank cross section. The tank was filled with water up to a height of $0.25\;\mbox{m}$. The wave height and characteristics were controlled by varying the frequency and stroke of the vertical plate. Oil was placed on a patch at the surface. The wave impingement and subsequent breakup processes were recorded using 3 high-speed cameras. The droplet size distribution was measured using digital inline holography. A sketch of the setup is depicted in figure $\ref{fig:sketch_wave}$. The oil patch on the surface was broken up into droplets by the plunging wave. The size distribution generated due to this process was recorded at a depth of $11.1\;\mbox{cm}$ from the free surface. A simplified sketch of the evolution of the concentration of a particular droplet size is shown in the right panel of figure $\ref{fig:sketch_wave}$. 

As shown by 
\citet{li2017}, the time evolution of the droplet size distribution at the measurement location could be represented well by a simple model that includes the effects of turbulent diffusion and droplet buoyancy only. Since the dissipation rate was quite low at the measurement location, \citet{li2017} neglected the effect of droplet breakup in their model. Consequently, for the case of crude oil with dispersants, the model under-predicted the number of smaller size droplets generated. This is due to the fact that with the effect of dispersants, the surface tension of the oil droplets was significantly lowered, resulting in droplet breakup despite of the weak turbulent dissipation rate. The Weber number ($We = 2\rho (\epsilon d)^{2/3} d/\sigma$) based on the droplet diameter for the case with dispersants is approximately $We=3$, confirming that the effects of droplet breakup are important. Our goal is to expand the model of \citet{li2017} by including breakup and select a value of $K^*$ that can achieve improved agreement with their experimental data. 

Adding the effect of breakup and following \citet{li2017} in considering only vertical diffusion and droplet rise velocity, the ensemble averaged number density $\overbar{n}_i(z,t)$ for a droplet of size $d_i$ obeys
\begin{equation}\label{eqn:diffusion_eq}
    \frac{\partial \overbar{n}_i(z,t)}{\partial t} +w_r(d_i)\frac{\partial \overbar{n}_i(z,t)}{\partial z} = D(t)\frac{\partial^2 \overbar{n}_i}{\partial z^2} + \sum_{j=i+1}^n P(d_i,d_j)g(d_j)\overbar{n}_j(z,t)-g(d_i) \overbar{n}_i(z,t) ,
\end{equation}
where $z$ is the vertical coordinate, $w_r(d_i)$ is the buoyancy-induced rise velocity of droplets of size $d_i$ and $D(t)$ is the turbulent diffusion coefficient. The daughter droplet probability function $P(d_i,d_j)$ and the breakup frequency $g$ are evaluated following the model presented in  \S \ref{sec:PBE}. The droplet rise velocity is calculated as a balance between the drag and buoyancy force acting on a droplet.
\begin{equation}
   w_r(d) =
    \begin{cases}
      w_{r,S}, & \text{if}\ Re_d < 0.2, \\
      w_{r,S}(1+0.15Re_d^{0.687})^{-1}, & 0.2 <Re_d < 750,
    \end{cases}
  \end{equation}
  where
  \begin{equation}
  w_{r,S} = \frac{(\rho_0-\rho_d)g\ d^2}{18\mu_f},
  \end{equation}
  and $Re_d = \rho_0 w_r d/\mu_f$ is the droplet rise velocity Reynolds number (not to be confused with the eddy Reynolds number $Re_i$ used to express the breakup frequency). 
  The time-dependent diffusion coefficient can be estimated using 
  $D(t)= k_D u'(t) L(t)$, where $u'(t)$ is the time-dependent turbulent fluctuation (root-mean-square) velocity as 
  measured in the experiment  and $L(t)$
  is the corresponding integral length scale, also measured. The constant $k_D$ is known to be
  between $0.23$ and $0.6$ for diffusion of droplets in isotropic turbulence 
  \citep{sato_yamamoto_1987,Gopalan2008}. We chose a value of $k_D=0.3$ in accordance with \citet{li2017} for 
  this study. \citet{li2017} fit the values of $u'(t)$ (in m/s) and $\epsilon$ (in m$^2$s$^{-3}$) with a power law in time. The data can be represented as $(\epsilon/\epsilon_0) = (t/t_0)^{p}$ and $(u'/u'_0) = (t/t_0)^{q}$ where $\epsilon_0 \approx 0.2$ $m^2 s^{-3}$, $u'_0 \approx 0.2\;\mbox{m/s}$ and $t_0 \approx 7\;\mbox{s}$.
  The exponents $p$ and $q$ can be related by  $p = 2q -1$, and the data was fitted  with $q=-0.89$ \citep{li2017}. The integral length scale $L(t)$ is then calculated as $u'(t)^3/\epsilon(t)$.
   \begin{figure}
      \hspace{-1cm}
    \centering
    \begin{subfigure}[t]{0.45\textwidth}
        \centering
        \includegraphics[width=1.15\linewidth,trim=4 4 4 4,clip]{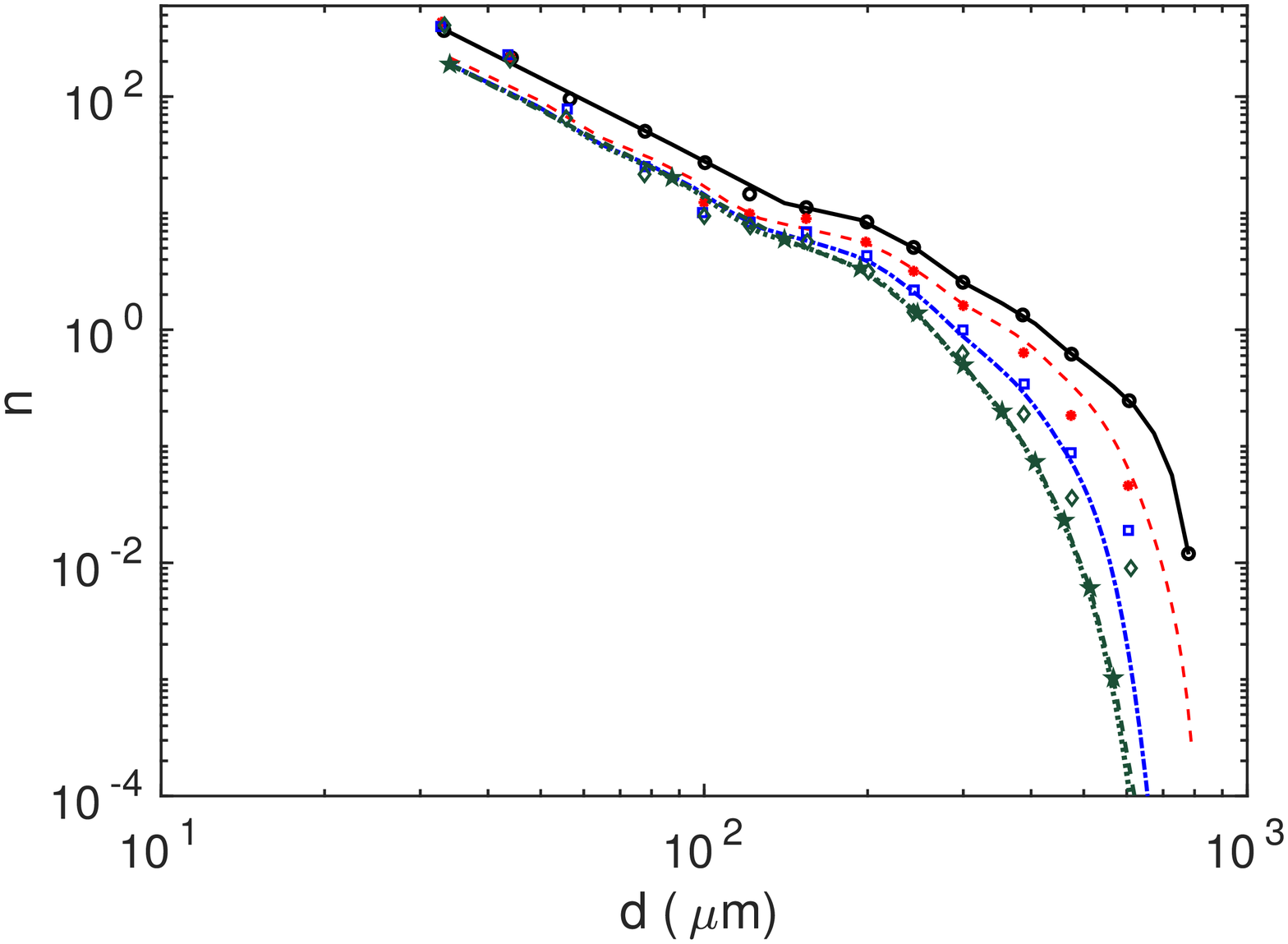}
        \caption{$ K^* =0.2$. } \label{fig:k0p2}
    \end{subfigure}
       \hspace{0.5cm}
    \begin{subfigure}[t]{0.45\textwidth}
        \centering
        \includegraphics[width=1.15\linewidth,trim=4 4 4 4,clip]{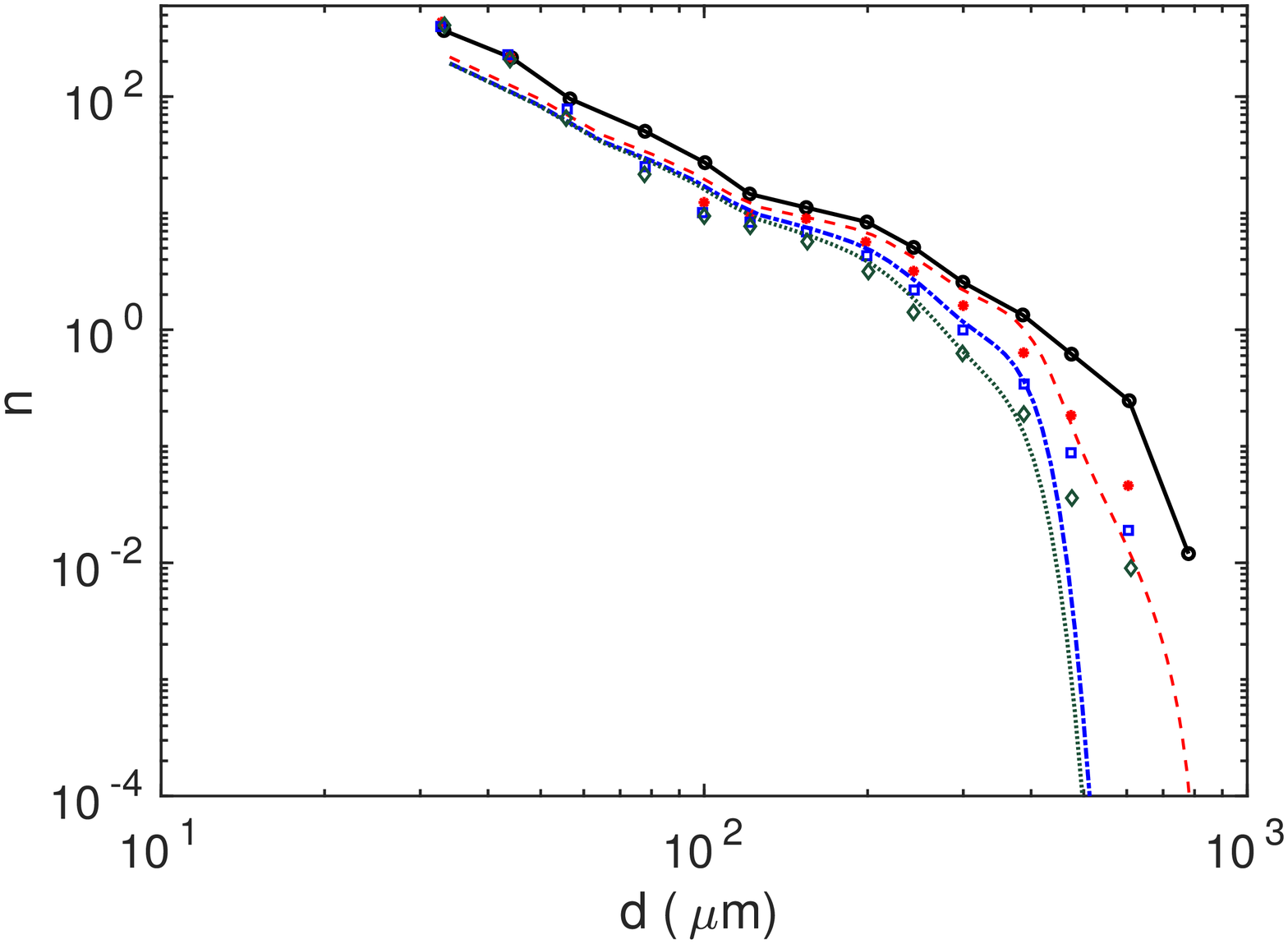} 
        \caption{$K^*=1$.} \label{fig:k0p1}
    \end{subfigure}
   \caption{Time evolution of the droplet size distribution 
   for two values of $K^*$, at the measurement location. The symbols correspond to the 
   experimental data, $t=5\;\mbox{s}$ (\protect\circleline) represents the initial condition where experiment and 
   simulation are matched, $t=15\;\mbox{s}$ ($\color{red} \ast$), 
   $t=35\;\mbox{s}$ (\protect\sq), $t=55\;\mbox{s}$
   (\protect\diamondline), while the lines correspond to the
   simulation, $t = 15 \;\mbox{s}$ (\protect\redline), $t=35\;\mbox{s}$ (\protect\blueline) and $t=55\;\mbox{s}$ (\protect\greenline). The green stars  (\protect\starline) in (\subref{fig:k0p2}) correspond to the size distribution at $t=55 \; \mbox{s}$ using $N_d=15$ bins . }
    \label{fig:dsd_wave}
\end{figure}

We solve equation $(\ref{eqn:diffusion_eq})$ numerically for the number density of the oil droplets. We discretize the size range into $N_d=70$ bins and assume that at the initial time all the concentrations are spatially homogeneous in the $z$ direction down to an initial intrusion depth of $z=13$ cm (see figure ($\ref{fig:sketch_wave}$)). 
The concentration equations are solved for each droplet size using a 
 second-order Crank--Nicholson temporal discretization method. The boundary condition at the bottom of the domain, at $z=-25\;\mbox{cm}$ is that of no flux, i.e. $\overbar{n}_i w_r-D\frac{\partial \overbar{n}_i}{\partial z}=0$. A Neumann boundary condition is applied at the top surface, i.e. $\frac{\partial \overbar{n}_i}{\partial z}=0$. We initialize the concentration of each diameter bin with the measured concentration at $z=-11.1\;\mbox{cm}$ which was recorded after $5\;\mbox{s}$ of impingement in the experiment. We integrate the model using different values of $K^*$ ranging from $0$ to $1$. 
 
 In figure $\ref{fig:dsd_wave}$ we compare the modeled size distributions (lines) to the experimental data (symbols) at various times, for $K^*=0.2$ and $K^*=1$, at the measurement location $z=-11.1\;\mbox{cm}$. Since the initial condition (black circles in figure $\ref{fig:dsd_wave}$) already includes the effects of significant initial breakup of the oil, a size distribution is already formed. Since the energy dissipation at the point of measurement, $z=-11.1\;\mbox{cm}$ is relatively low, the rate of further  breakup is not very large and thus the effect of $K^*$ in the model is subtle. Nevertheless, close inspection shows that there is too much breakup effect for the large droplets for $K^*=1$, as we can see that the number density of the larger droplets is lower than the experimental data, especially for later times. Qualitatively, it appears that $K^*=0.2$ captures the distribution slightly better, for both small and large droplets at the various time instants. We also calculate the size distribution in (\ref{eqn:diffusion_eq}) using a coarser discretization of $N_d=15$ bins. The resulting number density $\bar{n}$ is shown in figure~\ref{fig:dsd_wave}(\textit{a}) at $t=55 \; \mbox{s}$ using green stars. We see that for this coarser resolution there is a good agreement with the $N_d=70$ case.
  
  In order to make a quantitative comparison we define an error measure $\mathcal{E}$ as the integrated squared difference between the logarithmic experimental and modeled size distributions. The error is calculated for each droplet size, and integrated over all bins (the bin size varies logarithmically):
\begin{equation}\label{eqn:error}
 \mathcal{E} = \langle \,\, \sum_{i_{min}}^{i_{max}} \left[\log_{10}( \overbar{n}_{expt}(d_i))-\log_{10}(\overbar{n}_{mod}(d_i))\right]^2 \, \frac{\delta d_i}{d_i} \,\,\,\rangle_t,
\end{equation}
where $\delta d_i$ is the bin width,  ${n}_{expt}$ refers to the  experimental size distribution, and $n_{mod}$ is the modeled one. The maximum diameter at which the experimental data are reported is $d\approx 500\;\mu\mbox{m}$. Therefore, we select $i_{max}=52$ corresponding to $d=505\;\mu\mbox{m}$. And we use $i_{min}=4$ corresponding to $d=86\;\mu\mbox{m}$. The error is averaged over the three available times, $t=15\;\mbox{s}$, $35\;\mbox{s}$ and  $55\;\mbox{s}$.
  \begin{figure}
  \centering
  \includegraphics[width=0.6\textwidth]{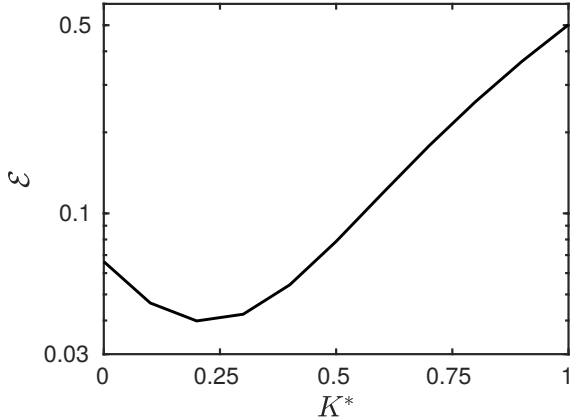}
  \caption{Average square error between predicted and measured logarithms of number densities averaged over 3 times during the evolution, at the measurement position, as a function of the breakup constant $K^{*}$ assumed in the model.}
  \label{fig:err}
  \end{figure}
  
As can be seen from figure $\ref{fig:err}$, the absolute value of the error is smallest at $K^* \approx 0.2$ and hence this value is chosen as the fitted parameter for future applications of the model.  Note that $K^*=0$ corresponds to the case without breakup. We can see from figure $\ref{fig:err}$ that the error is larger, showing the improved agreement when including the breakup term in the model.

\begin{figure}
   \hspace{-1cm}
    \centering
    \begin{subfigure}[t]{0.45\textwidth}
        \centering
        \includegraphics[width=1.15\linewidth,trim=4 4 4 4,clip]{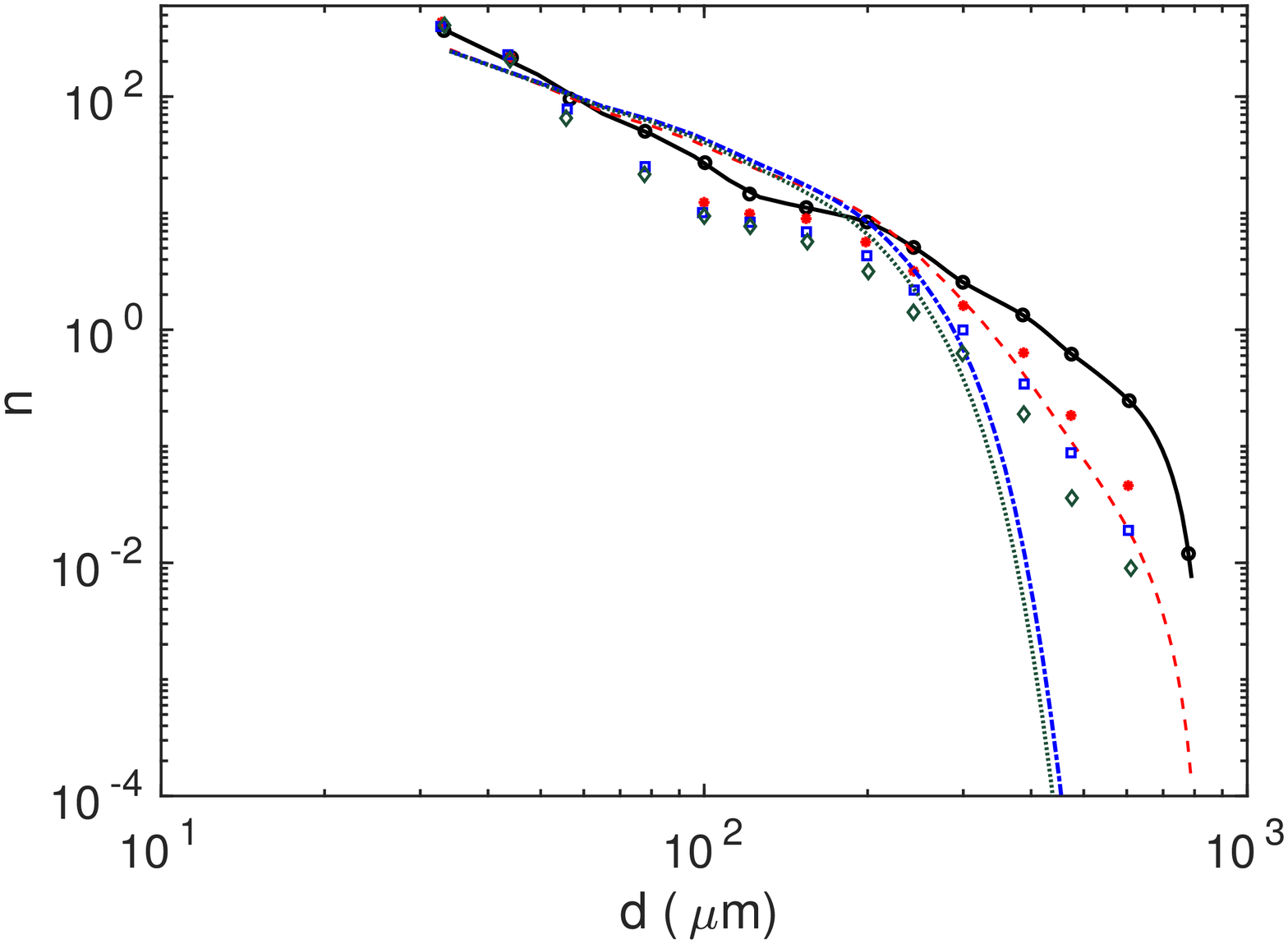}
        \caption{$ K^* =0.2$. } \label{fig:k0p2_inert}
    \end{subfigure}
    \hspace{0.5cm}
    \begin{subfigure}[t]{0.45\textwidth}
        \centering
        \includegraphics[width=1.15\linewidth,trim=4 4 4 4,clip]{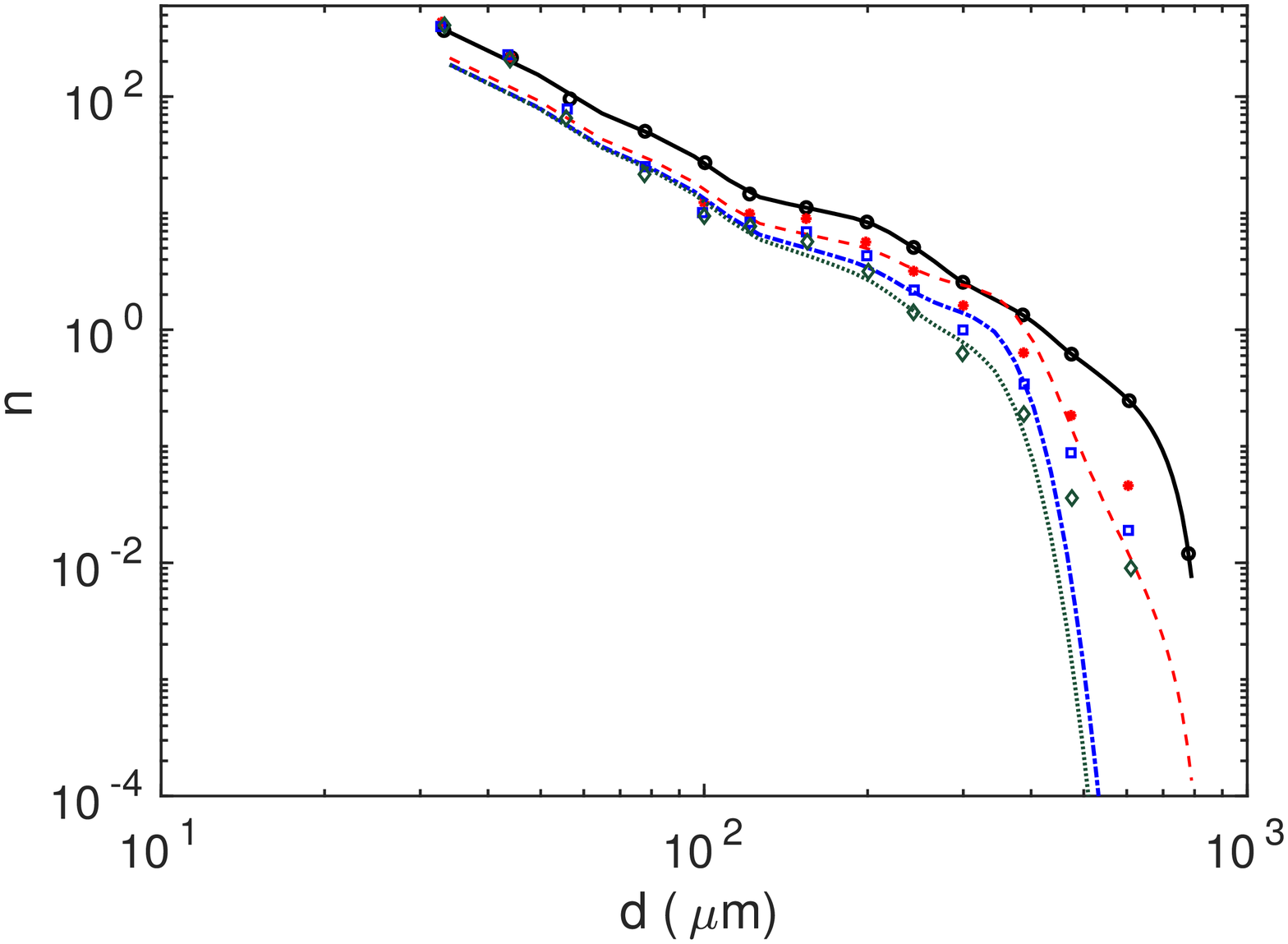}
        \caption{$K^*=0.2$.} \label{fig:k0p2n}
    \end{subfigure}
   \caption{Time evolution of the droplet size distribution 
   for $K^* =0.2$. ($\subref{fig:k0p2_inert}$) uses an inertial scaling of $u_e$. In ($\subref{fig:k0p2n}$) we plot the effect of using a normal distribution proposed by \citet{Coulaloglou1977}. The symbols correspond to the 
   experimental data, $t=5\;\mbox{s}$ (\protect\circleline) represents the initial condition where experiment and 
   simulation are matched, $t=15\;\mbox{s}$ ($\color{red} \ast$) , 
   $t=35\;\mbox{s}$ (\protect\sq), $t=55\;\mbox{s}$
   (\protect\diamondline), while the lines correspond to the
   simulation, $t = 15\;\mbox{s}$ (\protect\redline), $t=35\;\mbox{s}$ (\protect\blueline) and $t=55\;\mbox{s}$ (\protect\greenline).  }
    \label{fig:dsd_wave_inert}
\end{figure}

As stated earlier in the text, previous breakup models use an inertial-range scaling for the eddy fluctuation velocity for the entire range of scales,
\begin{equation}\label{eqn:inert_eddy}
    u_e^2 = 2.1(\epsilon d_e)^{2/3}.
\end{equation}
This approach results in overestimated velocities of eddies in the viscous range. In order to illustrate the net effects of this overestimation of turbulence at small scales in the overall model predictions, we can use equation $\ref{eqn:inert_eddy}$ in ($\ref{eqn:br_freq}$) to compute the breakup frequency using $K^*=0.2$ for this scaling of eddy velocity (through numerical integration). We solve equation ($\ref{eqn:diffusion_eq}$) and plot the resulting size distributions in figure $\ref{fig:dsd_wave_inert}$.
The left panel, figure $\ref{fig:k0p2_inert}$ compares the computed size distributions with the experimental data. We see that there is too much breakup effect resulting in too few of the larger droplets and an increase in the concentration of the intermediate-size droplets. In order to obtain a better agreement with the experiment we would need to set $K^*\sim O(10^{-2})$. Thus, we conclude that in order to maintain reasonable range of value for the parameter $K^*$ (which is expected to be of order unity), it is important to capture both the inertial and viscous range scalings as in equation ($\ref{eqn:structure_function}$) for the eddy fluctuation velocity.  

We can also study the effect of using a different breakup probability model $\beta(d_i,d_j)$. We use a truncated normal distribution of \citet{Coulaloglou1977} for simplicity. In this case it is assumed that the daughter droplet sizes for a parent drop of diameter $d_j$ are normally distributed about a mean value, $\bar{d} = d_j/2^{1/3}$, i.e.
\begin{equation}
    \beta(d_i,d_j) = \frac{1}{s \sqrt{2\pi}}\exp{\left[- \frac{(d_i-\overbar{d})^2}{2s^2}\right]},
\end{equation}
where $s = d_j/(3\times \ 2^{1/3})$. This 
gives us a maximum probability for equal volume 
breakup. We plot the resulting size distributions 
for $K^*=0.2$ in figure $\ref{fig:k0p2n}$. We observe a bump in the size distribution at $d 
\approx 300\;\mu\mbox{m}$ and a more rapid cutoff of the large droplets as compared to figure~\ref{fig:dsd_wave}(\textit{a}). Additionally, we do not find
an optimum value of $K^*$ that minimizes equation ($\ref{eqn:error}$), i.e. the error grows with increasing $K^*$. Hence, we may conclude that the form of the particular droplet breakup probability distribution is also important, though it plays a weaker role as compared to the effect of including the viscous range for this particular wave breaking case.

\begin{figure}
  \hspace{-1cm}
    \centering
    \begin{subfigure}[t]{0.45\textwidth}
        \centering
        \includegraphics[width=1.15\linewidth,trim=4 4 4 4,clip]{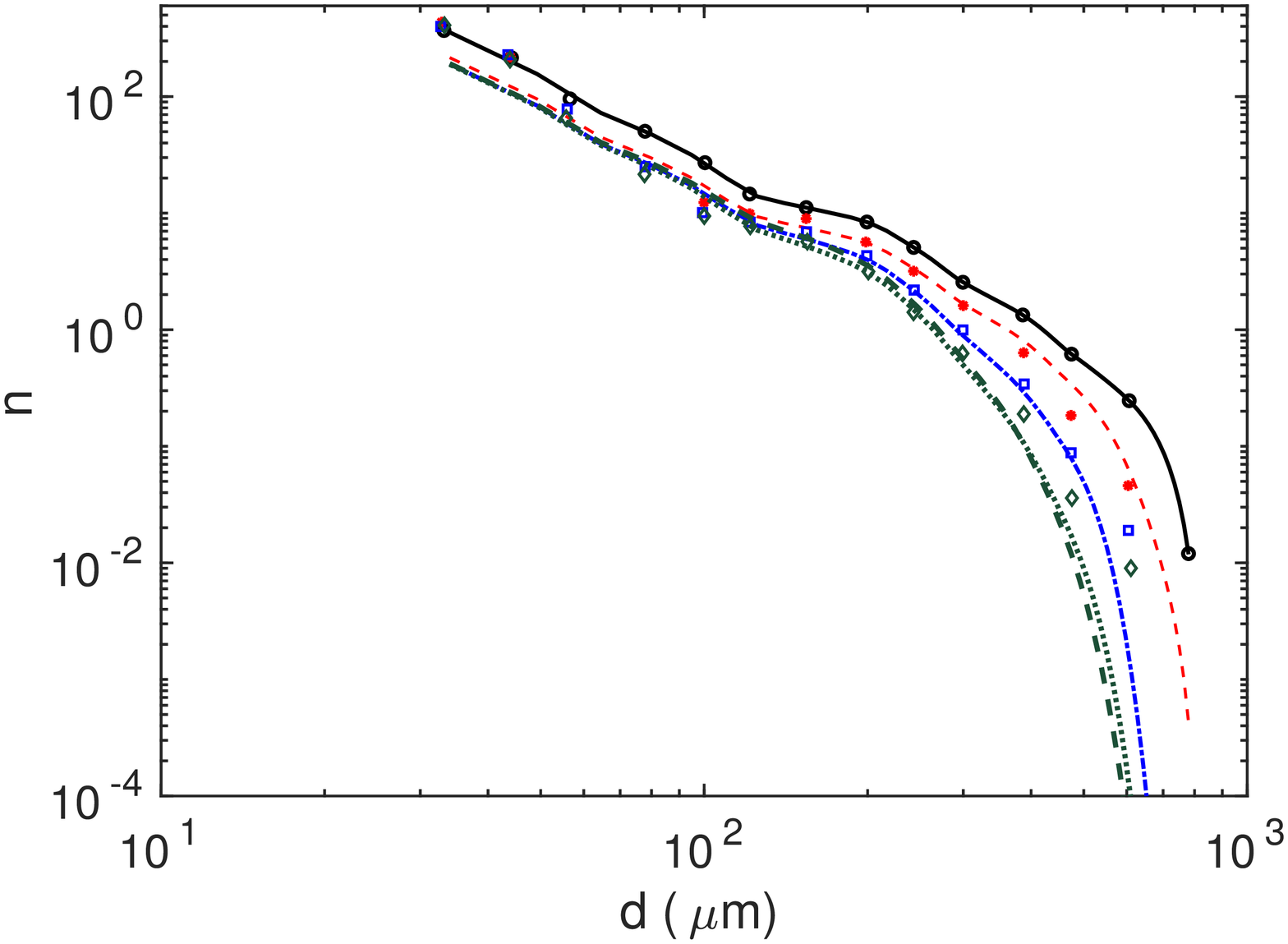}
        \caption{ } \label{fig:dsdc11p2}
    \end{subfigure}
    \hspace{0.5cm}
    \begin{subfigure}[t]{0.45\textwidth}
        \centering
        \includegraphics[width=1.15\linewidth,trim=4 4 4 4,clip]{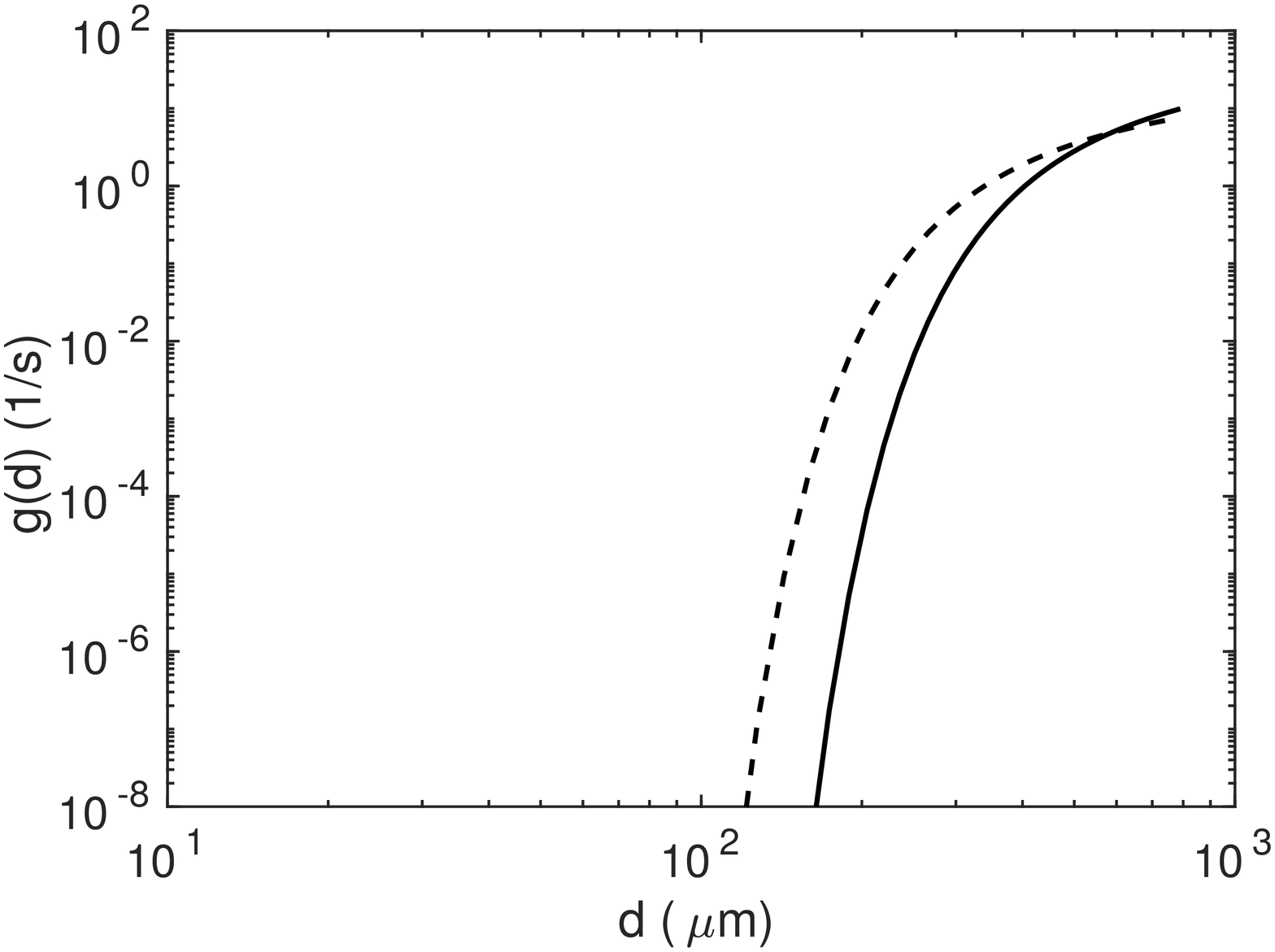}
        \caption{ } \label{fig:gcomp}
    \end{subfigure}
  \caption{We depict the effect of changing the maximum diameter size on the size distribution and breakage frequency. In ($\subref{fig:dsdc11p2}$) the symbols correspond to the 
  experimental data, $t=5\;\mbox{s}$ (\protect\circleline) represents the initial condition where experiment and 
  simulation are matched, $t=15\;\mbox{s}$ ($\color{red} \ast$) , 
  $t=35\;\mbox{s}$ (\protect\sq), $t=55\;\mbox{s}$
  (\protect\diamondline), while the lines correspond to the
  simulation using $K^*=0.2$, $t = 15\;\mbox{s}$ (\protect\redline), $t=35\;\mbox{s}$ (\protect\blueline) and $t=55\;\mbox{s}$ (\protect\greenline). The green dashed line (\protect\greendashline) corresponds to the case where $d_{e,max}=1.2 d_i$ and $K^*=0.1$ was used to calculate $g(d_i)$. In ($\subref{fig:gcomp}$)  we compare the resulting breakup frequency for the two cases, $d_{e,max}=1.2d_i$(\protect\blackdashline) and $d_{e,max}=d_i$(\protect\blackline).}
    \label{fig:dsd_eddy_max}
\end{figure}

We demonstrate the effect of an increase in the assumed maximum eddy size allowed to break the droplets in figure \ref{fig:dsd_eddy_max}. We find that for a maximum eddy size 20\% larger than the droplet size ($d_{e,max}=1.2 d_i$), the fitted prefactor is reduced to $K^*=0.1$. Using
this value of $K^*$ we see from figure \ref{fig:dsdc11p2} that the steady state size distribution is in good agreement with case
where $d_{e,max} = d_i$ and $K^* = 0.2$. Though clearly the resulting breakup frequency shows dependence on the upper cutoff of the integral.
   
\subsection{Breakup of oil in a turbulent jet \citep{eastwood2004}.}
In order to begin testing the model when 
applied to a system with different flow properties than the case for which $K^*$ was fitted, we consider the  experiment by \citet{eastwood2004}. Oil droplets of
varying density, viscosity and interfacial tension are injected continuously at the centerline in the fully developed region of a turbulent water jet. The downstream evolution of the centerline velocity and dissipation rate was well characterized and found to obey classic scalings of a turbulent round jet,
\begin{equation}\label{eqn:dissip_east}
   \frac{U_0}{U(x)} = \frac{1}{C_u} \left(\frac{x}{D_j} - \frac{x_0}{D_j}\right), \,\,\,\,\,\,\,\,\, \frac{\epsilon D_j}{U_0^3} = C \left(\frac{x}{D_j} - \frac{x_0}{D_j}\right)^{-4},
\end{equation}
where $C_u = 4.08$ and $C=36$ are empirical constants. The virtual origin $X_0/D_j=5.47$ was found by fitting the experimental data with equation (\ref{eqn:dissip_east}).
The breakup and downstream evolution of oil droplets were tracked using digital image processing techniques \citep{eastwood2004}. The authors defined a characteristic droplet size $d_{max}$ whose number density $n(d_{max})$ can only change due to its own breakup but cannot increase due to breakup of other (larger) droplets. This condition isolates the effect of the breakup frequency on the evolution of the number density. Mathematically the evolution of the size distribution can be tracked using equation (\ref{eqn:pbe_discrete}), where for the size $d_{max}$ we can drop the first term in equation (\ref{eqn:break_source}). Additionally for
the quasi-one-dimensional steady state jet flow considered, we can write the PBE for the largest size as
\begin{equation}\label{eqn:numd_eastwood}
    \nabla_x \cdot [ \vv(d_{max},\xx) n(d_{max},\xx)  ] = -g(d_{max},\xx) n(d_{max},\xx),
\end{equation}
where $n(d_{max},\xx) = N(d_{max},\xx)/V_w$, with $V_w$ being the volume of the interrogation window. The droplet velocity $\vv(d_{max},\xx)$ can be approximated by the local mean velocity of the turbulent jet, $U(x)$. Equation (\ref{eqn:numd_eastwood}) can then be written for the number of droplets $N(d_{max},\xx)$ as,
\begin{equation}\label{eqn:num_east}
     \frac{\mathrm{d}}{\mathrm{d}x}[U(x) \, N(d_{max},x) ]=-N(d_{max},x)g(d_{max},x).
\end{equation}

\begin{table}
\centering
\begin{tabular}{ c  c  c  c c c } 
 $Oil$ & $d_{max}$(mm) & $We$ & $\rho_d$ (kg/m$^3$) & $\mu_d$ (Pa s) & $\sigma$ (N/m)  \\ [0.5ex] 
 \midrule
  Heptane & 1.9  & 10 & $684$ & $5.00\times 10^{-4}$ & $4.8 \times 10^{-2}$ \\ 
   Olive Oil & 1.91 & 30 & $881$ & $7.19\times 10^{-2}$ & $2.0 \times 10^{-2}$ \\ 
      10 cSt silicone oil & 1.92 & 20 & $936$ & $9.70\times 10^{-3}$ & $3.5 \times 10^{-2}$ \\ 
        50 cSt silicone oil & 1.81 & 20 & $970$ & $5.09\times 10^{-2}$ & $3.7 \times 10^{-2}$ \\ 
\end{tabular}
\caption{Dispersed fluid properties.}
\label{tab:eastoil_prop}
\end{table}

The maximum diameter 
$d_{max}$ represented the size where at least $80\%$ of the volume of the distribution was contained in droplets smaller than $d_{max}$. The overall decay of $N(d_{max})$ with downstream distance was found to be similar when this criteria was enforced, thereby ensuring that the evolution of the largest size class is being captured. In order to validate our model with the experiments, we solve equation (\ref{eqn:num_east}) using a fourth-order explicit Runge-Kutta method (ode45 in MATLAB). Equation (\ref{eqn:br_freq_nd}) is used to calculate the breakup frequency $g(d_{max},x)$ with $K^*=0.2$ for all the oils considered.
A summary of the 
physical properties of the different oils used for the simulation along with the corresponding size $d_{max}$ is provided in Table \ref{tab:eastoil_prop}.
\begin{figure}
   \hspace{-1cm}
    \centering
\includegraphics[width=0.8\linewidth]{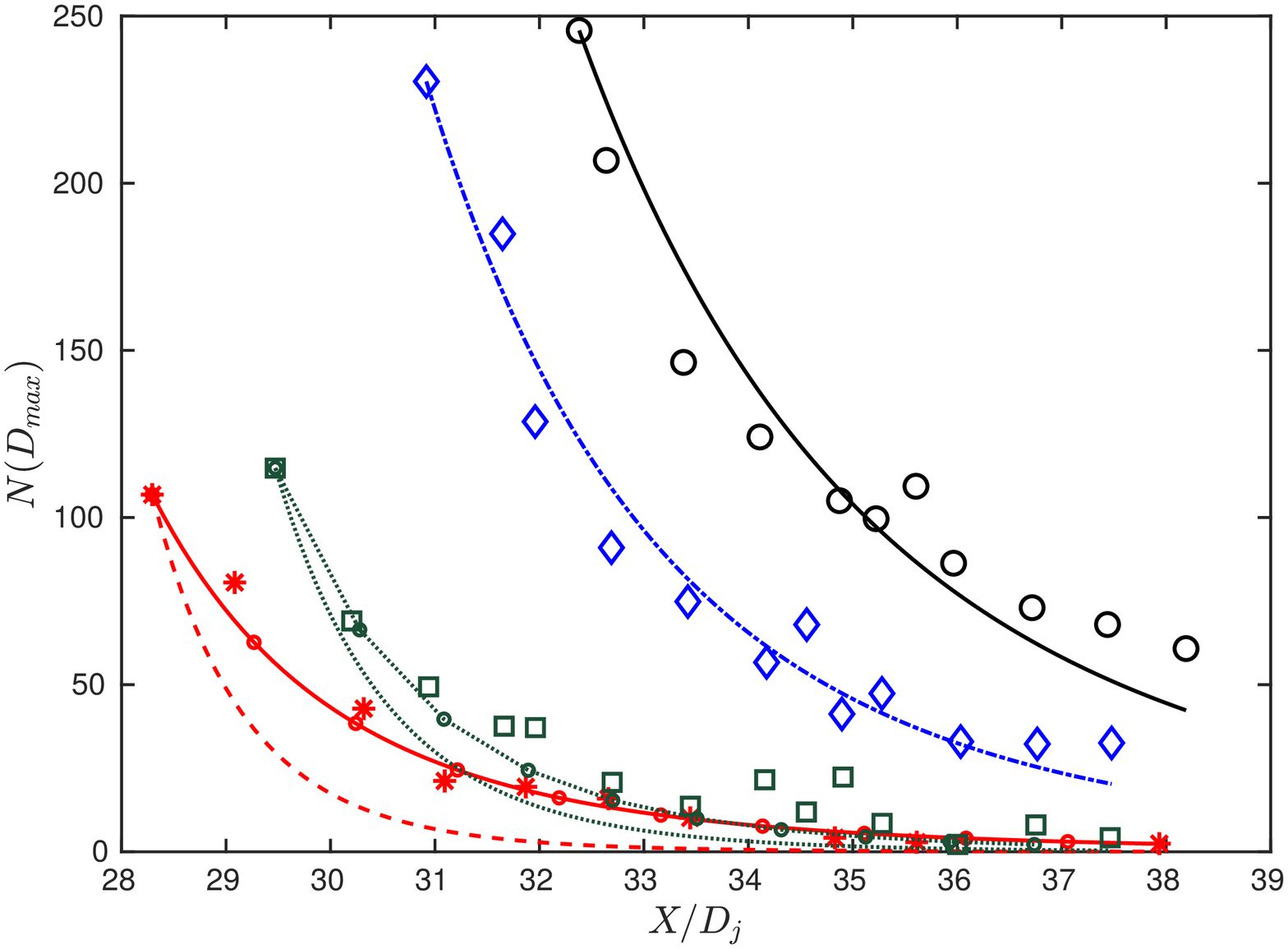}
   \caption{Evolution of the $N(d_{max})$ with downstream distance for the four different oils in \ref{tab:eastoil_prop}. The symbols correspond to the experimental data of the different oils, 50 cSt Silicone oil (\protect\figone), Olive oil (\protect\figtwo), 10 cSt Silicone oil (\protect\figthree) and Heptane ($\color{red} \ast$). The different lines correspond to the model, 50 cSt Silicone oil (\protect\blackline), Olive oil (\protect\blueline), 10 cSt Silicone oil (\protect\greenline) and Heptane (\protect\redline) with $K^*=0.2$.  Also shown are the model for Heptane (\protect\circlelinered) with $K^* =0.1$ and 10 cSt Silicone Oil (\protect\circlelinegrn) with $K^*=0.15$.   }
    \label{fig:east_comp}
\end{figure}

We compare the model predictions with the experimental data in figure \ref{fig:east_comp}. We see that the model does a good job of capturing the decay of the number of droplets for the 50 cSt silicone oil and the olive oil case. For heptane and 10 cSt silicone oil the predicted decay is too rapid and $K^*=0.2$ appears not to be the optimal value, while $K^* =0.1$ and $K^*=0.15$ are seen to give better agreement with the data. We have not found any obvious parameter dependencies that could explain the different $K^*$. So based on the argument that $K^* =0.2$ gives a good match for a majority of the applications considered, we proceed to use this value for the LES test case in the next section.

\section{Application to Large Eddy Simulation of Jet in Crossflow} \label{sec:LES_model}
Large eddy simulations capture the large and intermediate scale turbulent motions (depending on the grid resolution), and only require modeling of the unresolved subgrid-scale  turbulence effects. While the cost of LES is higher than Reynolds-Averaged Navier--Stokes simulations, LES provides the ability to resolve unsteady spatially fluctuating phenomena at least down to scales on the order of the grid scale. In this work we will highlight this strength of LES in the context of simulations of the evolution and transport of size distributions of polydispersed oil droplets in water.  We first describe the LES model equations, present the simulation setup for LES modeling of jet in crossflow and then present results and comparisons with experimental data. 
  
\subsection{LES equations for Eulerian description of scalar transport}
We adapt the  code used by \citet{Yang2014} and \citet{Yang2015} for simulations of hydrocarbon plume dispersion in ocean turbulence to simulate here a turbulent round jet with an imposed crossflow and with oil droplets being released near the source of the jet. Let $\xx = (x,y,z)$ with $x$ and $y$ being the horizontal coordinates and $z$ the vertical direction, and let $\uu = (u,v,w)$ be the corresponding velocity components. The jet and surrounding fluid are governed by the three--dimensional incompressible filtered Navier--Stokes equations with a Boussinesq approximation for buoyancy effects:
\begin{equation}\label{eqn:div}
\nabla \cdot \tuu =0,
\end{equation}
\begin{equation}\label{eqn:Navier_stokes}
\frac{\partial \tuu}{\partial t} + \tuu \cdot\nabla\tuu = -\frac{1}{\rho_c}\nabla\tp - \nabla\cdot \ttau^d +  \left(1 - \frac{\rho_d}{\rho_c}\right)\sum_i(V_{d,i}\tn_i)g\ee_3.
\end{equation}
\begin{equation}\label{eqn:conc}
\frac{\partial \tn_i}{\partial t} + \nabla \cdot (\tvv_i\tn_i) + \nabla \cdot \ppi_{i} = \widetilde{S_{b,i}}, \,\,\,\, i=1,2.. N_d.
\end{equation}
A tilde denotes a variable resolved on the LES grid, $\tuu$ is the filtered fluid velocity, $\rho_d$ is the density of the 
droplet, $\rho_c$ is the carrier fluid (seawater) density, $V_{d,i}=\pi d_i^3/6$ is the volume of a spherical 
oil droplet of size $d_i$, $\ttau = (\widetilde{\uu\uu} - \tuu\tuu)$ is the subgrid-scale stress tensor with deviatoric part, $\ttau^d=\ttau-[tr(\ttau)/3]\ii$ where $\ii$ 
is the identity tensor, $\tp=\tilde{p}/\rho_r + tr(\ttau)/3 + |\tuu|^2/2$ is the pseudo-pressure, with $\tilde{p}$ being the resolved dynamic pressure, $\tn_i$ is the resolved 
number density of the droplet of size $d_i$, and $\ee_3$ is the unit vector in the vertical direction. 
The filtered version of the transport equation for the number density $\tilde{n}_i(\xx,t;d_i)$ is given by equation (\ref{eqn:conc}). The term $\ppi_{i} = (\widetilde{\vv_in_i} - \tvv_i\tn_i)$ is the subgrid-scale concentration flux of oil droplets of size $d_i$.

Closure for the SGS stress tensor $\ttau^d$ is obtained from the Lilly-Smagorinsky eddy viscosity model \citep{SMAGORINSKY1963}, $\tau^{d}_{ij} = 
-2\nu_{\tau}\tS_{ij} = -2(c_s\Delta)^2|\tS|\tS_{ij}$, where $\tS_{ij}=(\partial \tu_i/\partial 
x_j + \partial \tu_j/\partial x_i)/2$ is the resolved strain rate tensor, $\nu_{\tau}$ is the 
SGS eddy viscosity, and $\Delta$ is the grid (filter) scale. The Smagorinsky coefficient 
$c_s$ is determined dynamically during the simulation using the Lagrangian averaging scale-dependent dynamic (LASD) SGS model \citep{Bou-Zeid}, which accounts for spatial 
inhomogeneity. The SGS scalar flux $\ppi_{i}$ is modeled using an eddy-diffusion SGS model. We use the approach of prescribing a turbulent Schmidt and Prandtl number, $Pr_{\tau} = Sc_{\tau} =0.4 $ \citep{Yang2016}. The SGS flux can then be parameterized as $\pi_{n,i} = -(\nu_{\tau}/Sc_{\tau})\nabla\tn_{i}$ 

The LES code with the LASD model has been used in a number of prior LES studies \citep[see][]{Porte-Agel2000,Tseng2006,Yang2014}. With the evolution of 
oil droplet concentrations being simulated, their effects on the fluid velocity field are 
modeled and implemented in  $(\ref{eqn:Navier_stokes})$ as a buoyancy force term (the last term on the right-hand side of the equation) using the 
Boussinesq approximation. A basic assumption for treating the oil droplets as an active 
scalar field being dispersed by the fluid motion is that the volume and mass fractions of the 
oil droplets are small within a computational grid cell.
   
The droplet transport velocity $\tvv_i$ is calculated by an expansion in the droplet time 
scale $\tau_{d,i}=(\rho_d+\rho_c/2)d_i^2/(18\mu_f)$ \citep{Ferry2001}. The 
expansion is valid when $\tau_{d,i}$ is much smaller than the resolved fluid time scales, 
which requires us to have a grid Stokes number $St_{\Delta,i} = \tau_{d,i}/\tau_{\Delta} \ll 
1$, where $\tau_{\Delta}$ is the turbulent eddy turnover time at scale $\Delta$. 
The transport velocity of droplets of size $d_i$, $\tvv_i$,  is given by \citep{Ferry2001}
\begin{equation}\label{eqn:rise_vel}
\tvv_i = \tuu + w_{r,i}\ee_3 + (R-1)\tau_{d,i}\left(\frac{D\tuu}{D t} + \nabla \cdot \ttau\right) +O(\tau_{d,i}^{3/2}),
\end{equation}
where $w_{r,i}$ is the droplet terminal (rise) velocity, $\ee_3$ is the unit vector in the 
vertical direction, and $R = 3\rho_c/(2\rho_{d}+\rho_c)$ is the acceleration parameter. A more 
detailed discussion of the droplet rise velocity in equation $(\ref{eqn:rise_vel})$ can be found in \citet{Yang2016}.

 The term $\widetilde{S_{b,i}}$ in equation ($\ref{eqn:conc}$) represents the rate of change of droplet number density due to breakup and is modeled using equations ($\ref{eqn:br_prob}$) and ($\ref{eqn:br_final}$)  described in \S\ref{sec:PBE}. In implementing the model, when evaluating the filtered source term, we use the filtered parameters (e.g. grid-scale dissipation rate, etc.), that is to say, we assume $\widetilde{g n} \approx \tilde{g}\tilde{n}$, and further that $\widetilde{g(\epsilon,..)} \approx  g(\tilde{\epsilon},..)$. This means that we neglect the subgrid correlations between locally fluctuating dissipation rates at scales smaller than the grid scale and subgrid fluctuations in the concentration field. The neglect of such subgrid-scale contributions to the modeled source terms must be kept in mind, especially for applications at very high Reynolds numbers when the Kolmogorov scale is much smaller than the grid scale where further refinements and new subgrid models may be required.
 
 The breakup rate $g_i$ is evaluated using the fits that depend on the local Reynolds number expressed in terms of the local rate of dissipation. From the SGS model, the local rate of dissipation at the LES grid scale is given by
 \begin{equation}
 \epsilon({\xx},t) = 2 (c_s\Delta)^2|\tS| \tilde{S}_{ij}\tilde{S}_{ij}.
 \label{eq:dissipLES}
 \end{equation}

In order to capture a range of sizes the number density is discretized into $N_d = 
15$ bins for droplets between $d_1 =  20\;\mu\mbox{m}$ up to $d_{N_d} =  1\;\mbox{mm}$. 
The droplet size range is discretized according to
\begin{equation}
    d_{i+1} = 2^p d_i, \quad i=1,2,...,N_d ,
\end{equation}
where, $p=0.403$. We 
then solve $N_d$ separate transport equations for the number densities $\tilde{n}_i(\xx,t;d_i)$  with $i=1,2,...,N_d$.

The equations ($\ref{eqn:div}$) and ($\ref{eqn:Navier_stokes}$) are discretized using a pseudo-spectral method on a collocated grid in the horizontal directions and using centered finite differencing on a staggered grid in the vertical direction \citep{Albertson1999}. The velocity field is advanced using a 
fractional-step method with a second-order Adams--Bashforth scheme for the time integration. 
The transport equations for the droplet number densities, equation ($\ref{eqn:conc}$),
are discretized as in \citet{Chamecki2008}, \citet{Yang2014} and \citet{Yang2015},   by a finite-volume algorithm with a bounded third-order upwind scheme for the advection term \citep{Gaskell1988}. This approach prevents the appearance of negative local concentrations that tend to occur when using spectral methods. Information between the finite-volume and pseudo-spectral grids is exchanged using a conservative interpolation scheme developed by 
\citet{Chamecki2008}.

\subsection{Simulation Setup}

\begin{figure}
\centering
\includegraphics[width=0.8\textwidth]{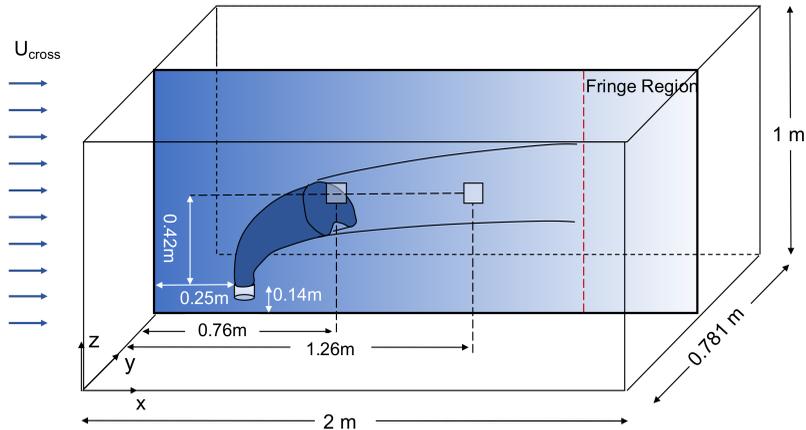}
\caption{Sketch of the simulation domain and dimensions.}
\label{fig:sketch_sim}
\end{figure}

A sketch of the simulation domain is shown in figure $\ref{fig:sketch_sim}$. We simulate a turbulent jet with imposed crossflow aiming to reproduce the experiments of \citet{Murphy2016}, who studied a turbulent oil jet in crossflow and measured droplet size distributions using a submersible digital inline holography technique. 
The experiments were carried out in a $2.5\;\mbox{m} \times 0.9\;\mbox{m} \times 0.9\;\mbox{m}$ acrylic tank. The injection nozzle was connected to a carriage, driven by a stepper motor to generate desired crossflow speeds. The injection nozzle had a $D_{expt}=4\;\mbox{mm}$ diameter orifice and was located at a distance of $0.14\;\mbox{m}$ from the bottom of the tank. The oil was injected at a flow rate of $Q=1.9\;\mbox{L/min}$, (i.e. an injection velocity of $U_{j,expt} = 2.5 \;\mbox{m/s}$) and the carriage was towed at a speed of $U_c = 0.15 \;\mbox{m/s}$. In the experiments performed, the number of droplets was measured using a holocam fixed at the center of the tank in the horizontal plane, and at a distance of $0.47\;\mbox{m}$ from the nozzle exit. It thus sampled different sections of the jet in the cross stream direction in the course of its evolution. Numbers of droplets in various size bins were measured at two times, at $t=3.4\;\mbox{s}$ and $t=6.9\;\mbox{s}$. Additionally, the total oil distribution calculated by summing over five time measurements was recorded. As the nozzle in the experiments moved with a constant speed of $0.15\;\mbox{m/s}$, this corresponds to a downstream location of measurement at $x=0.76\;\mbox{m}$ and $x=1.3\;\mbox{m}$ respectively, in a frame moving with the jet nozzle (as will be done in the simulation). The total oil concentration at the measurement location height corresponds to the sum of the number of droplets measured as a function of downstream distance. In the experiments, the droplet size distributions were measured in 3 realizations of the experiments, hence the resulting size distribution and droplet concentrations were not fully statistically converged but the shape of the size distribution (relative size distribution) was well captured in the measurements.

As shown in  figure $\ref{fig:sketch_sim}$, the simulations are performed in a rectangular box of size $(L_x,L_y,L_z) = (2,0.781,1)\;\mbox{m}$ in which the jet is stationary but a constant inflow (crossflow) velocity is prescribed. 
The domain size is chosen to mimic the experimental setup and the length is sufficiently large to capture the complete turning of the jet. The crossflow is imposed along the $x$ direction while the jet is pointed in the $z$ direction. In order to handle the inflow and outflow conditions at the two boundaries in the $x$ direction within the code that uses a Fourier-series based pseudo-spectral method, we specify a fringe zone that starts at $x=1.666\;\mbox{m}$, which damps out the velocity towards the inflow value.
 The simulations use a grid with $N_x \times N_y \times N_z = 384 \times 150 \times 384$ points for spatial discretization, 
and a timestep $\Delta t=0.0002\;\mbox{s}$ for time integration. The resolution in the horizontal directions is $\Delta x = \Delta y = 5.2\;\mbox{mm}$, and is $\Delta z = 2.6\;\mbox{mm}$ in the vertical direction for the finite difference method. Since the latter can be regarded as requiring about twice as much resolution for the same accuracy, we regard the overall numerical resolution to be about $5.2\;\mbox{mm}$.

The injected jet is modeled in the LES using a locally applied vertically upward pointing body force, since at the LES resolution used in the simulation it is not possible to resolve the small-scale features of the injection nozzle. 
The applied force is spatially smoothed in a region over three grid points in $x,y$ and $z$  using a super-Gaussian smoother (of order 5 and width $\sigma_G = \Delta x$) centered at $(x,y,z) = (0.25,0.39,0.85)\;\mbox{m}$.

The resulting injection velocity is controlled by the strength of the imposed body force. Since the details of the nozzle cannot be resolved, the body force is applied at a location downstream of the real nozzle, where the jet in the experiment is expected to have grown to a scale at which the numerical grid is sufficient to resolve at least the mean velocity profile. Using a classical round jet without crossflow for calibration,  the force is chosen such that the resulting centerline velocity in the LES matches the mean centerline velocity expected for the experiment at that location. More details are provided in Appendix B.  The location where the body force is applied is found to be $38.2\;\mbox{mm}$ downstream of the experimental nozzle's virtual origin (see Appendix B), while the injection velocity in the simulation is determined to be $U_{j,sim} = 1.6\;\mbox{m/s}$, i.e. lower than that in the experiment owing to the fact that the centerline velocity has already decayed at the simulated injection point.
A uniform crossflow of $U_{cross}=0.15\;\mbox{m/s}$ is imposed at the inflow boundary in order to simulate a jet in crossflow.
The droplet number density fields are initialized to zero everywhere. Oil droplets of size $d =1\;\mbox{mm}$ are injected at the jet source after a delay of $1\;\mbox{s}$ to allow the flow to be established. The number density transport equation for the bin corresponding to the largest droplets ($i=N_d$ with $d_i = 1\;\mbox{mm}$) contains a source term on the RHS which represents injection with a specified volume flow rate that matches the experimental value of Q= $1.9\;\mbox{L/min}$ as in \citet{Murphy2016}. The source is distributed over two grid points in the z direction with weights $0.25$ and $0.75$ and over three grid points in the x and y directions centered at $x = 0.245\;\mbox{m}$, $y=0.385\;\mbox{m}$ with weights $0.292$ for the center and $0.177$ for the neighbouring points.
The physical properties of the oil and the simulation parameters are given in tables $\ref{tab:oil_prop}$ and $\ref{tab:Sim_param}$. The Weber number ($We$) has been calculated using the near nozzle dissipation $\langle\epsilon\rangle = 30\;\mbox{m}^2\mbox{s}^{3}$ and the injection droplet size $d=1\;\mbox{mm}$.
 
\begin{table}
\centering
\begin{tabular}{ c  c  c  c c} 
 $\rho_d$ (kg/m$^3$) & $\nu_d$ (m$^2$/s) & $\sigma$ (N/m) & $\rho_c$ (kg/m$^3$) & $\Gamma$ \\ [0.5ex] 
 \midrule
   $864$ & $1.02\times 10^{-5}$ & $1.9 \times 10^{-2}$ & $1018.3$ & 10.5\\ 
\end{tabular}
\caption{Physical properties of fluids used in the simulation.}
\label{tab:oil_prop}
\end{table}

\begin{table}
\centering
\begin{tabular}{ c c  c c c } 
 \multicolumn{1}{p{3cm}}{\centering $Re=\frac{U_JD_J}{\nu_c}$}
 & \multicolumn{1}{p{3cm}}{\centering $We=\frac{2\rho_c\langle\epsilon\rangle^{2/3}d^{5/3}}{\sigma}$} & $U_J$ (m/s) & $D_J$ (mm) &$D_{expt}$ (mm)\\ [1ex]
\midrule
   12500 & 10 & 1.6 & 7.8  & 4\\ 
\end{tabular}
\caption{Simulation parameters.}
\label{tab:Sim_param}
\end{table}

The jet in crossflow is simulated for a total of $26\;\mbox{s}$, corresponding to $13\times 10^4$ timesteps. The oil droplets are released after $t=1\;\mbox{s}$ giving sufficient time for the jet to become fully turbulent. Starting from $t=12\;\mbox{s}$, 350 three-dimensional snapshots of the entire simulation domain are recorded for statistical analysis with an interval of $0.04\;\mbox{s}$ (200 timesteps) between each snapshot. 

\subsection{Results}\label{sub:results}
 \begin{figure}
    \centering
    \begin{subfigure}[t]{0.45\textwidth}
        \centering
        \includegraphics[width=1.2\linewidth,trim=4 4 4 4,clip]{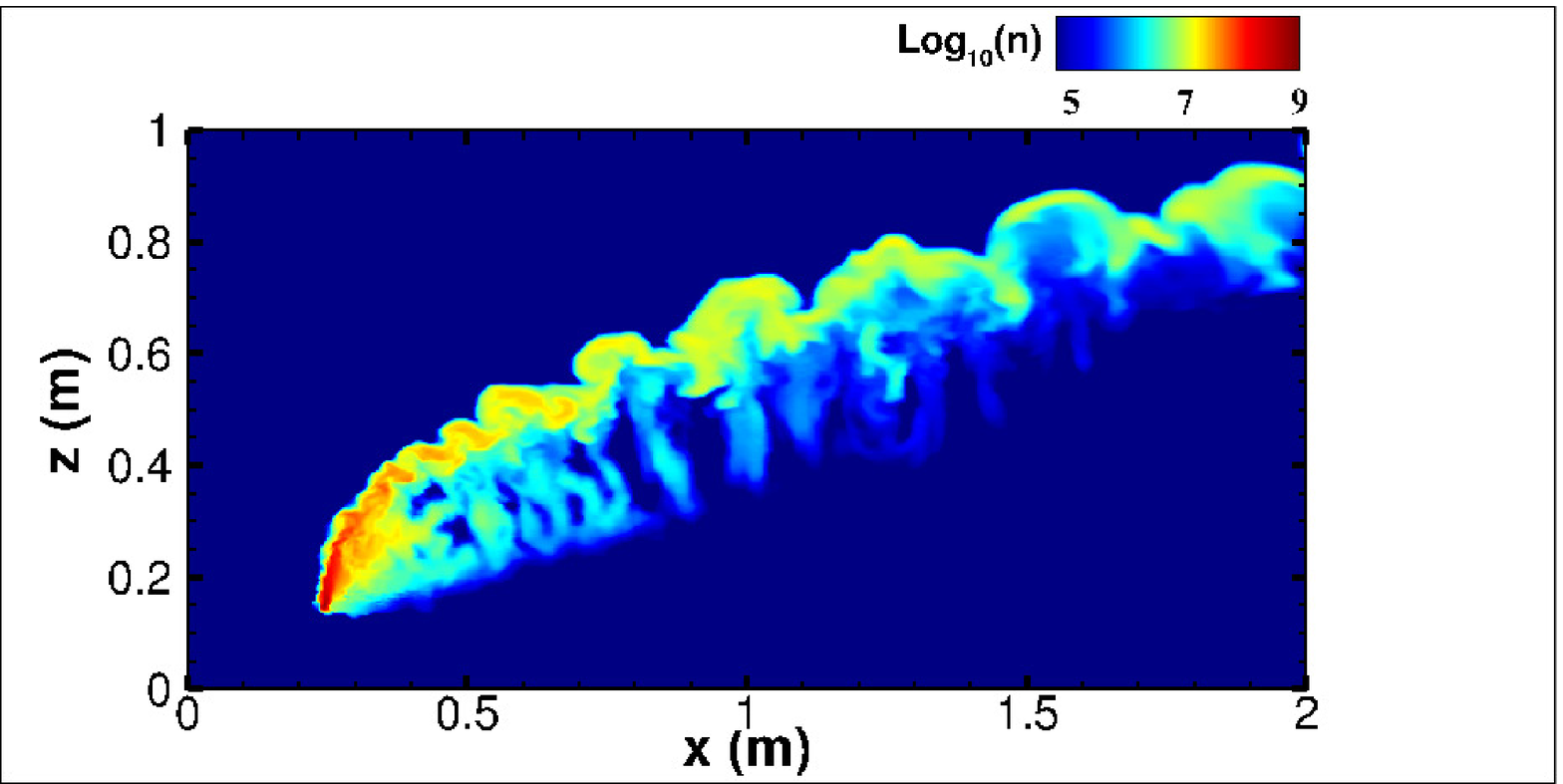}
        \caption{$d = 1000\;\mu\mbox{m}$. } \label{fig:cont15}
    \end{subfigure}
    \hfill
    \begin{subfigure}[t]{0.45\textwidth}
        \centering
        \includegraphics[width=1.2\linewidth,trim=4 4 4 4,clip]{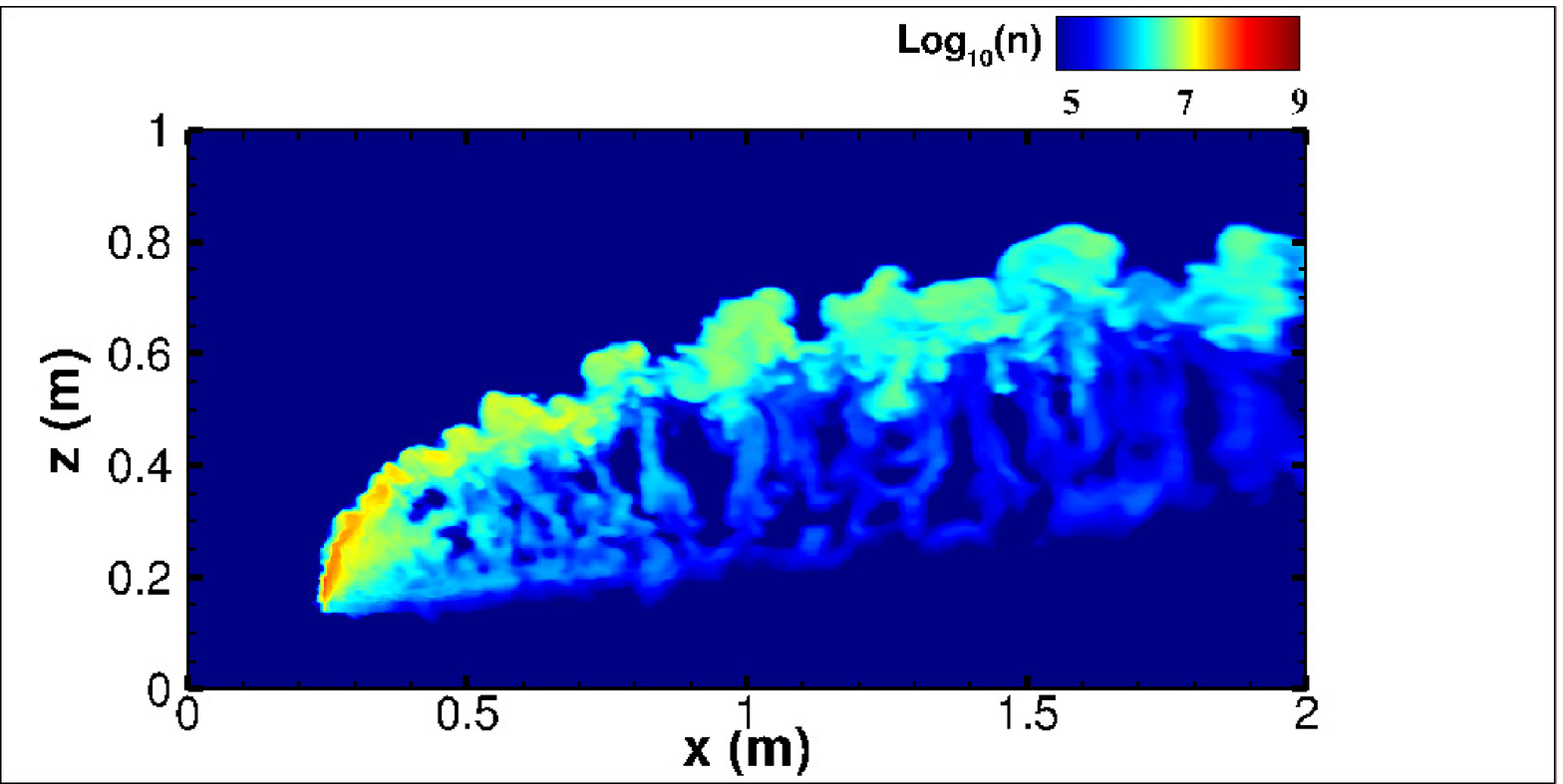} 
        \caption{$d = 432\;\mu\mbox{m}$} \label{fig:cont12}
    \end{subfigure}

    \vspace{1cm}
    \begin{subfigure}[t]{0.45\textwidth}
        \centering
        \includegraphics[width=1.2\linewidth,trim=4 4 4 4,clip]{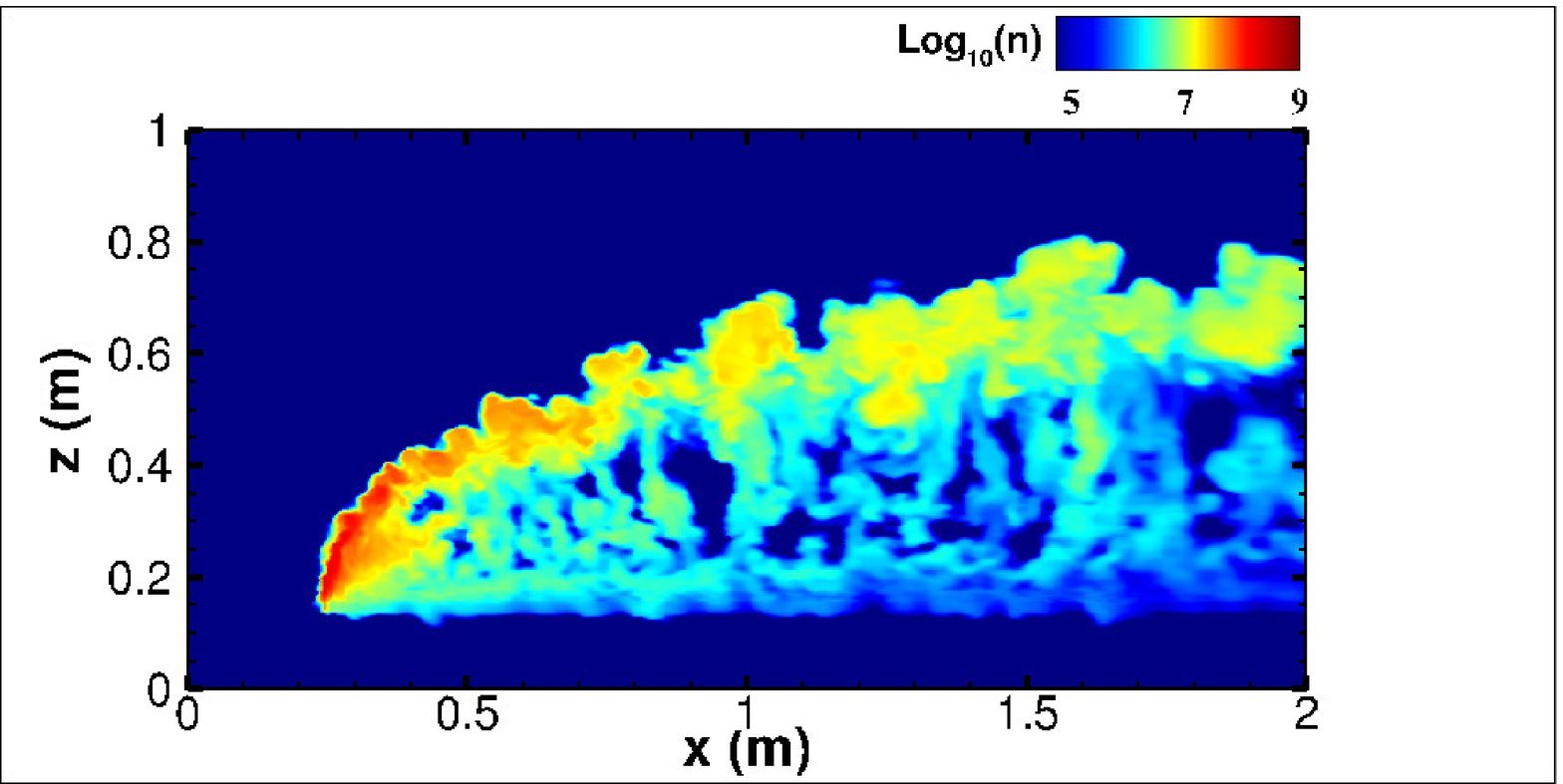} 
        \caption{$d = 107\;\mu\mbox{m}$} \label{fig:cont7}
    \end{subfigure}
    \hfill
    \begin{subfigure}[t]{0.45\textwidth}
       \centering
        \includegraphics[width=1.2\linewidth,trim=4 4 4 4,clip]{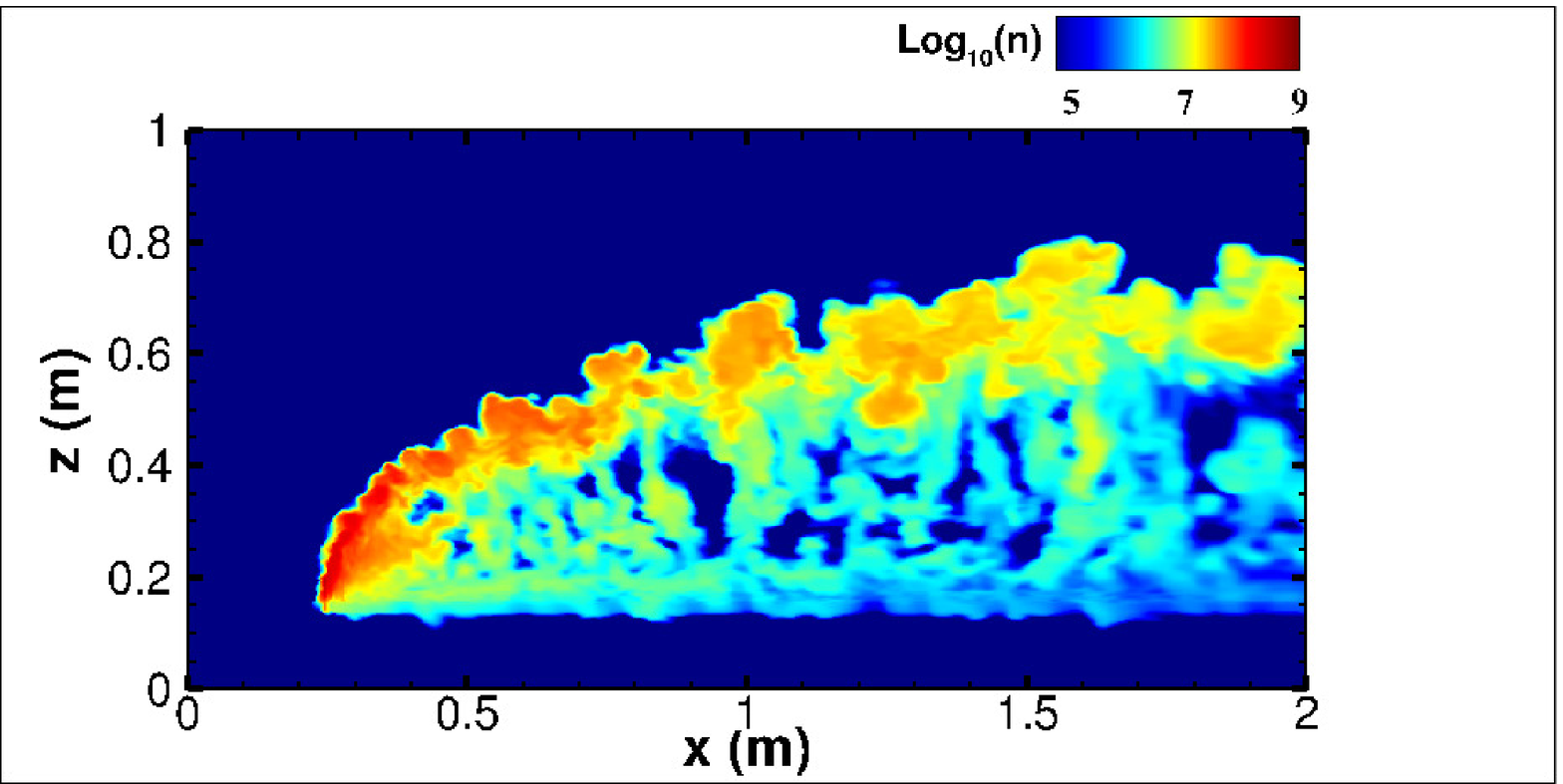}
        \caption{$d = 20\;\mu\mbox{m}$} \label{fig:cont1}
    \end{subfigure}
    \caption{Contour plots of instantaneous concentration fields at the midplane of the jet. The concentration is plotted in logarithmic scale. (\subref{fig:cont15}) is the concentration of the droplet of size $1000\;\mu\mbox{m}$, (\subref{fig:cont12}) for $d=432\;\mu\mbox{m}$, (\subref{fig:cont7})   for $d=107\;\mu\mbox{m}$, and (\subref{fig:cont1}) for $d=20\;\mu\mbox{m}$. }
    \label{fig:inst_fields}
\end{figure}
In figure $\ref{fig:inst_fields}$ we show  contour plots of instantaneous number density in logarithmic scale ($\log_{10}(\tilde{n}_i)$) for four representative droplet sizes on the mid y-plane as function of $x$ and $z$. The largest droplet size is in the top left panel and the
smallest one is in the bottom right. The spatial distributions of the number densities for different droplet sizes show distinct qualitative behaviours. The plumes of the smaller droplets appear significantly more dispersed than those of the larger sizes, showing some presence also in the bottom portions of the domain and clear effects of vertical column vortices. The largest droplets are more concentrated towards the upper portions of the overall plume (consisting of all size bins) as expected from their larger rise velocity. 

\begin{figure}
    \centering
    \begin{subfigure}[t]{0.45\textwidth}
        \centering
        \includegraphics[width=1.2\linewidth,trim=4 4 4 4,clip]{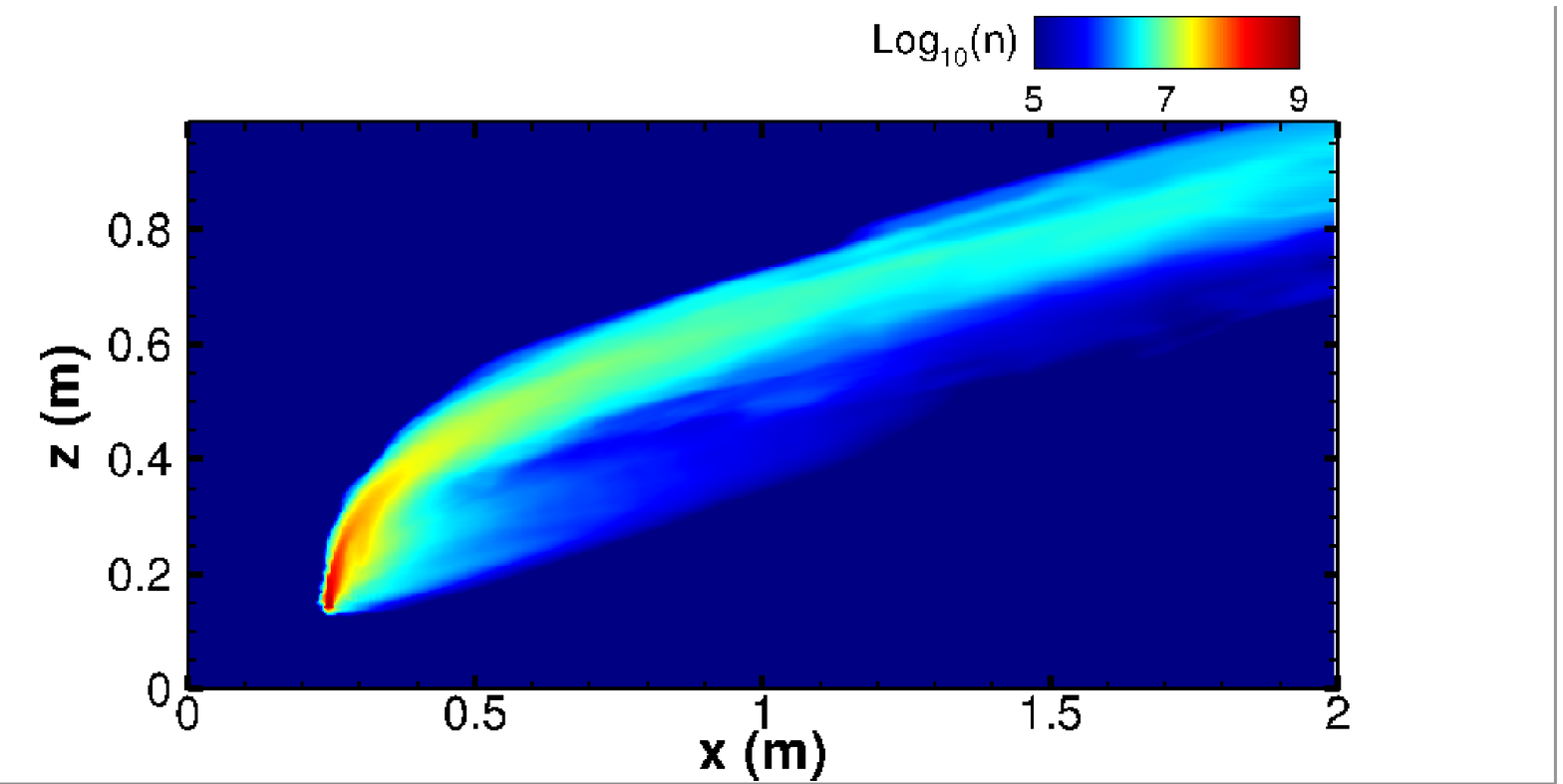}
        \caption{$d = 1000\;\mu\mbox{m}$. } \label{fig:av_cont15}
    \end{subfigure}
    \hfill
    \begin{subfigure}[t]{0.45\textwidth}
        \centering
        \includegraphics[width=1.2\linewidth,trim=4 4 4 4,clip]{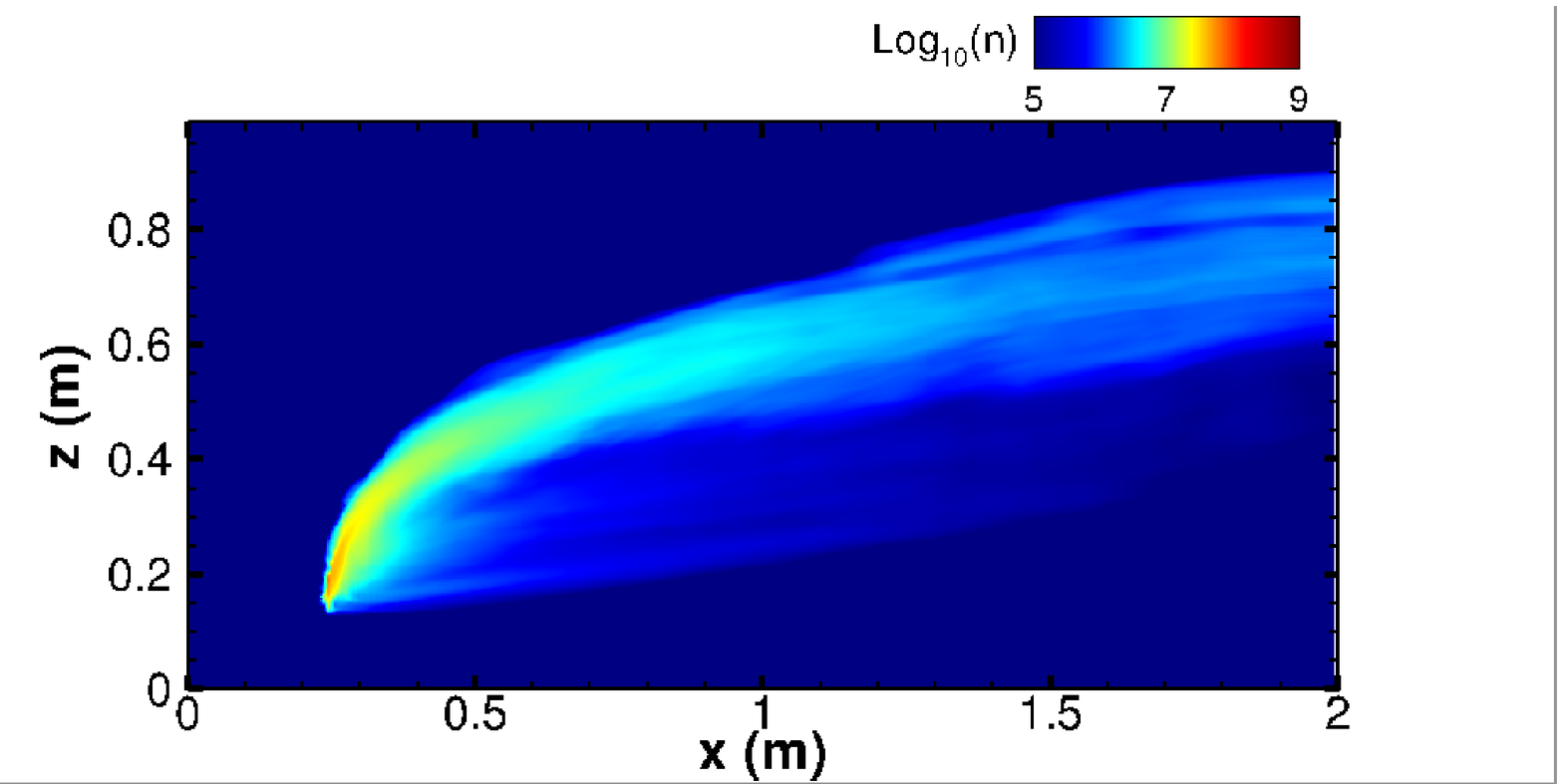} 
        \caption{$d = 432\;\mu\mbox{m}$} \label{fig:av_cont12}
    \end{subfigure}

    \vspace{1cm}
    \begin{subfigure}[t]{0.45\textwidth}
        \centering
        \includegraphics[width=1.2\linewidth,trim=4 4 4 4,clip]{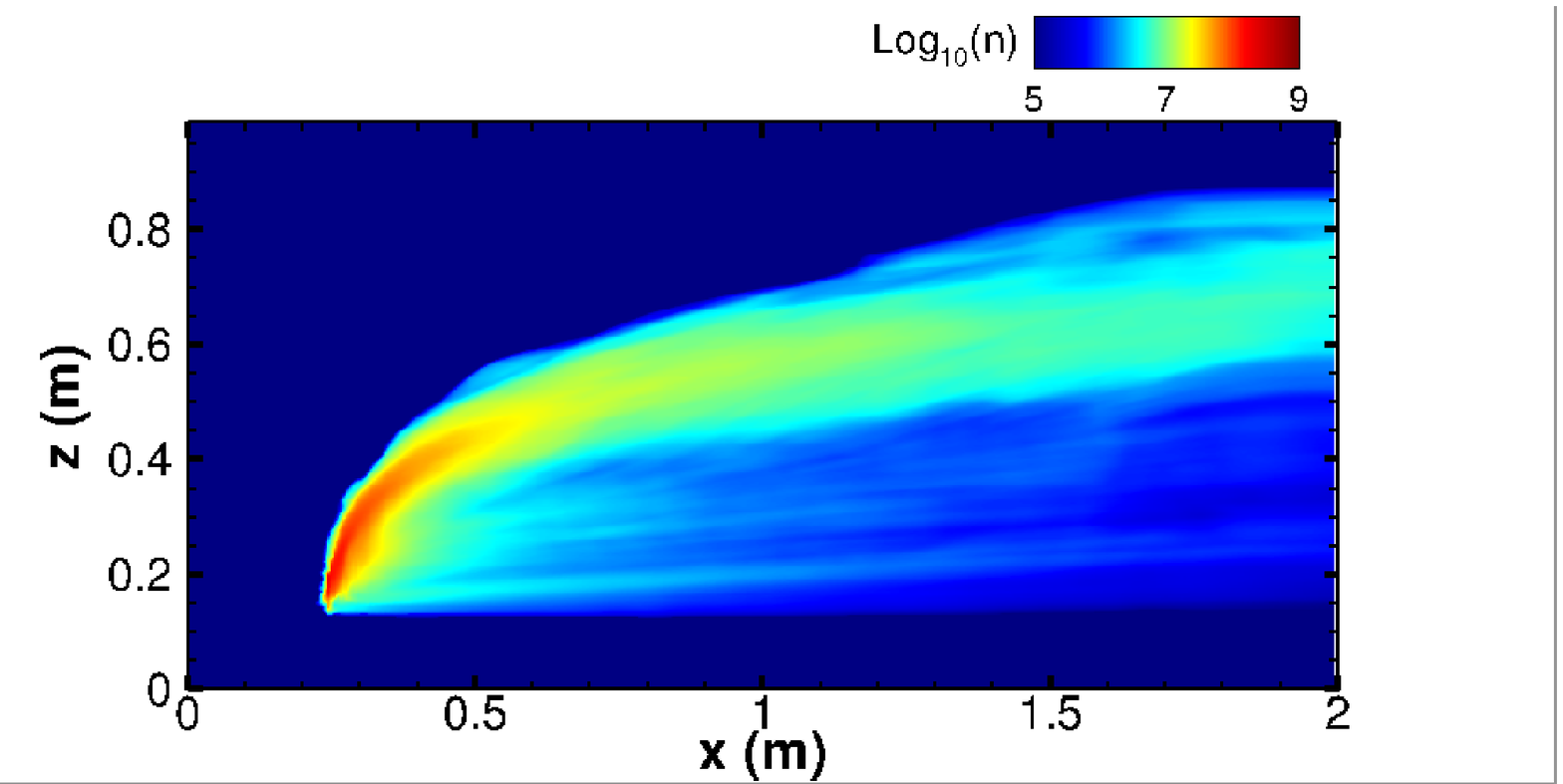}
        \caption{$d = 107\;\mu\mbox{m}$} \label{fig:av_cont7}
    \end{subfigure}
    \hfill
    \begin{subfigure}[t]{0.45\textwidth}
       \centering
        \includegraphics[width=1.2\linewidth,trim=4 4 4 4,clip]{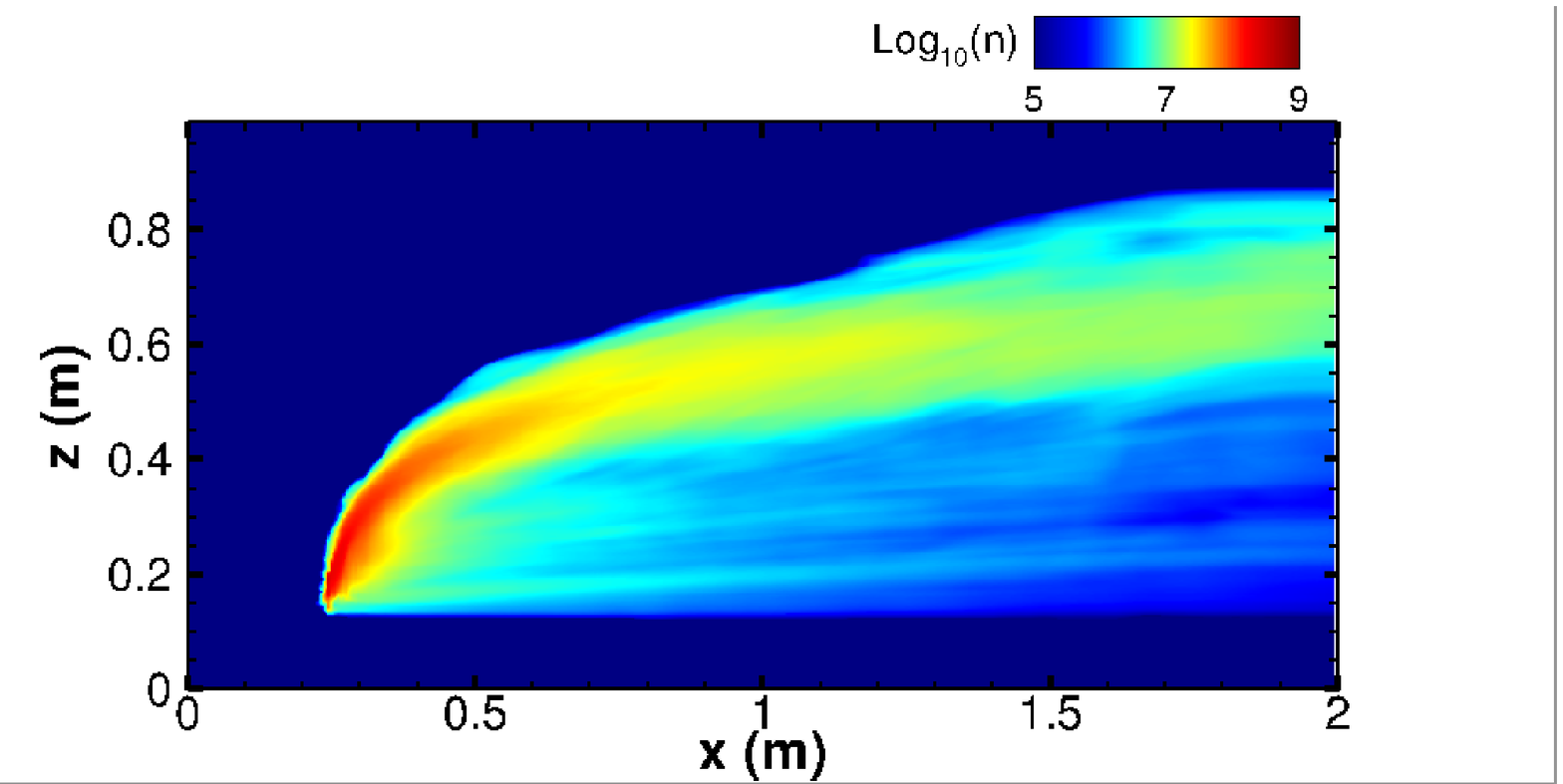}
        \caption{$d = 20\;\mu\mbox{m}$} \label{fig:av_cont1}
    \end{subfigure}
    \caption{Time averaged concentration fields at the midplane of the jet plotted in logarithmic scale. Panel (\subref{fig:cont15}) is the concentration of droplets of size $1000\;\mu\mbox{m}$, (\subref{fig:cont12}) shows the concentration field for $d=432\;\mu\mbox{m}$, (\subref{fig:cont7})  shows the concentration field for $d=107\;\mu\mbox{m}$, and (\subref{fig:cont1})  shows the concentration field for $d=20\;\mu\mbox{m}$. } 
    \label{fig:conc_c}
\end{figure}

Figure \ref{fig:conc_c} shows the time averaged number density fields of the various 
droplet sizes at the mid y-plane.  The maximum concentration for each bin 
size occurs in the near-nozzle region where the energy dissipation is also the 
highest and thus the breakup rate is fastest. 
The larger droplets are more buoyant, having a larger rise velocity 
and thus their average plume has a higher inclination angle with respect to the crossflow 
direction.

To make a quantitative comparison with the experiments we compare droplet size distributions measured at the two cross-stream locations indicated in figure $\ref{fig:sketch_sim}$. The measurement locations are centered at a vertical distance of $42\;\mbox{cm}$ above from the nozzle (see Appendix B). The width of bins used in the experiments are not necessarily the same as that used in the simulation. In order to make comparisons with the experiments, we define a number density normalized by the bin width, i.e.
\begin{equation}
    n_i^* = \frac{\tilde{n}_i}{\delta d_i},
\end{equation}
here $\tilde{n}_i$ is the number of droplets per $\mathrm{m^3}$ of fluid in bin $i$,  $\delta d_i = \frac{1}{2}(d_{i+1}-d_{i-1})$ for $i=2$ to $14$, $\delta d_1=d_2-d_1$ and $\delta d_{15} = d_{15}-d_{14}$.  This normalization ensures that the result is conceptually independent of the discretization of the size range, i.e. the bin width.  The simulated number density fields are averaged in time and the normalized time-averaged number density $\overline{n}_i^*$ in each bin is obtained. 

\begin{figure*}
    \centering
        \includegraphics[width=0.8\linewidth,trim=4 4 4 4,clip]{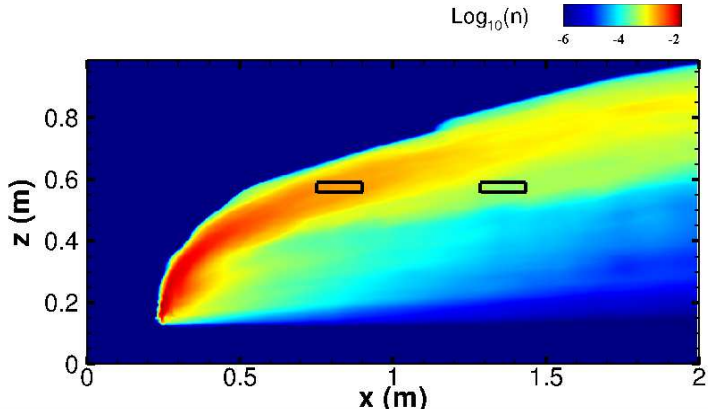}
    \caption{Contour plot of the logarithm of the averaged total volume concentration of oil showing the measurement location at $x=0.76\;\mbox{m}$ and $x=1.3\;\mbox{m}$.}
     \label{fig:cont_vol}
\end{figure*}

\begin{figure}
 \hspace{-1cm}
\centering
\begin{subfigure}{0.45\textwidth}
        \centering
        \includegraphics[width=1.15\linewidth,trim=4 4 4 4,clip]{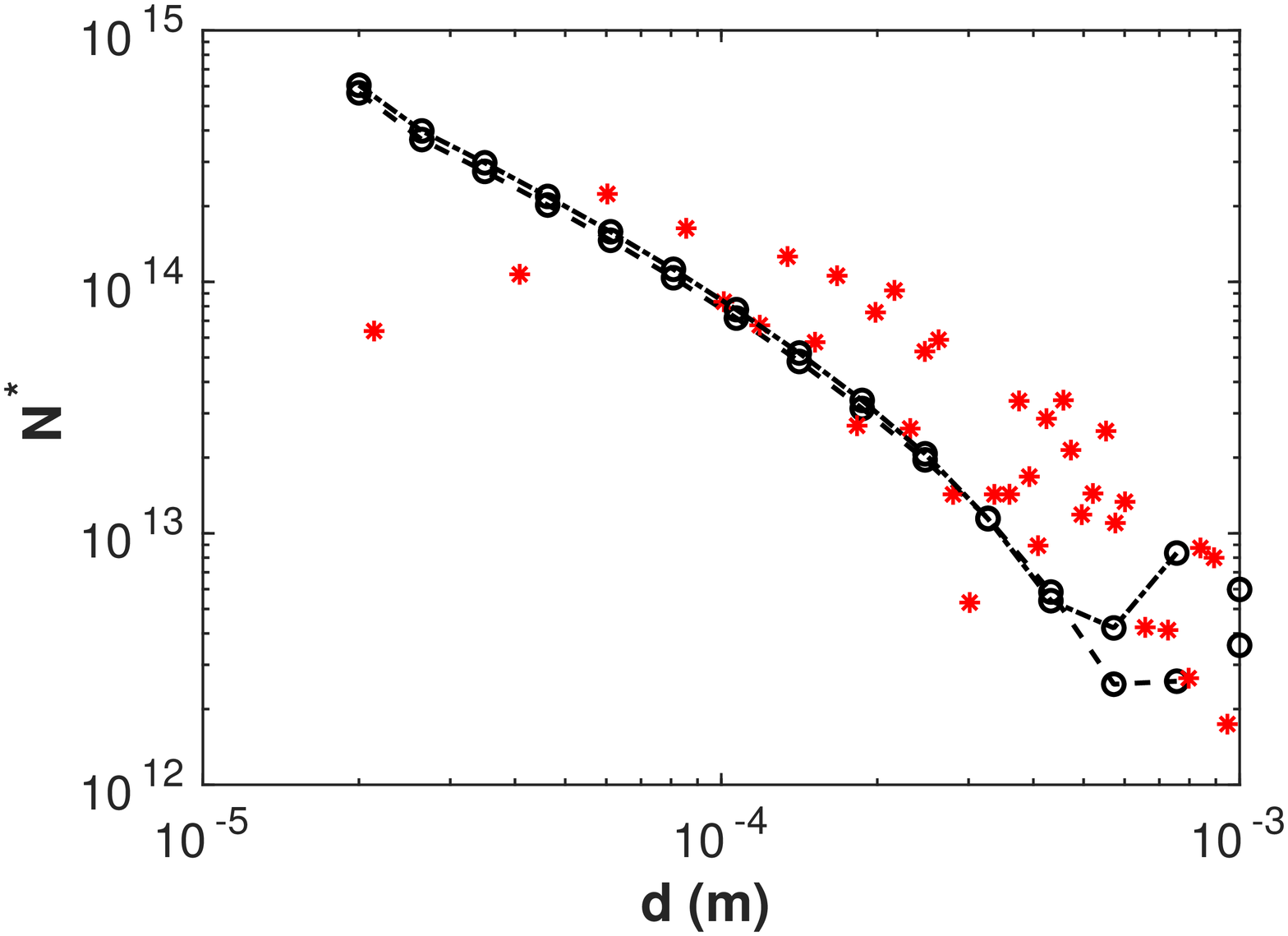}
    \label{fig:cont1_1}
    \end{subfigure}
    \hspace{0.5cm}
\begin{subfigure}{0.45\textwidth}
\centering
\includegraphics[width=1.15\linewidth,trim=4 4 4 4,clip]{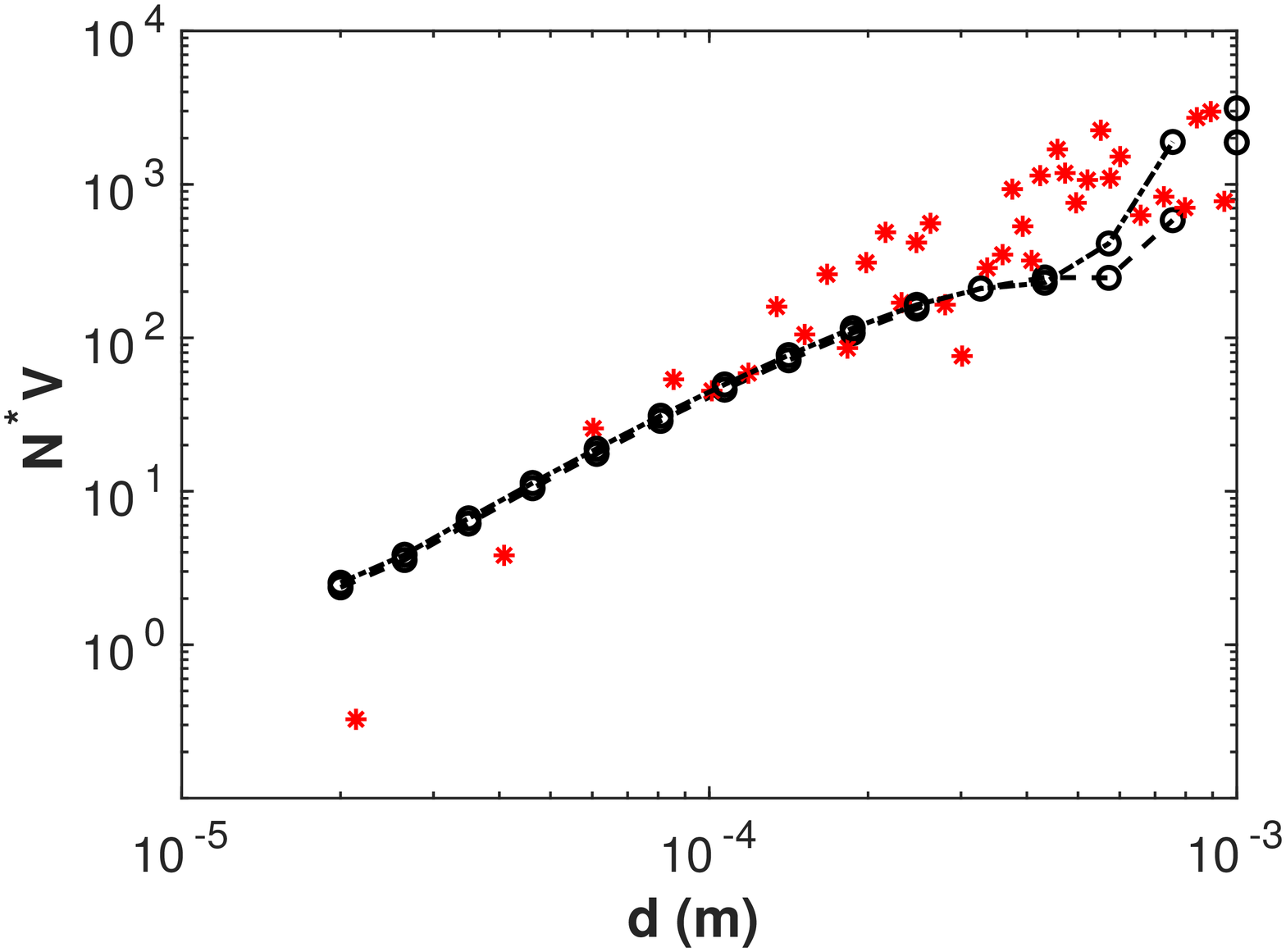}
\end{subfigure}
\caption{Comparison of LES model at $x=0.76\;\mbox{m}$ for mono-dispersed injection (\protect\circlenline) and bi-dispersed injection (\protect\circlelinedot), and experimental data from \citep{Murphy2016} measured at the corresponding time ($\color{red} \ast$). Left panel: Relative size distribution from LES, Right panel: Relative volume distribution.}
\label{fig:dsd3p4}
\end{figure}

\begin{figure}
 \hspace{-1cm}
\centering
\begin{subfigure}{0.45\textwidth}
        \centering
        \includegraphics[width=1.15\linewidth,trim=4 4 4 4,clip]{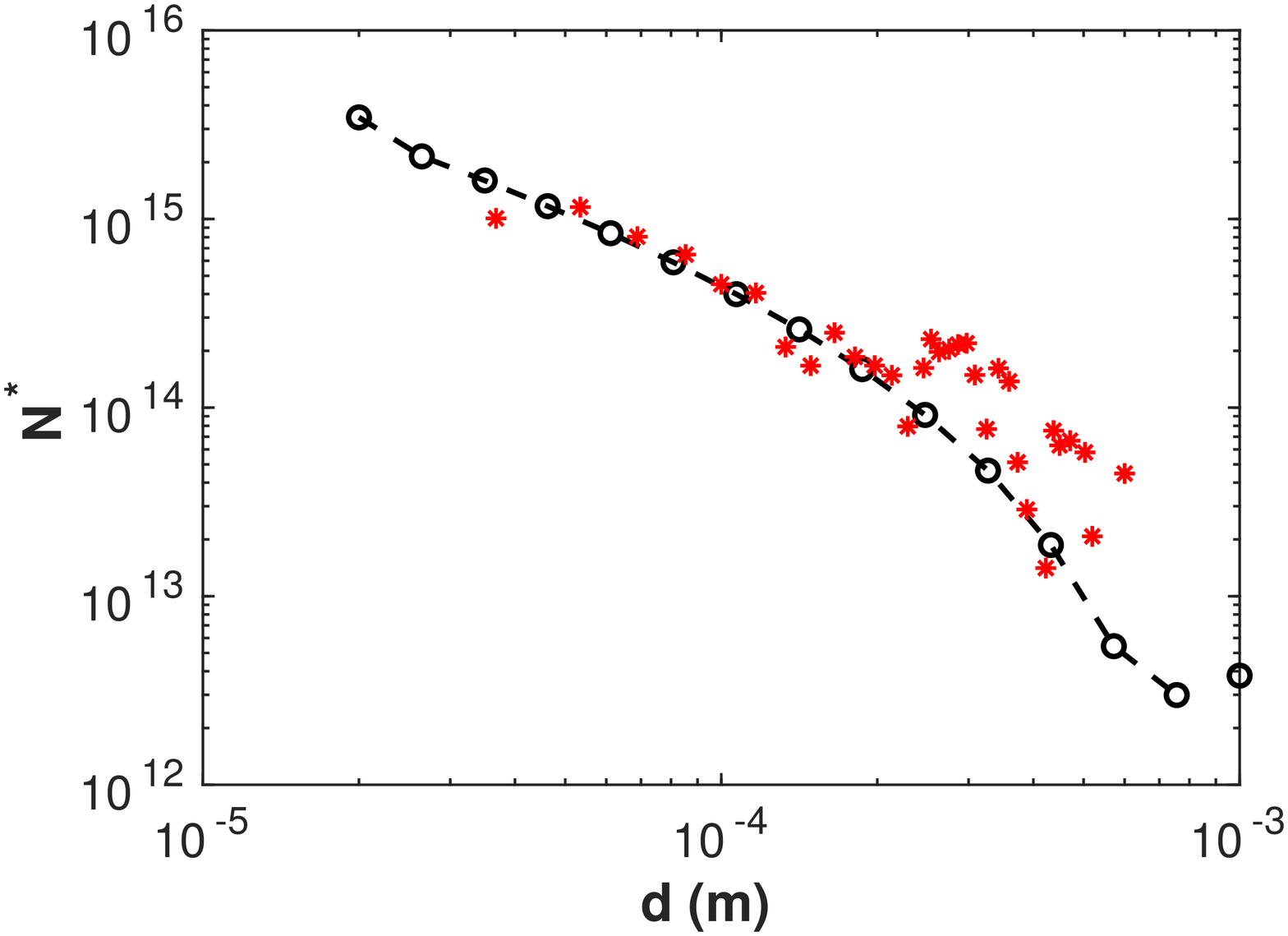}
    \label{fig:cont1_2}
    \end{subfigure}
        \hspace{0.5cm}
    \begin{subfigure}{0.45\textwidth}
 \centering
\includegraphics[width=1.15\linewidth]{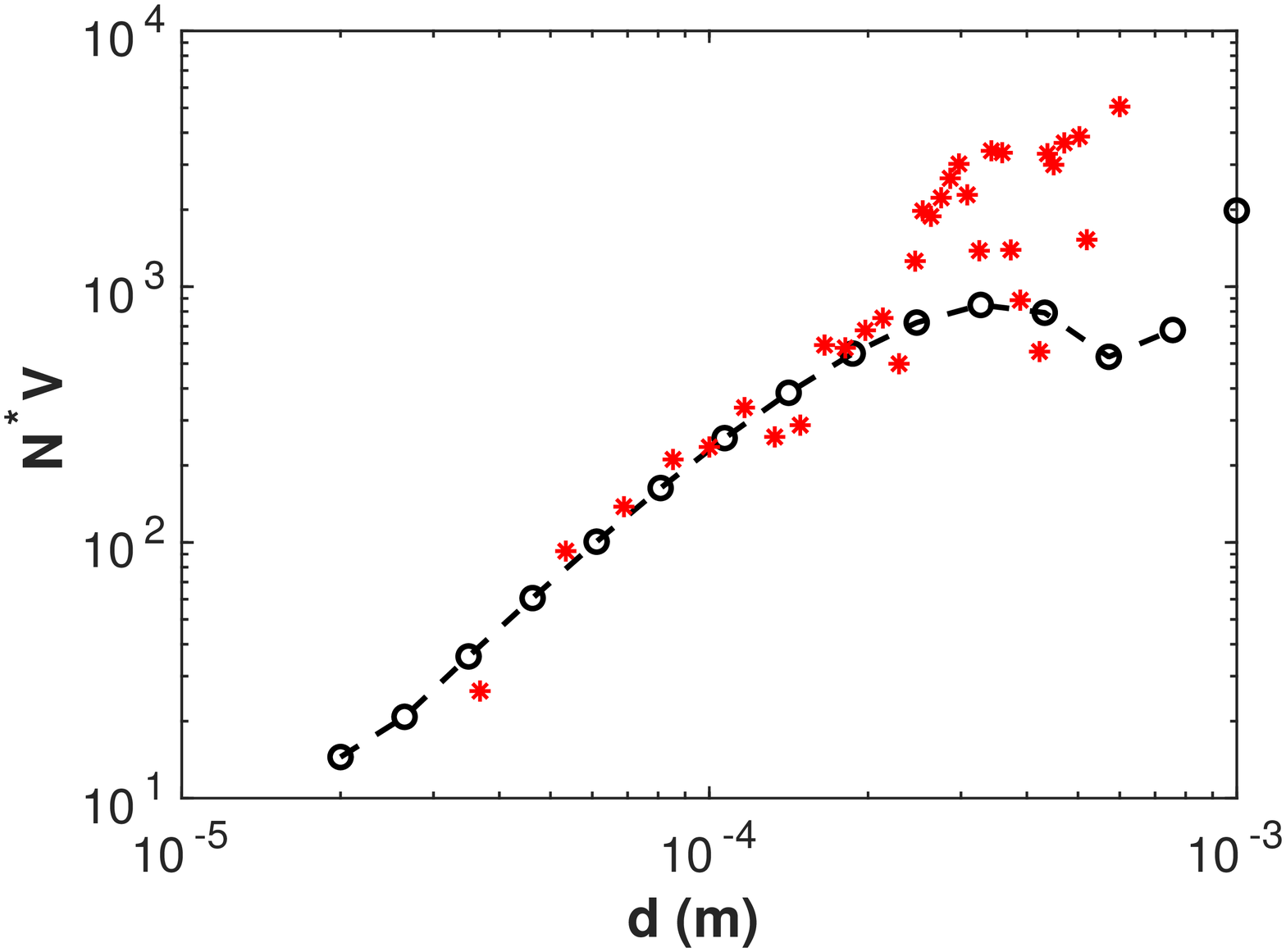}
\end{subfigure}
\caption{Comparison of LES model at $x=1.3\;\mbox{m}$ (\protect\circlenline) , and experimental data from \citep{Murphy2016} measured at the corresponding time ($\color{red} \ast$). Left panel: Relative size distribution from LES, Right panel: Relative volume distribution.}
\label{fig:dsd6p9}
\end{figure}

\begin{figure}
 \hspace{-1cm}
\centering
\begin{subfigure}{0.45\textwidth}
        \centering
        \includegraphics[width=1.15\linewidth,trim=4 4 4 4,clip]{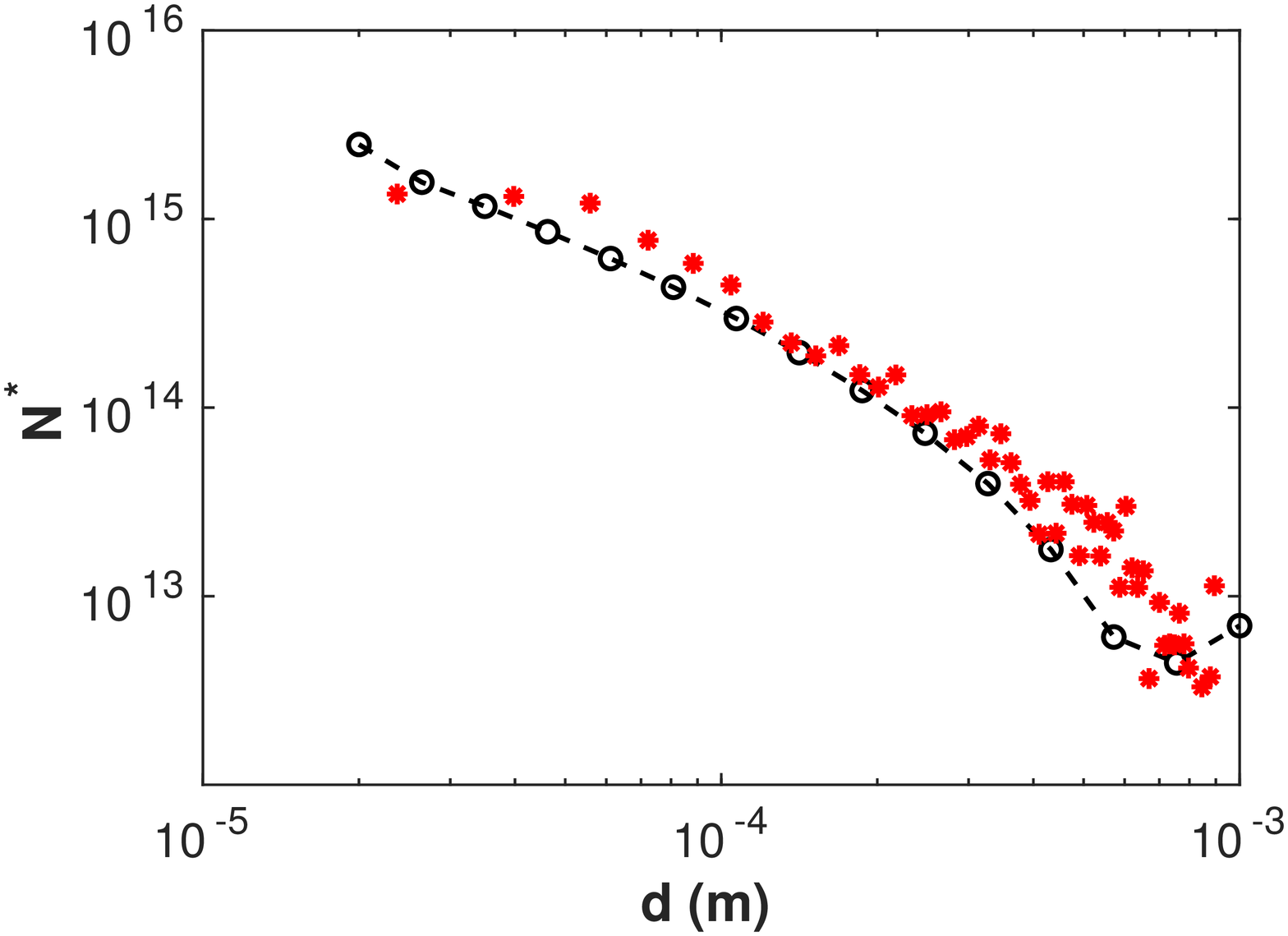}
    \label{fig:dsd_num}
    \end{subfigure}
        \hspace{0.5cm}
    \begin{subfigure}{0.45\textwidth}
 \centering
\includegraphics[width=1.15\linewidth]{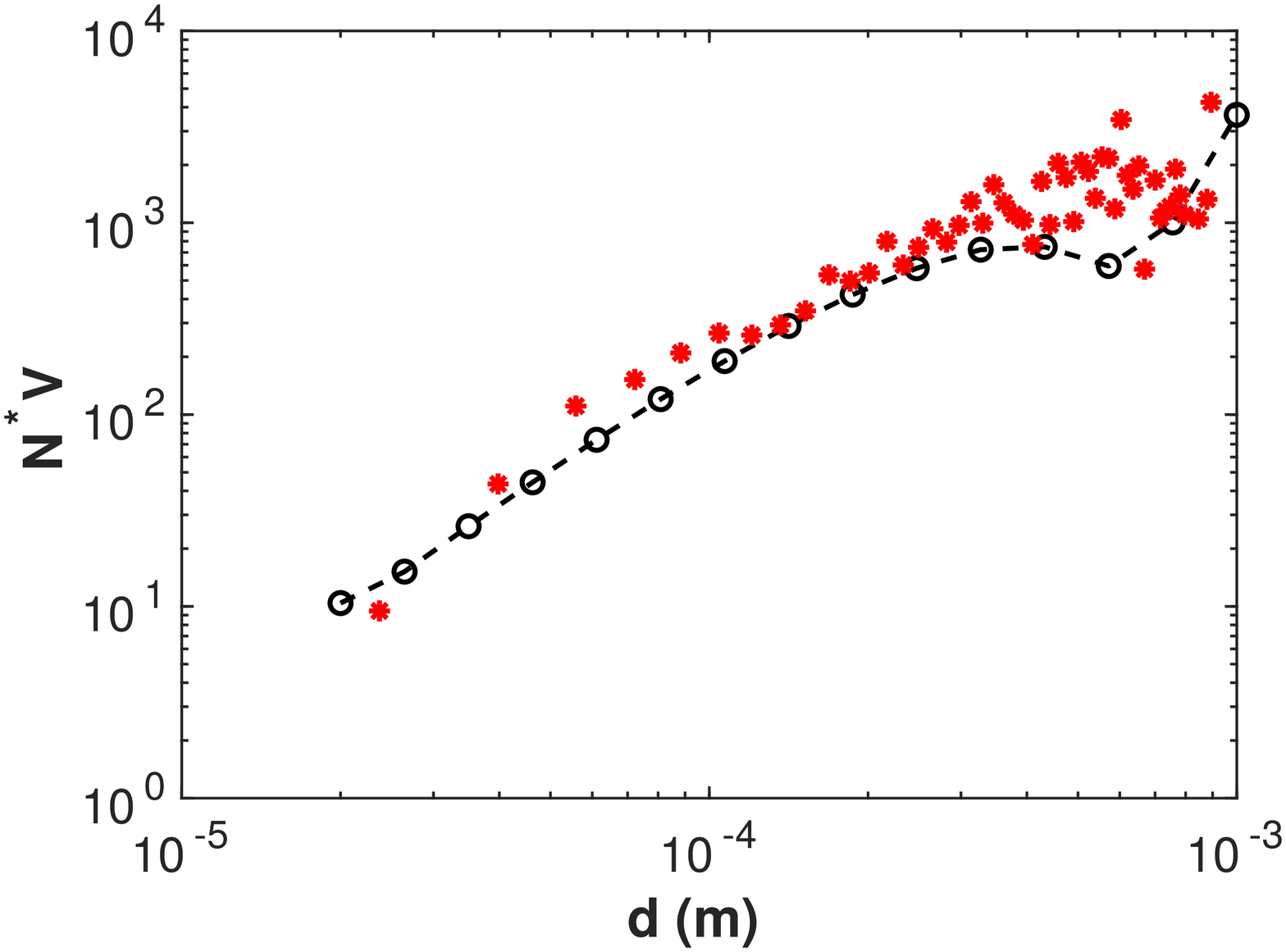}
\end{subfigure}
\caption{Comparison of LES model at $x=1.3\;\mbox{m}$ (\protect\circlenline) , and experimental data from \citep{Murphy2016} measured at the corresponding time ($\color{red} \ast$). Left panel: Relative size distribution from LES, Right panel: Relative volume distribution.}
\label{fig:dsdtot}
\end{figure}

Since the experimental data are not fully converged statistically, a comparison of the average total oil concentration (integrated over all bins) between experiment and simulation  yielded differences of factors of $1.4$ and $3.7$ at downstream locations corresponding to $x=0.76\;\mbox{m}$ and $x=1.3\;\mbox{m}$ respectively, in this case. Therefore, here we focus the comparison between experiment and simulation on the shape of the resulting size distribution rather than the total concentration. In particular, we normalize the size distribution for both the experiment and the simulation by the  total volume concentration ($\overbar{n}_i^*\times V_i$) summed over the entire size range, where $V_i$ is the volume of a droplet of size $d_i$, as defined earlier. We define the relative size distribution, $N_i^*$ according to
\begin{equation}
    N_i^* =  \frac{\overline{n}_i^*}{\sum_j (\overline{n}_j^* V_j)\delta d_j}.
\end{equation}
 Figure \ref{fig:cont_vol} depicts the log of the total volume-averaged concentration of oil, $\log_{10}(\sum_i \overline{\tilde{n}}_i V_i)$. The black squares show the locations of the measurement volumes used to obtain the size distributions.  We compare the relative size and volume distributions obtained from the simulations with the experimental data in figures \ref{fig:dsd3p4} , \ref{fig:dsd6p9} and \ref{fig:dsdtot}. The left panel depicts the relative size distribution and the right panel  shows the relative volume defined by $N^*V_i$, where $V_i$ is the volume of a particular bin.
The data reported in the experiment represents an average over three realizations recorded during $1\;\mbox{s}$. In the simulation, the nozzle is fixed, and so the measurement at $t=3.4\;\mbox{s}$ for the experiment translates to a window between $x=0.76\mbox{--} 0.91\;\mbox{m}$. We chose a region from $y_1 =0.37\;\mbox{m}$ to $y_2=0.4\;\mbox{m}$ and $z_1 =0.56\;\mbox{m}$ to $z_2 = 0.59\;\mbox{m}$ for our measurement volume.
We can see that the simulation captures the overall relative size volume distribution at this location, although the experimental data have large scatter. The number density for the smallest droplet sizes are higher in the simulation than in the experiment, and we do not observe the dip seen in the experimental data. The higher number density for the smaller sizes seen in the LES results may be due to the fact that the breakup probability density function favors the formation of small droplets according to the model equation used for $P(d_i,d_j)$. The reported data is not statistically converged and the experiment can only measure droplets with diameter larger than about $20\;\mu\mbox{m}$.

In order to explore the sensitivity of the results to the assumed initial size distribution at the injection point, we perform a second simulation in which instead of placing the entire volume injection rate into a single bin at $1 \; \mbox{mm}$, it is distributed equally among the two largest bins. As shown in figure \ref{fig:dsd3p4} (circles with dot dashed line), the results for droplets smaller than $400 \; \mu \mbox{m}$ are the same and are quite robust to details of the injection distribution at the large droplets.

Figure $\ref{fig:dsd6p9}$ shows the normalized size distribution at $t=6.9\;\mbox{s}$ for the experiment corresponding to a window of $x=1.285\mbox{--}1.435\;\mbox{m}$ for the simulation.  We see that the relative size and volume distribution is well matched for this later time, now also including the smaller droplets. Finally, figure $\ref{fig:dsdtot}$ shows the total normalized size distributions for the experiment. The total size distribution was measured in the experiment using data from 5 time instances corresponding to a spatial window $x= 0.76\mbox{--}1.66 \; \mbox{m}$ in the simulation. We see that the model captures the relative size and volume distributions well.


\begin{figure} 
\centering
\begin{subfigure}{0.5\textwidth}
        \centering
        \includegraphics[width=1.1\linewidth,trim=4 4 4 4,clip]{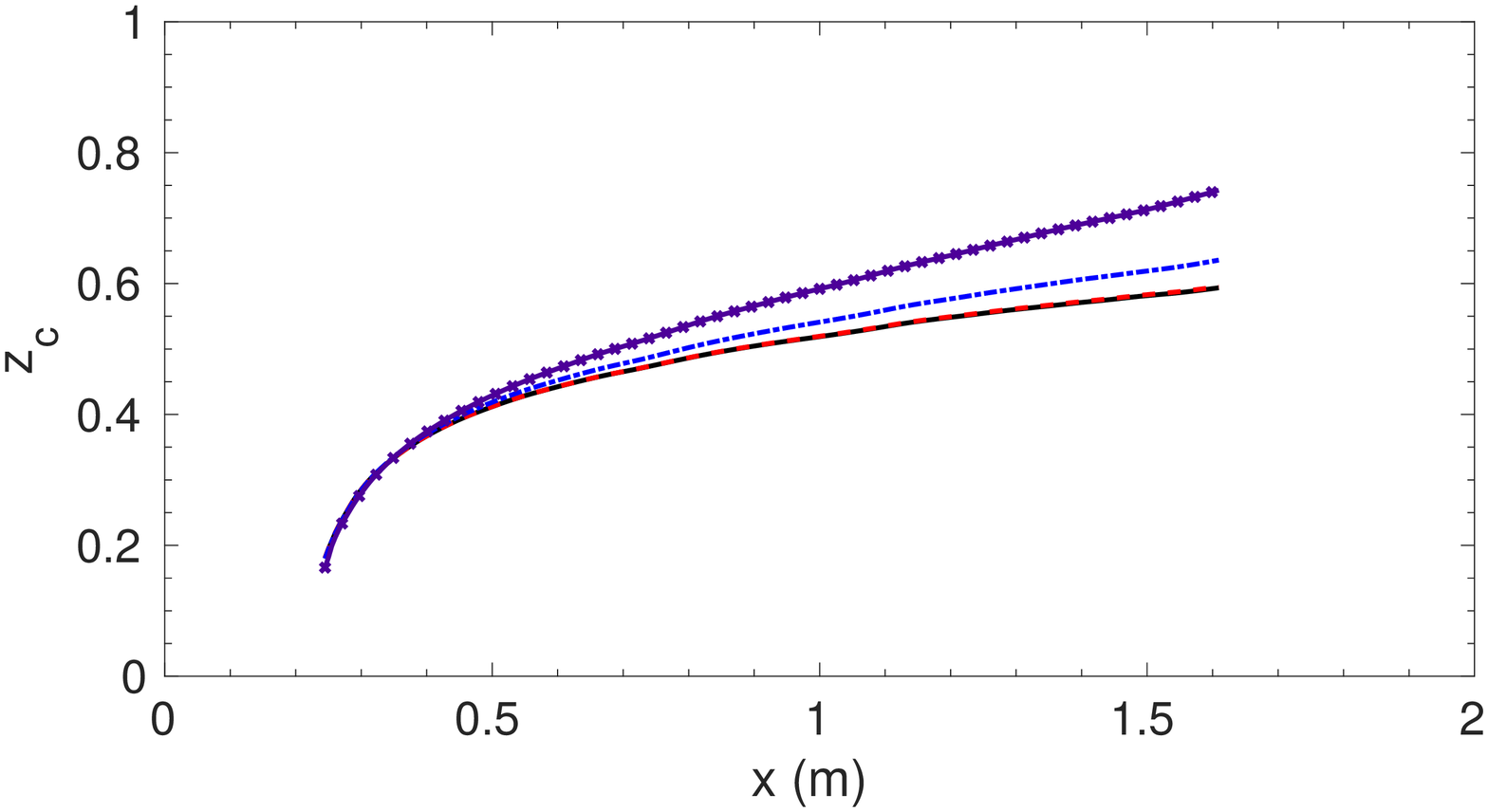}
    \end{subfigure}
    \hfill
    \begin{subfigure}{0.45\textwidth}
 \centering
\includegraphics[width=1.1\linewidth,trim=4 4 4 4,clip]{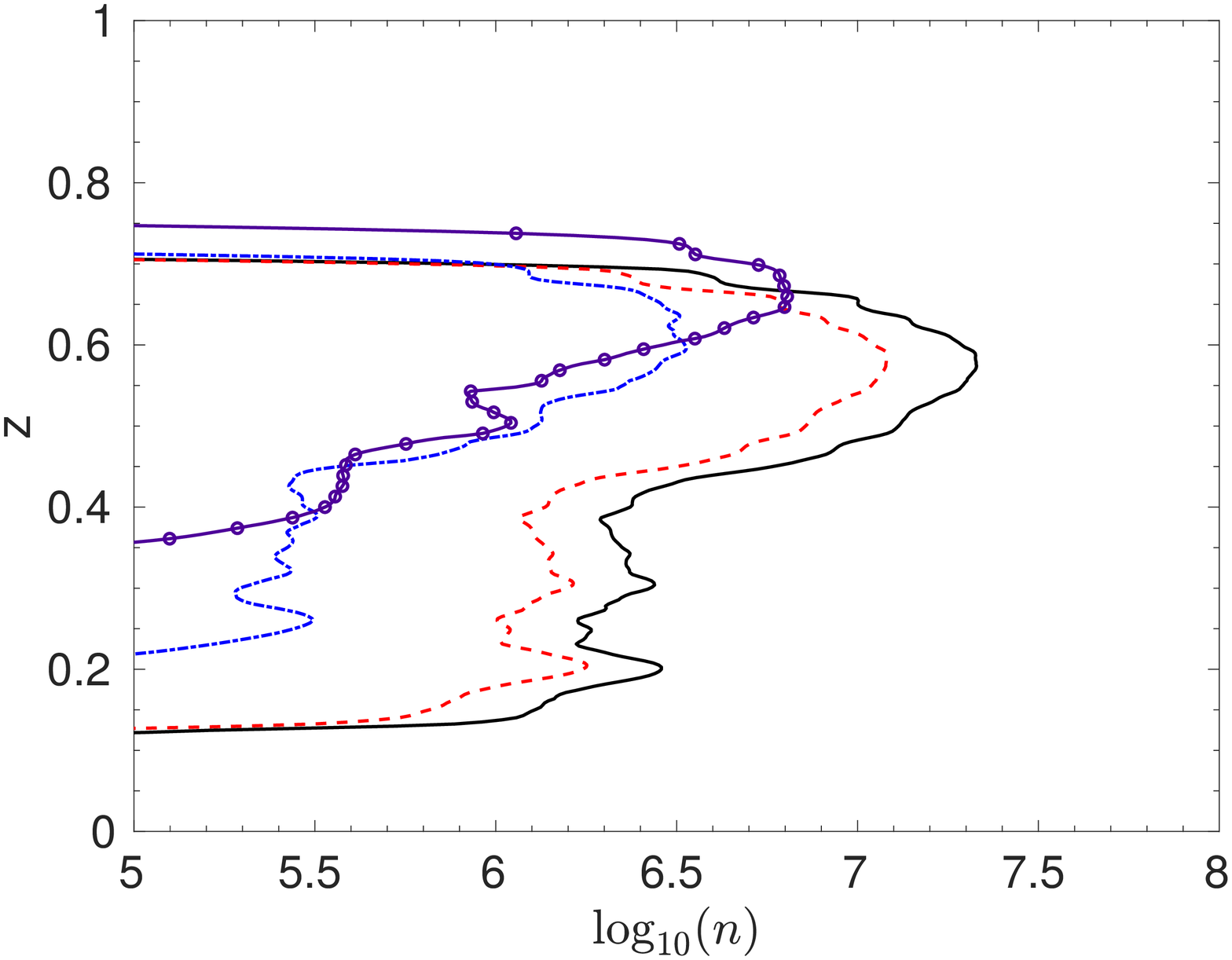}
\end{subfigure}
\caption{Left panel: evolution of centroid of various droplet plumes. Right panel: logarithm of the concentration profile as function of height at a downstream distance, $x=1\;\mbox{m}$ and transverse position $y=0.385\;\mbox{m}$. The lines are  $d=1000\;\mu\mbox{m}$(\protect\pcircleline), $d=432\;\mu\mbox{m}$(\protect\blueline), $d=107\;\mu\mbox{m}$(\protect\redline)  and  $d=20 \;\mu\mbox{m}$(\protect\blackline).}
\label{fig:z_cmconc}
\end{figure}

We can track the plume paths of the different droplet sizes by calculating the centroid of the plume in the axial direction for each droplet size as a function of cross-stream distance. We can see from the left panel of figure $\ref{fig:z_cmconc}$ that as the jet moves 
farther downstream,  the centroid for the larger droplets 
moves above the smaller ones, with the difference in height being related to the
difference in rise velocities as noted already. The centroid evolution for the smallest droplets shown (20 and 107 $\mu$m) are indistinguishable. The rise velocities for these droplet sizes are very small ($3 \times 10^{-5}$ and $9 \times 10^{-4}$ m/s, respectively) and their trajectories and plume centroids are thus dominated (equally for both droplet sizes) by mean flow and turbulence, but not appreciably by buoyancy.

We can also examine the concentration distribution of the different sizes along the axial direction at different cross-stream locations. In the right panel of figure \ref{fig:z_cmconc} we plot the concentration distribution at $x=1\;\mbox{m}$. We can see that the concentration is
peaked more towards the top end of the plume. This trend can be attributed to the counter-rotating vortex pair generated due to the jet in crossflow \citep{Cortelezzi2001}.
This results in droplets being moved from
the bottom of the plume towards the top, leading to a higher concentration at the top end. We can also confirm that the plume of smaller droplets is wider than that of larger droplets, showing that the smaller droplets are more dispersed by the turbulence.

The simulation showcases the importance of including the viscous range of scales in the formulation of the breakup frequency. The Kolmogorov scale in the near nozzle region, close to the injection location ($x=0.125\;\mbox{m}$ and $z = 0.14\;\mbox{m}$) where $\langle \epsilon\rangle=30 \; \mbox{m}^2\mbox{s}^{-3}$ can be calculated as $\eta = (\nu^3/\langle\epsilon\rangle)^{1/4} \approx 13\;\mu\mbox{m}$. Droplets smaller than $\approx 10\eta = 130\;\mu\mbox{m}{}$ would lie in the viscous range. Further downstream at the $y$-mid-plane, where the average dissipation has decayed to $\langle \epsilon\rangle \approx 0.1\;\mbox{m}^2\mbox{s}^{-3}$, $\eta \approx 60\;\mu\mbox{m}$. Thus most of the droplet size range is in the viscous range. Earlier models that assumed that all the sizes were in the inertial range would predict incorrect breakup frequencies for these droplets as it would overestimate the eddy fluctuation velocity at the scale of the droplet. This highlights the importance of having a  framework that can smoothly transition between droplets in the inertial and viscous range.

\begin{figure*}
    \centering
        \includegraphics[width=0.8\linewidth]{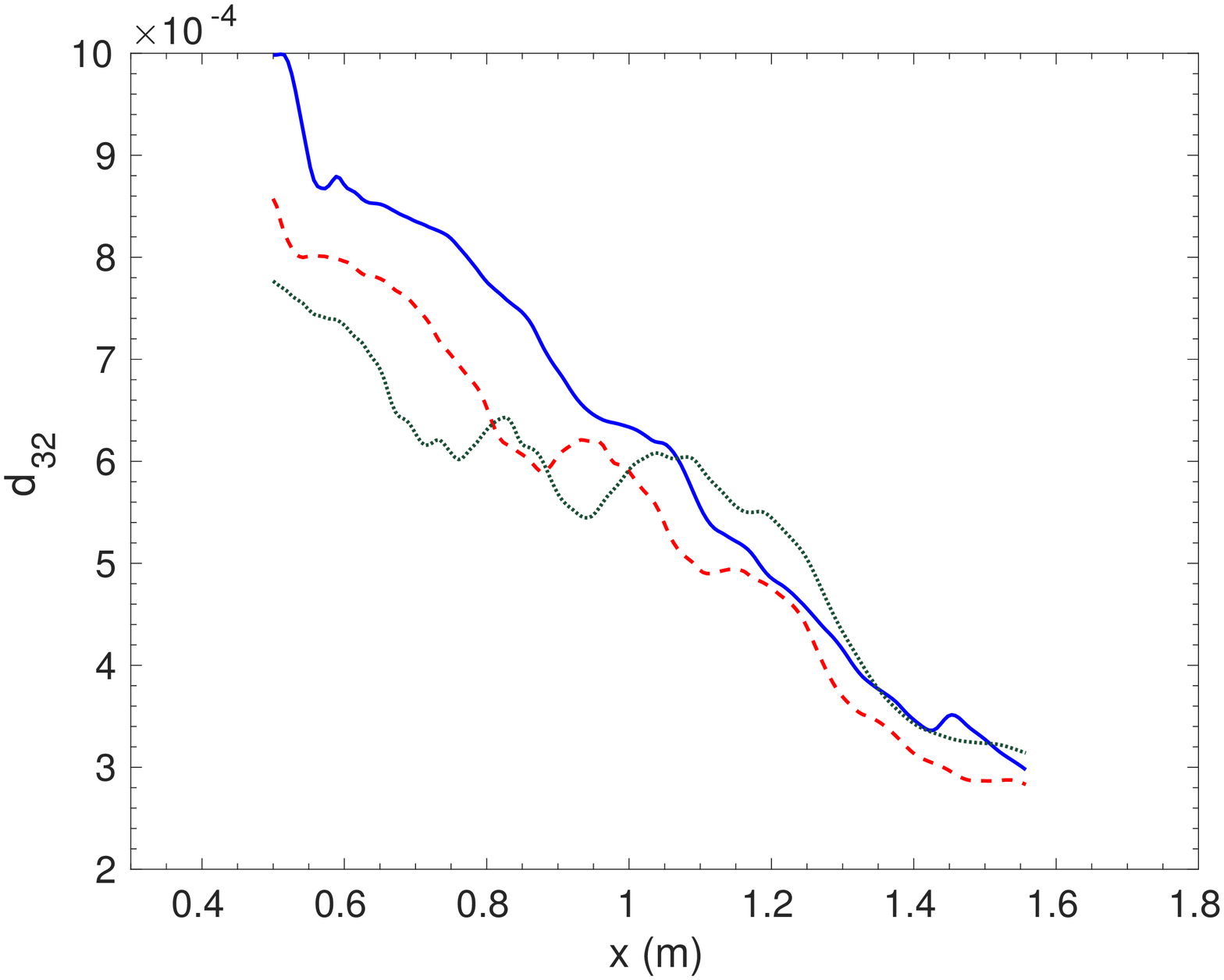}
    \caption{Average $d_{32}$ diameter as function of downstream distance $x$ at various plume heights. The lines correspond to $z=0.55\;\mbox{m}$ (\protect\bluesline), $z=0.50\;\mbox{m}$(\protect\redline) and $z=0.45\;\mbox{m}$(\protect\greenline)}
     \label{fig:d32avg}
\end{figure*}

In order to characterize the `typical size' of droplets, one may evaluate the widely used Sauter mean diameter, denoted as $d_{32}$, that expresses the mean diameter of the polydisperse oil by taking into account the volume to surface area ratio of the distribution. It is calculated directly from the distribution using the formula,
\begin{equation}
    d_{32} = \frac{\sum_i \tilde{n}_i d_i^3}{\sum_i \tilde{n}_i d_i^2}.
\end{equation}
Using the mean concentrations from the LES, the $d_{32}$ value can be computed at various locations of the flow. Figure 
\ref{fig:d32avg} shows the downstream evolution of $d_{32}$ as a function of the downstream distance $x$ at three heights. Clearly the mean droplet size decreases as the jet flow evolves along the downstream direction of the crossflow due to droplet breakup, though the rate of change diminishes and appears to reach a stationary scale of about $d_{32} \approx 300\;\mu\mbox{m}$ at large distances away from the nozzle.

Another scale often used is the Hinze (maximum) diameter, $d_{max} \sim \langle\epsilon \rangle^{-2/5} (\sigma/\rho_c)^{3/5}$, by assuming that droplet coalescence does not occur \citep{Kolmogorov,hinze1955} and using Kolmogorov scaling. Since here the dissipation rate varies greatly from one location to the next, it requires us to first compute the average dissipation. It is computed as the time average of $\epsilon$  according to equation ($\ref{eq:dissipLES}$) from the LES. Typical values are
$\langle \epsilon\rangle \approx 0.6\;\mbox{m}^2/\mbox{s}^{3}$ at $x=0.3\;\mbox{m}$ and $z = 0.29\;\mbox{m}$ near the nozzle, and $\langle \epsilon\rangle \approx 0.001\;\mbox{m}^2/\mbox{s}^{3}$ at $x=0.75\;\mbox{m}$ and $z = 0.56\;\mbox{m}$ further downstream.
Accordingly, using $\rho_c = 1018.3\;\mbox{kg}/\mbox{m}^{3}$ and $\sigma = 1.9 \times 10^{-2}\;\mbox{N}/\mbox{m}$  (see table 1), we obtain $d_{max} = 1\;\mbox{mm}$ near the nozzle while $d_{max} = 18\mbox{mm}$ far from the nozzle. The latter value is consistent with the fact that far from the nozzle breakup becomes far less frequent and the distribution has acquired an equilibrium value.
The results show that the Hinze scale at a particular location in which the flow has large differences in dissipation rates from one location to another (as is usually the case in turbulent shear flows) cannot be used to determine the typical local droplet scale that is, instead, influenced mostly by upstream events. Note that at a few grid points from the nozzle exit, where the dissipation $\langle \epsilon\rangle \approx 30\;\mbox{m}^2/\mbox{s}^{3}$ the Hinze diameter, $d_{max} = 300\;\mu\mbox{m}$. The dissipation in a turbulent flow is highly intermittent, a property that is captured in the current study and is discussed further subsequently. Hence, we prefer to continue the discussion of the median diameter $d_{32}$ and its variability in the next section.

\subsection{Analysis of droplet size distribution variability in LES}

As mentioned earlier, LES enables us to diagnose variability of the droplet size distributions and number density transport that RANS cannot obtain, since the latter only predicts time or ensemble average values. In order to illustrate this capability of LES, we now ask what is the inherent variability of typical droplet sizes as well as that of other practically relevant quantities.

\begin{figure*}
 \hspace{-2cm}
    \centering
    \begin{subfigure}[t]{0.45\textwidth}
        \centering
        \includegraphics[width=1.2\linewidth]{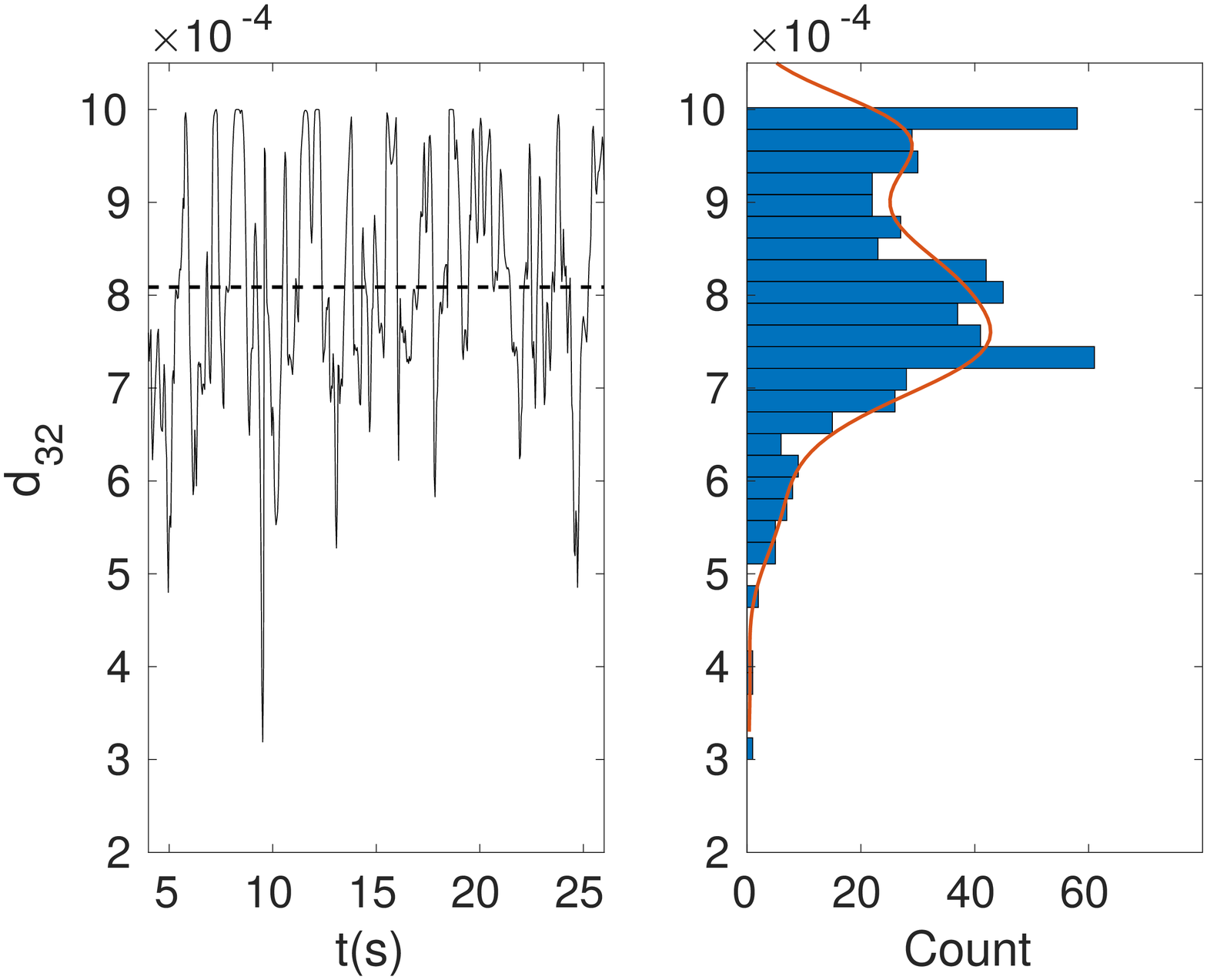}
        \caption{$x=0.75\;\mbox{m}$, $z=0.56\;\mbox{m}$.}
    \end{subfigure}
     \hspace{1cm}
    \begin{subfigure}[t]{0.45\textwidth}
        \centering
        \includegraphics[width=1.2\linewidth]{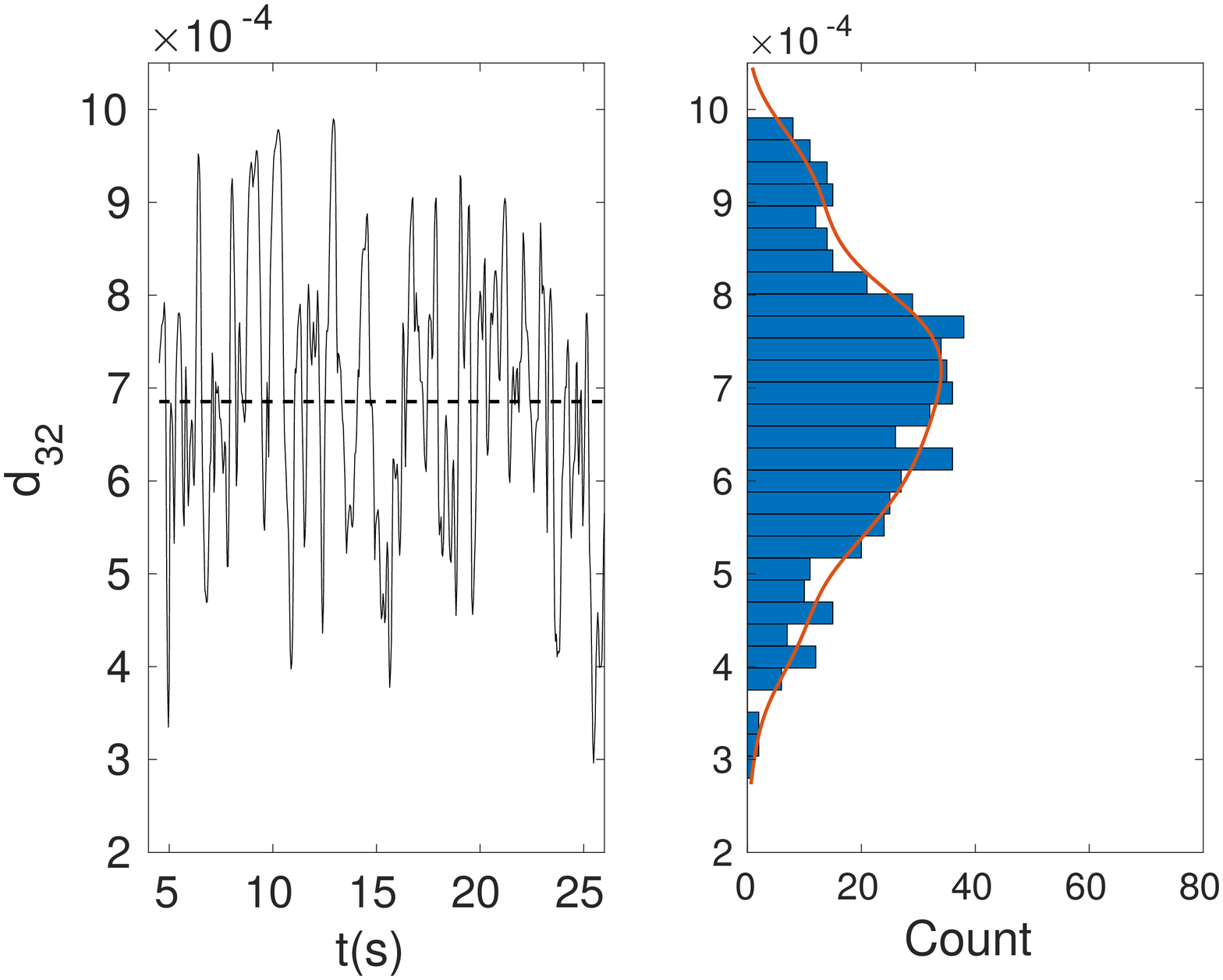}
        
        \caption{$x=0.9\;\mbox{m}$, $z=0.56\;\mbox{m}$.}
    \end{subfigure}
    \caption{Representative time signals (left panels) and histograms (right panels) for the Sauter mean diameter, $d_{32}$ at two downstream locations on the centerline. Dotted lines denote mean values.}
     \label{fig:d32_pdf}
\end{figure*}

We plot a time signal and histograms of the $d_{32}$ diameters at different plume locations in figure $\ref{fig:d32_pdf}$. We can see that there is a high variability of the diameter about the time averaged mean value. This variability can be observed in LES since we are solving for 3D time-dependent number density fields for each bin and so $d_{32}$ can be evaluated at any grid point at any timestep.  
At $x=0.75\;\mbox{m}$, where the averaged $d_{32}$ diameter is about $800\;\mu\mbox{m}$, we see a variability from $300\;\mu\mbox{m}$ to $1000\;\mu\mbox{m}$. The PDF also shows non-Gaussianity, with two peaks, clearly showing that the average values of $d_{32}$ do not provide a complete description. The peak near $900\;\mu\mbox{m}$ is affected by the discrete bins. This scale corresponds to the first bin considered in the summation corresponding to $d=1\;\mbox{mm}$.  

\begin{figure*}
 \hspace{-2cm}
    \centering
    \begin{subfigure}[t]{0.45\textwidth}
        \centering
        \includegraphics[width=1.25\linewidth]{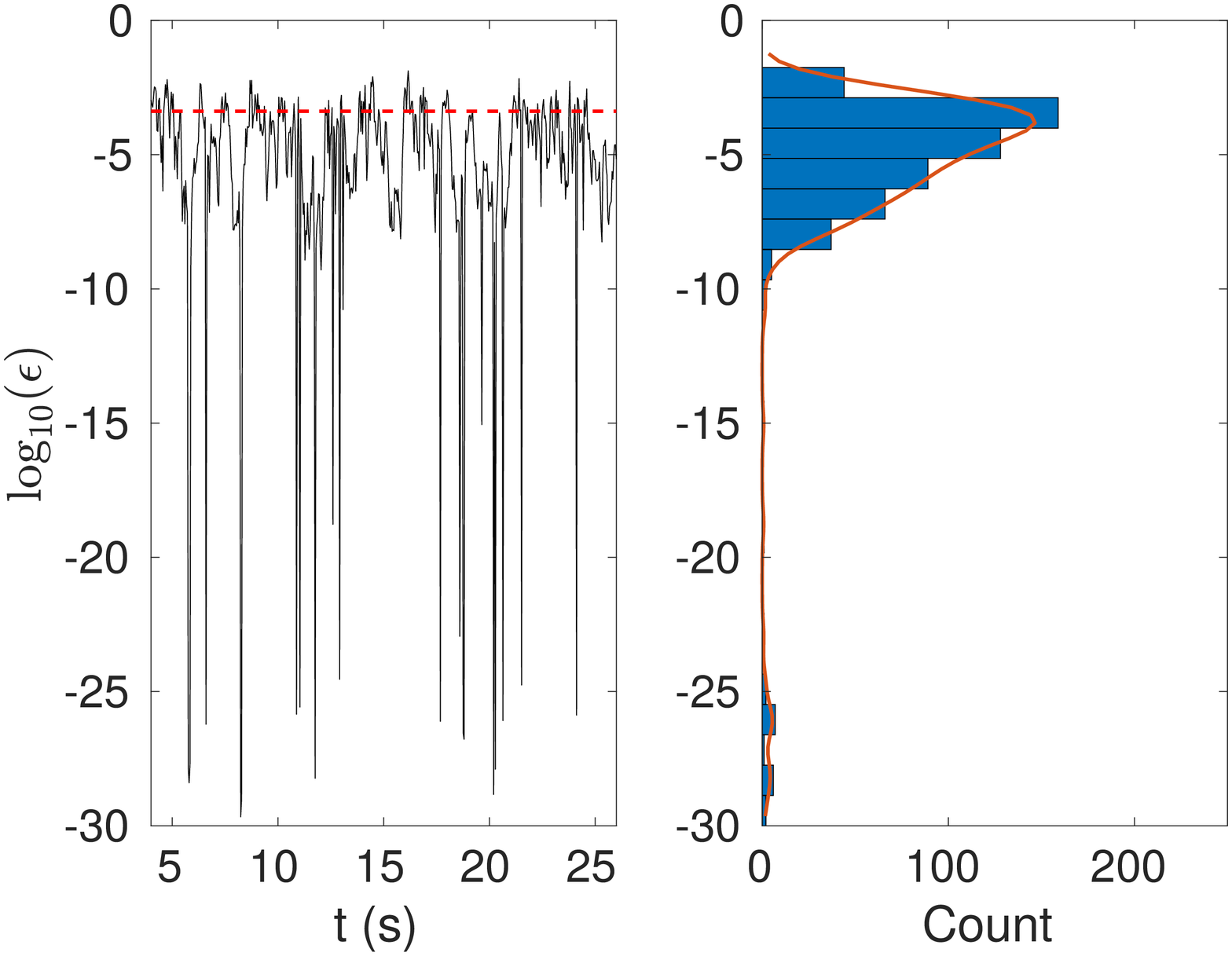}
        \caption{$x=0.75\;\mbox{m}$, $z=0.56\;\mbox{m}$.}
    \end{subfigure}
     \hspace{1cm}
    \begin{subfigure}[t]{0.45\textwidth}
        \centering
        \includegraphics[width=1.25\linewidth]{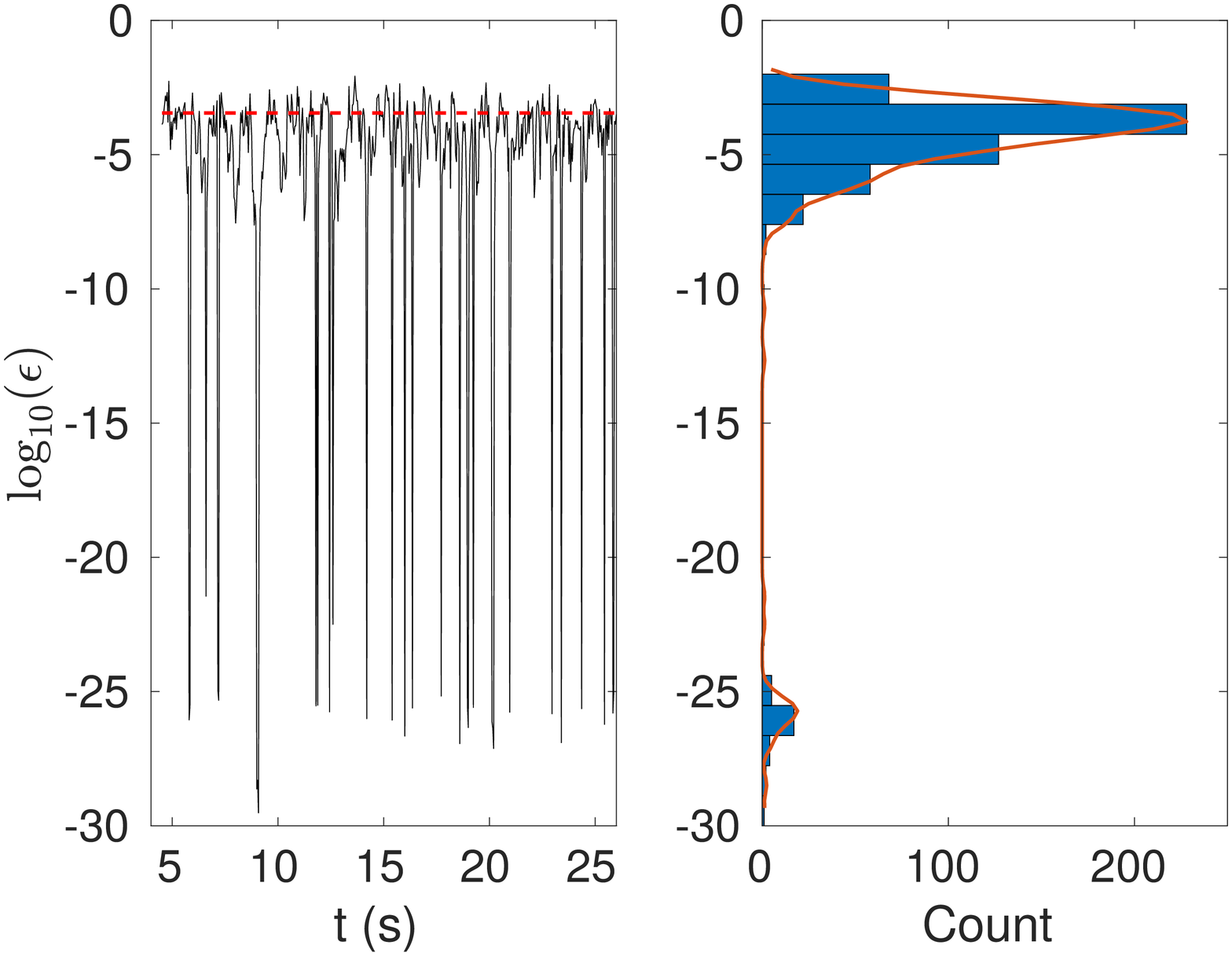}
        \caption{$x=0.9\;\mbox{m}$, $z=0.56\;\mbox{m}$.}
    \end{subfigure}
    \caption{Representative time signals (left panels) and histograms (right panels)  of $\log_{10}(\epsilon)$ at two downstream locations on the centerline. $\epsilon$ is in m$^2$/s$^3$. Dotted lines denote mean value.}
     \label{fig:eps_pdf}
\end{figure*}

Recall that in LES the breakup time scale depends upon the local value of dissipation, which is also known to be highly intermittent in turbulent flows. In order to illustrate the (grid-scale averaged) dissipation intermittency, in figure $\ref{fig:eps_pdf}$ we show time signals of the logarithm of dissipation as well as histograms.  The histogram of the logarithm of dissipation is reminiscent of Gaussian (log-normality) but with a non-Gaussian highly asymmetric tail and some outliers at very low dissipation, corresponding possibly to laminar regions outside the plume. This highly variable quantity then determines the local time scale of droplet breakup in the LES model.
 
\begin{figure*}
  \hspace{-2cm}
    \centering
    \begin{subfigure}[t]{0.45\textwidth}
        \centering
        \includegraphics[width=1.25\linewidth]{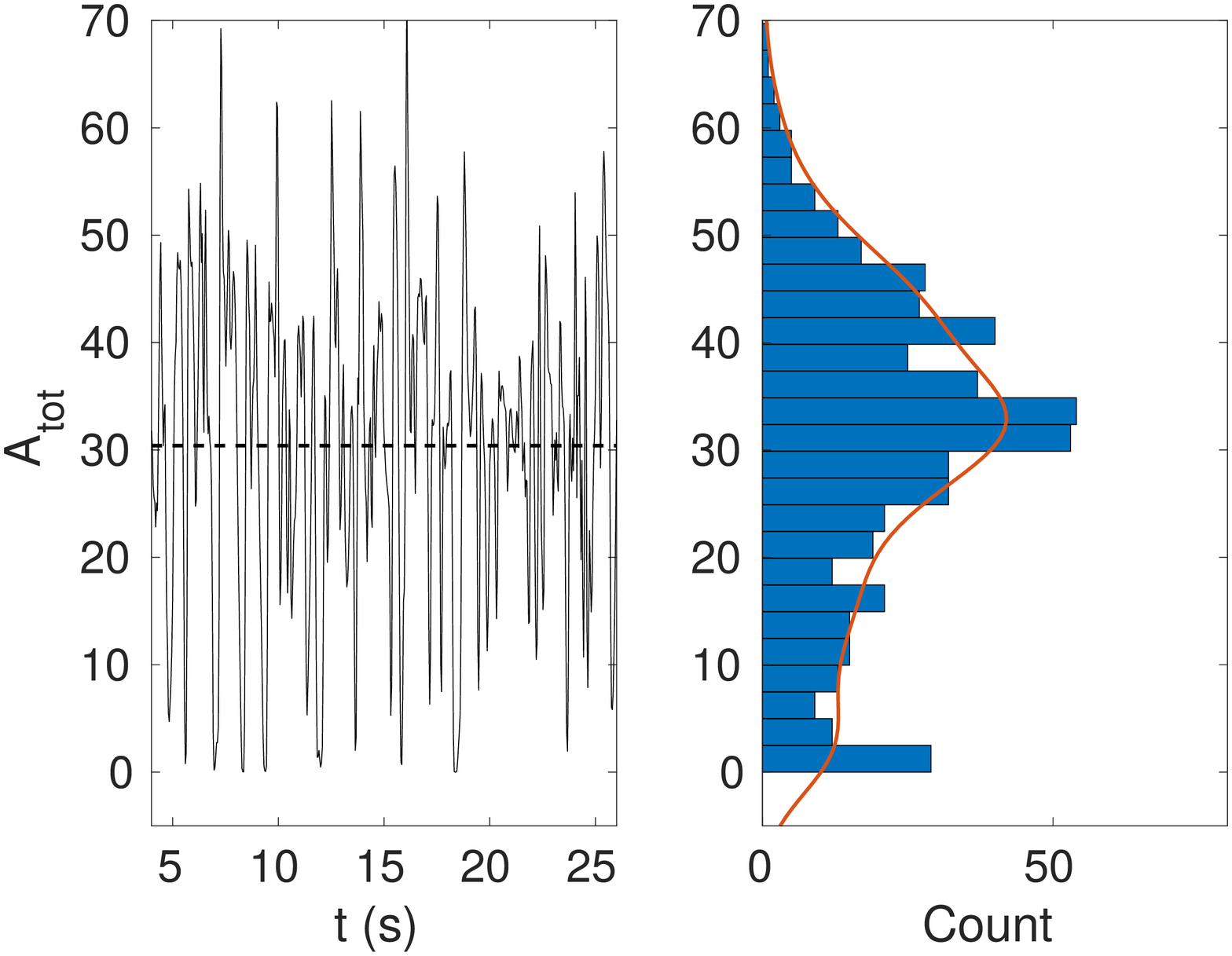}
        \caption{$x=0.75\;\mbox{m}$, $z=0.56\;\mbox{m}$.}
    \end{subfigure}
    \hspace{1cm}
    \begin{subfigure}[t]{0.45\textwidth}
        \centering
        \includegraphics[width=1.25\linewidth]{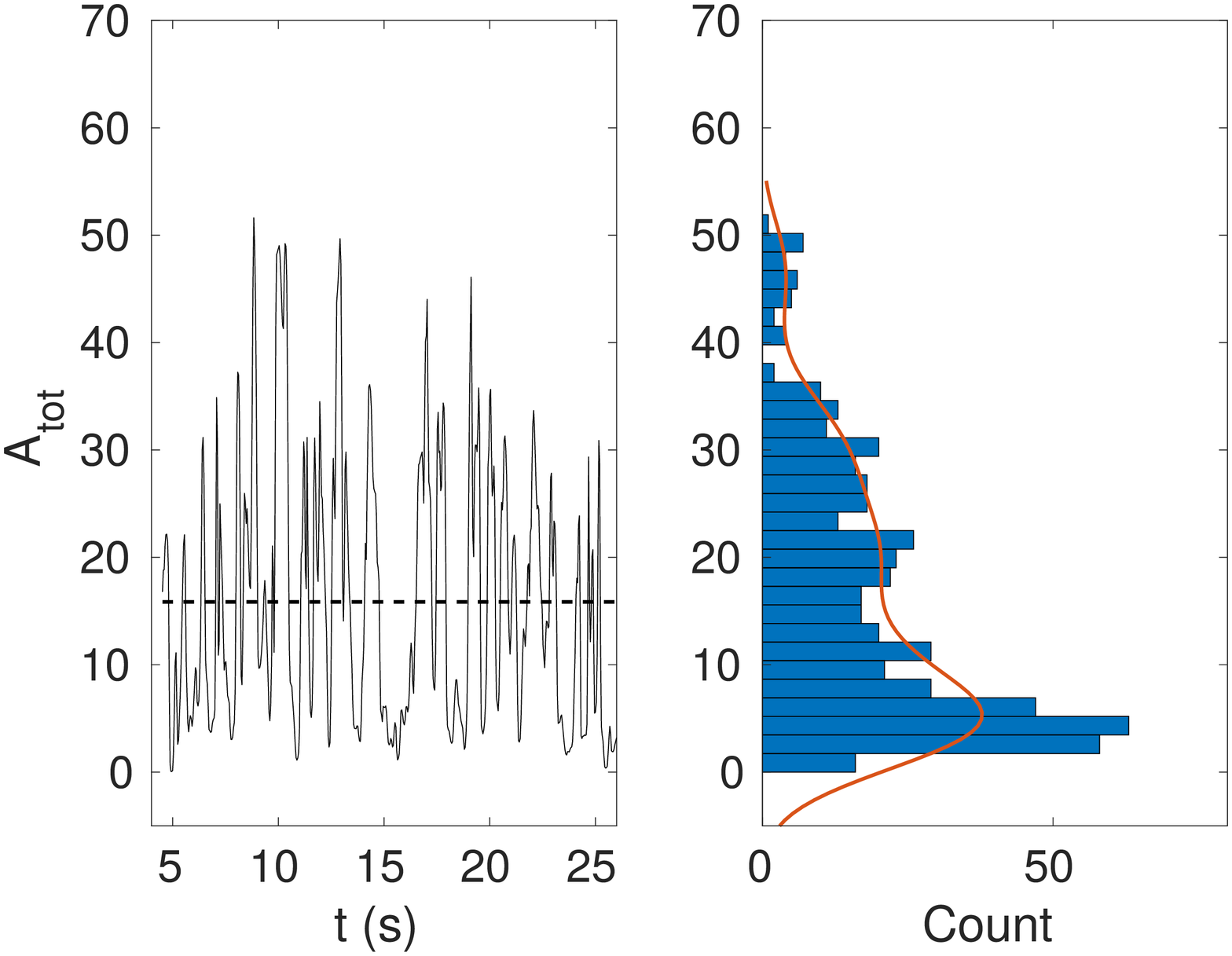}
        \caption{$x=0.9\;\mbox{m}$, $z=0.56\;\mbox{m}$.}
    \end{subfigure}
    \caption{Representative time signals (left panels) and histograms (right panels)  of total surface area of the oil per cubic meter of fluid at two downstream locations on the centerline. $ A_{tot}$ is in m$^2$/m$^3$. Dotted lines denote mean values.}
       \label{fig:area_pdf}
\end{figure*}

 Next, we consider a property that is crucial in determining reaction rates for processes that occur on the droplet surface such as bio-degradation. The total rate of bio-degradation will depend on the total surface area of the oil available for microorganisms to act upon. Given the instantaneous concentration of droplets in each bin, the instantaneous total surface area available for surface reactions can be evaluated according to 
 \begin{equation}
     A_{tot}({\bf x},t)  = \sum \limits_{i=1}^{N_d} \tilde{n}_i({\bf x},t)\, (\pi d_i^2).
 \end{equation}
 Representative signals and histograms of $A_{tot}({\bf x},t)$  are shown in figure $\ref{fig:area_pdf}$, again at the two locations $x=0.75\;\mbox{m}$ and 
 $x=0.9\;\mbox{m}$ at $z=0.56\;\mbox{m}$ and the plume center in the transverse direction.

 We can see from the panels in figure $\ref{fig:area_pdf}$ that there is a high variability of the total area about the mean of about $30\;\mbox{m}^2$ per cubic meter of water at $x=0.75\;\mbox{m}$ and about $16\;\mbox{m}^2$ per cubic meter of water at $x=0.9\;\mbox{m}$ (even though one may expect smaller droplet sizes to be associated with an increase in total surface area, further downstream the total oil concentration has also decreased due to turbulent transport thus leading to the smaller area there). The root mean square of the surface area distribution is quite significant, of similar order of magnitude to the mean area. 
 
\begin{figure*}
 \hspace{-2cm}
    \centering
    \begin{subfigure}[t]{0.45\textwidth}
        \centering
        \includegraphics[width=1.25\linewidth]{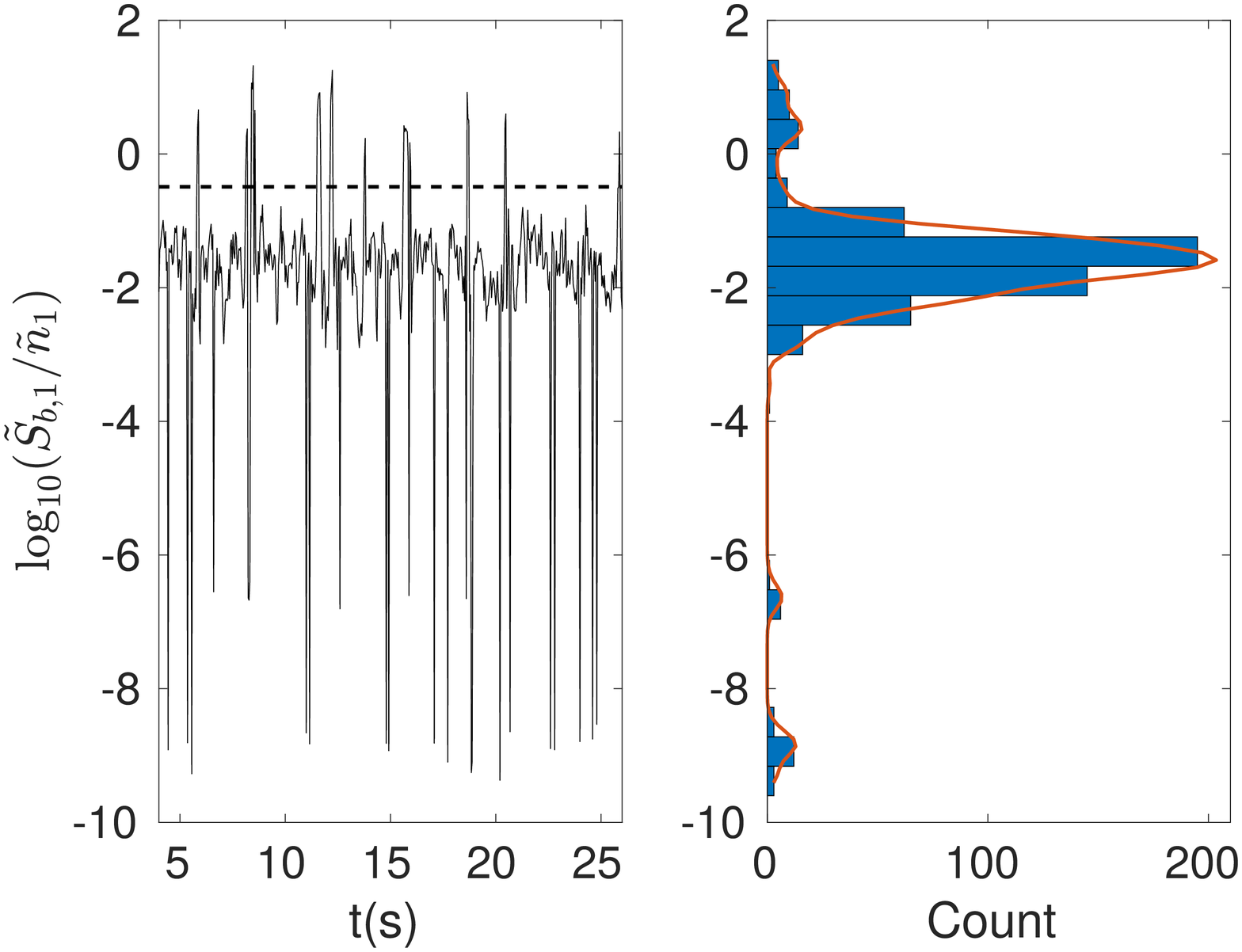}
        \caption{$x=0.75\;\mbox{m}$,
        $z=0.56\;\mbox{m}$.}
        \label{fig:rhs_pdfa}
    \end{subfigure}
   \hspace{1cm}
    \begin{subfigure}[t]{0.45\textwidth}
        \centering
        \includegraphics[width=1.25\linewidth]{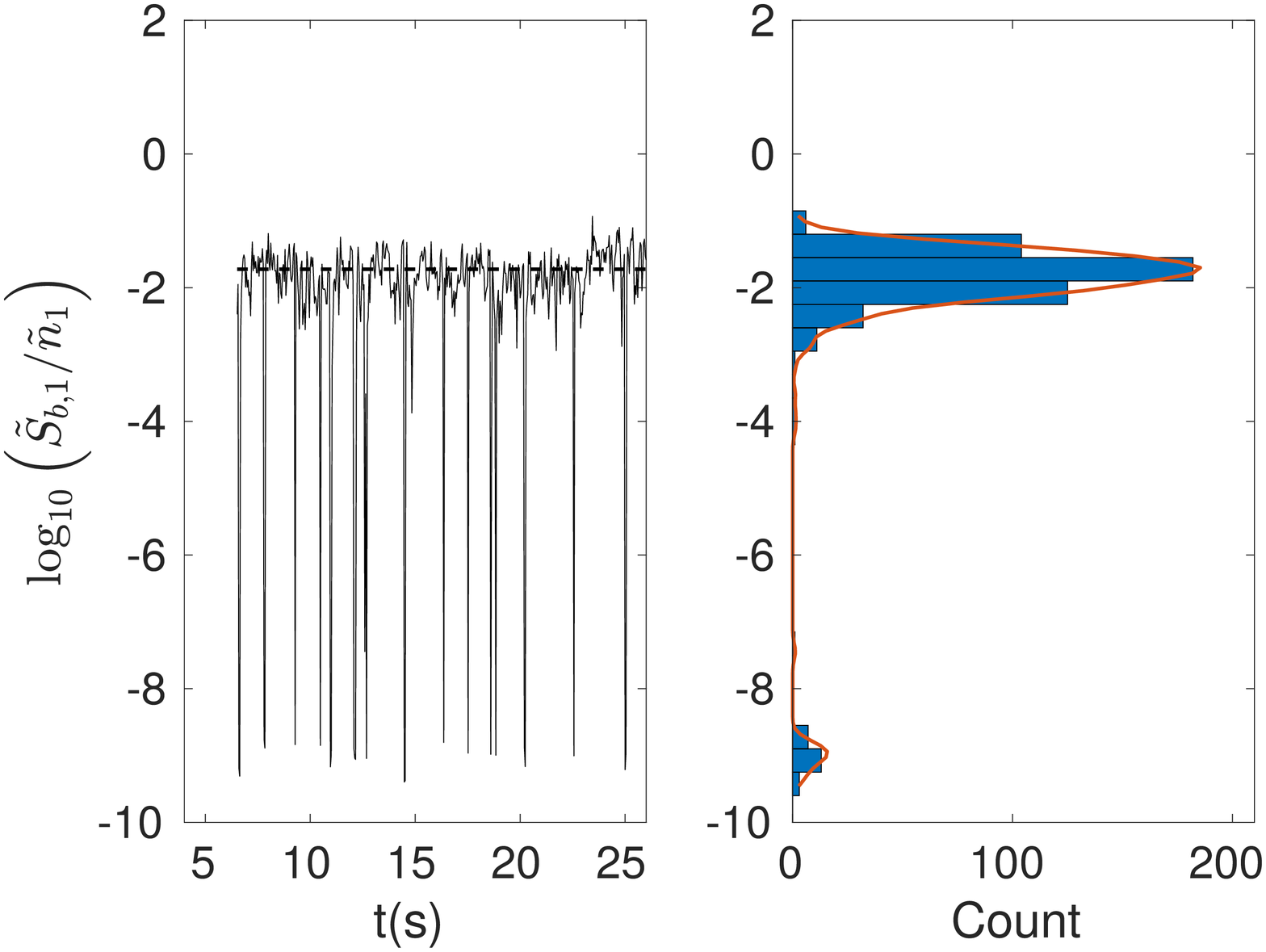}
        \caption{$x=1.21\;\mbox{m}$, $z=0.56\;\mbox{m}$.}
    \end{subfigure}
    \caption{Representative time signals (left panels) and histograms (right panels) of $\log_{10}(\tilde{S}_{b,i}/\tilde{n}_i)$ for $d=20\;\mu\mbox{m}$ plotted at two downstream locations of the plume at a fixed height. The values of $\tilde{S}_{b,i}/\tilde{n}_i$ are given in $1/s$.  Dotted lines denote mean values.}
       \label{fig:rhs_pdf}
\end{figure*}

  It is also instructive to examine time signals and statistics of the breakup source term for each droplet size, $\tilde{S}_{b,i}$, normalized by the concentration. This normalization can be interpreted as an inverse time scale for the droplet breakup, i.e. it tells us the inverse of the time taken for the number of droplets in any given bin to change appreciably over its existing value, at any given scale. In figure $\ref{fig:rhs_pdf}$ we show representative signals of $\tilde{S}_{b,i}({\bf x},t)/\tilde{n}_i({\bf x},t)$ in logarithmic scale, as well as its  histograms at two locations. As can be seen from the right panel of figure $\ref{fig:rhs_pdfa}$, the average values are around $0.5\; \mbox{s}^{-1}$, with very large variability about this value. It means that it takes about $2$ seconds for the local breakup rate to appreciably change the local concentration of droplets of size $20\;\mu\mbox{m}$ but occasionally the breakup can be far more rapid.
 
\begin{figure}
    \centering
    \begin{subfigure}[t]{0.48\textwidth}
        \centering
        \includegraphics[width=1\linewidth]{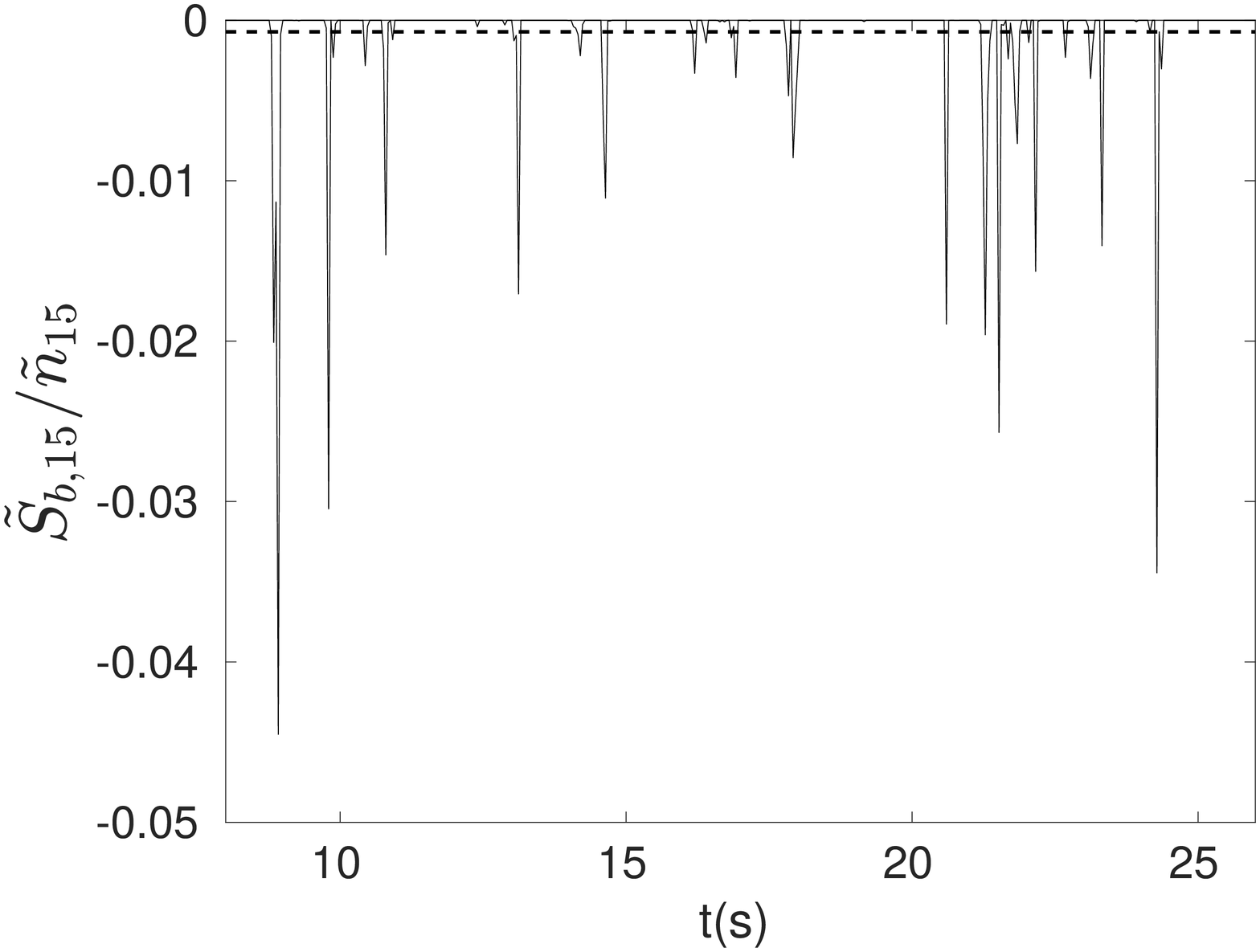}
        \caption{$x = 0.75\;\mbox{m}$. } \label{fig:rhsx115}
    \end{subfigure}
    \hfill
    \begin{subfigure}[t]{0.48\textwidth}
        \centering
\includegraphics[width=1\linewidth]{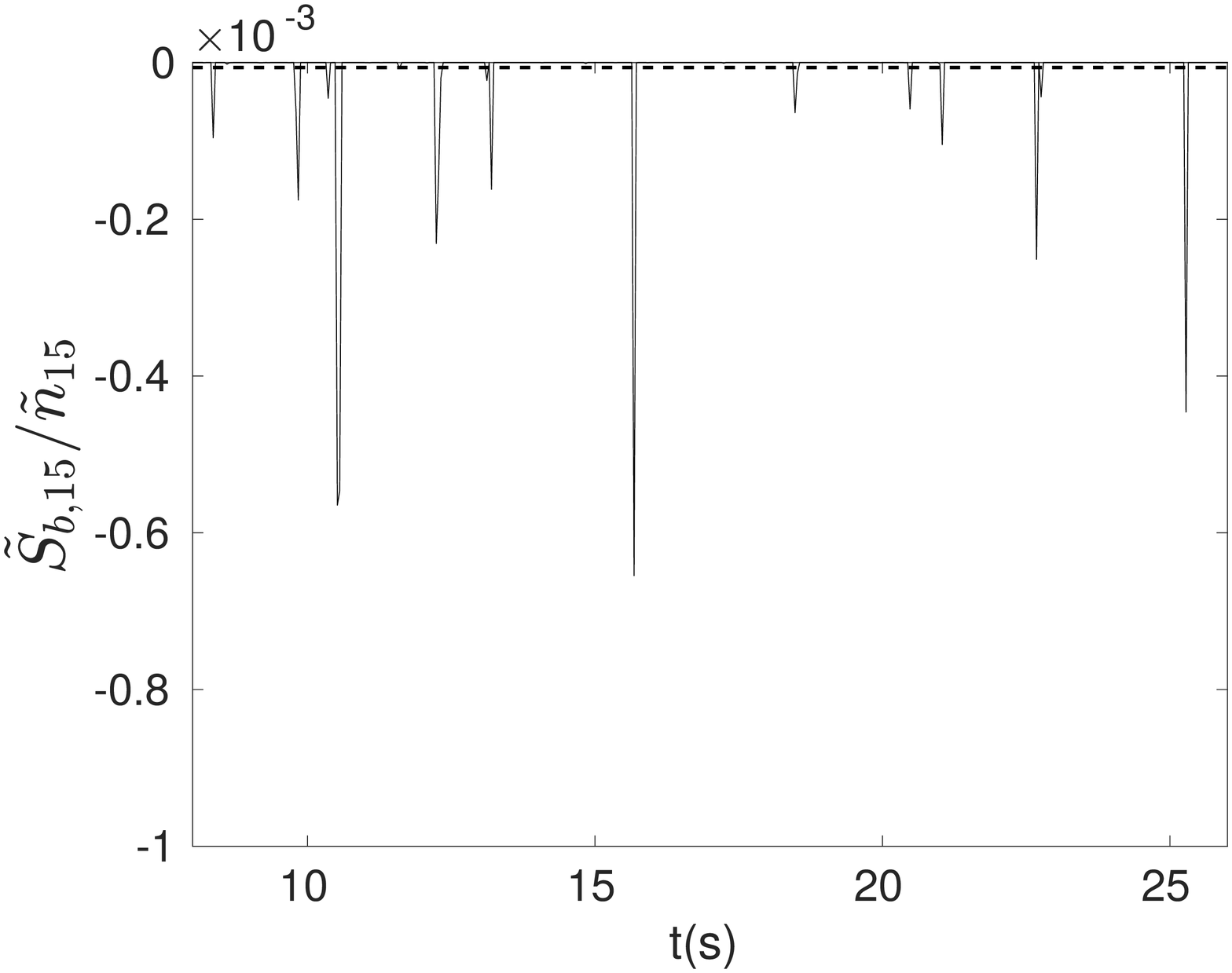} 
        \caption{$x=1.56\;\mbox{m}$} 
        \label{fig:rhsx615}
    \end{subfigure}
   \vspace{1cm}
    \begin{subfigure}[t]{0.48\textwidth}
        \centering
        \includegraphics[width=1\linewidth]{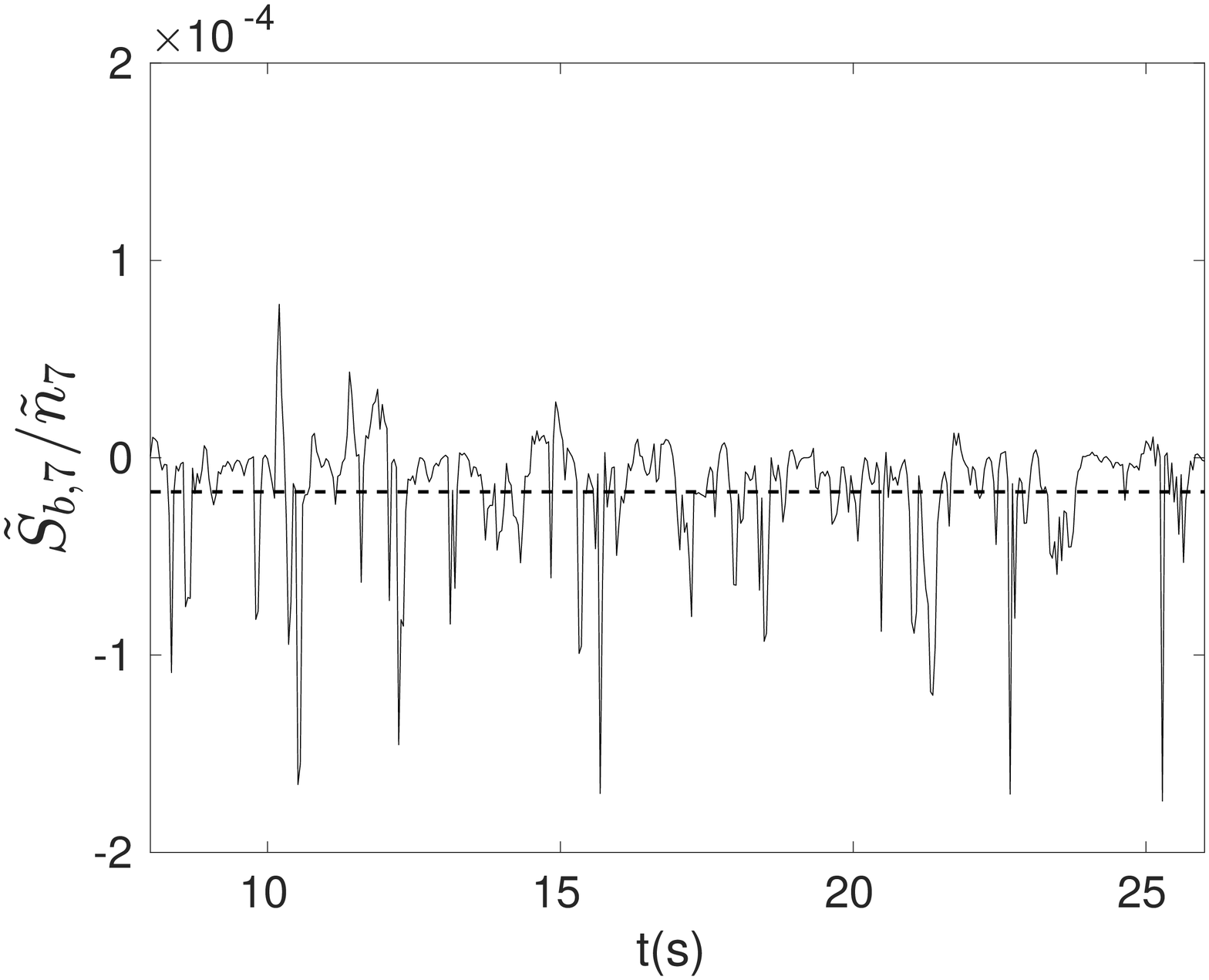} 
        \caption{$x =1.56\;\mbox{m}$} \label{fig:rhsx67}
    \end{subfigure}
    \hfill
    \begin{subfigure}[t]{0.48\textwidth}
      \centering
        \includegraphics[width=1\linewidth]{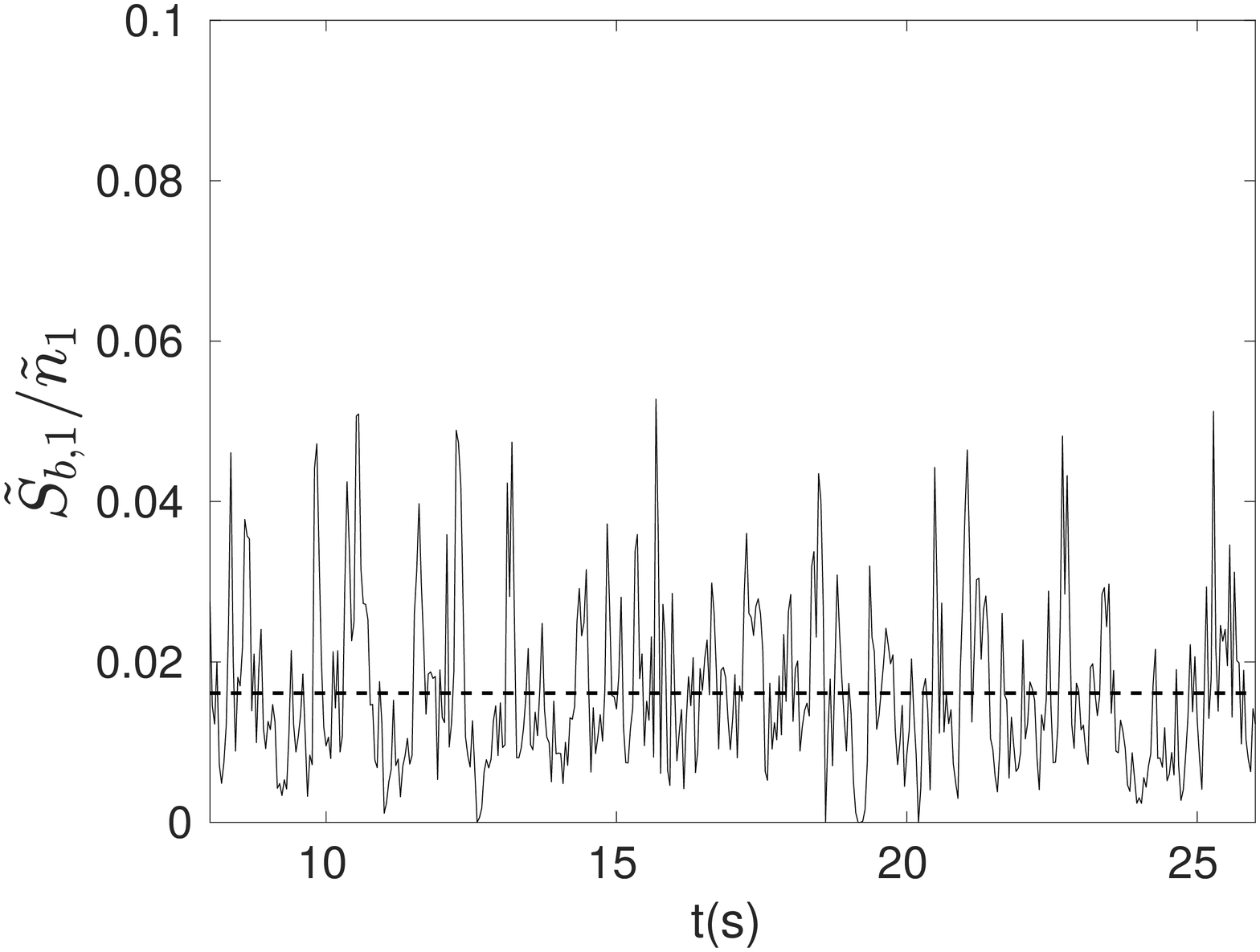}
        \caption{$x = 1.56\;\mbox{m}$} \label{fig:rhsx61}
    \end{subfigure}
    \caption{Time history of $\tilde{S}_{b,i}$ normalized by concentration for different droplet sizes at $z=0.56\;\mbox{m}$ on the centerline.  (\subref{fig:rhsx115}) and (\subref{fig:rhsx615}) represent the droplet of size $1000\;\mu\mbox{m}$ at two different $x$ locations, (\subref{fig:rhsx67}) is the time history for $d=432\;\mu\mbox{m}$ and (\subref{fig:rhsx61}) is for $d=20\;\mu\mbox{m}$. Dotted lines denote mean values. }
    \label{fig:t_rhs}
\end{figure}

 Finally, we document the breakup source term by plotting it in linear units for different droplet sizes, at two different locations as shown in figure $\ref{fig:t_rhs}$. We can see that the time signals for the source term  are highly intermittent, with the largest size (i.e. 15th bin) acting as a source for the smaller ones (negative source term in its transport equation).  Further downstream at $x=1.56\;\mbox{m}$, figure  $\ref{fig:rhsx615}$ shows much smaller frequencies indicating a decreased breakup of the largest droplets. Some of the intermediate bins display both positive and negative values,
 as some intermediate droplet sizes act as both sources and sinks at different locations along the plume (for example see panel for $S_{b,7}/n_7$ in figure $\ref{fig:rhsx67}$).
 
 We still see a significant variance in the volume median diameter ($d_{32}$) at this location despite the magnitude of the normalized source terms for the larger droplets being small. Clearly, the turbulent nature of the flow prevents us from solely relying on the averaged quantities to provide us with a complete view of the droplet size distribution in this flow, while LES contains significant amount of new information regarding the fluctuations, at least down to the grid scale.
  
\section{Conclusions}
\label{sec:conclusions}
Prediction of droplet size distributions in a turbulent flow is essential for understanding the dynamics of many types of multiphase flows.  
We have proposed a method to couple LES with a population balance equation to study the evolution of polydisperse oil droplets in turbulence. We use the method of classes to discretize the droplet size range into contiguous subclasses and consider the case of round droplets at relatively low volume fraction for which coalescence can be neglected. Using a jet in crossflow as a flow application inspired by previous studies on deep-water oil spills, the model can be used to predict the turbulent transport of droplets of various sizes while accounting for breakup due to the turbulent flow field.

 We follow the general procedure of \citet{Konno1982}, \citet{Prince1990} and \citet{Tsouris1994} in which the breakup is modeled as due to collision of turbulent eddies with droplets. Previous models assumed the droplet size to be in the inertial range of turbulence and use Kolmogorov scaling (K41) for the velocity increment  valid for the inertial range. For many applications the droplet size range can lie in the viscous subrange.
 We have thus proposed a model that includes the effect of the viscous range of scales of  turbulence using a generalized structure function approach to characterize the eddy fluctuation velocity. 
 The  formulation contains an adjustable parameter $K^*$ that has been fitted using experimental data.  
 To reduce computational cost, we parameterize the breakup frequency in 
 terms of the various (locally changing) nondimensional parameters, and provide practical fits that enable rapid calculation.
 
 The population equation along with the breakup 
 model is then implemented in an LES framework. LES enables us to accurately simulate turbulent shear flows like jets and plumes and study 
 the advection and  breakup of, for example, oil droplets in these flows.  We tested the formulation by comparing the size distributions of oil droplets obtained at different locations along the plume with the experimental data of \citet{Murphy2016} and obtained good agreement for the relative droplet size distribution. 
 
 Finally, we used the LES results to quantify various new properties of the distribution as it refers to the inherent variability of turbulence. We show how the LES provides information on the variability of the median diameter, the total area available for surface reactions and illustrate the highly non-Gaussian properties of the source (reaction) terms in the transport equations for each bin of droplet concentration fields. 
 
 Clearly, additional followup studies are required to explore in more detail various relevant aspects such as the possible effects of the initial size distribution assumed at the nozzle exit (here we assumed a single droplet size at the injection), the effects of different breakup probability models ($\beta(d_i,d_j))$, the effects of changing grid resolution in LES, the possible effects of various subgrid-scale models for the momentum and scalar fluxes, and many other possible extensions such as combining with Lagrangian models for the subgrid-scale velocity gradient fluctuations \citep{JohnsonMeneveau2018}.

\section*{Acknowledegements} 
The authors thank J. Katz, D. Murphy, X. Xue, R. Ni, B. Chen, S. Sokolofsky and M.C. Boufadel for useful conversations and insights. This research was made possible by a grant from the Gulf of Mexico Research Initiative. Computational resources were provided by the Maryland Advanced Research Computing Center (MARCC). Data are publicly available through Gulf of Mexico Research Inititative Information and Data Cooperative (GRIIDC) at \url{https://data.gulfresearchinitiative.org} (doi:10.7266/12C3ZMTD).

 \appendix 
 \section{Fits for breakup frequency integral}\label{appA}
\begin{figure}
\centering
\includegraphics[width=0.6\linewidth]{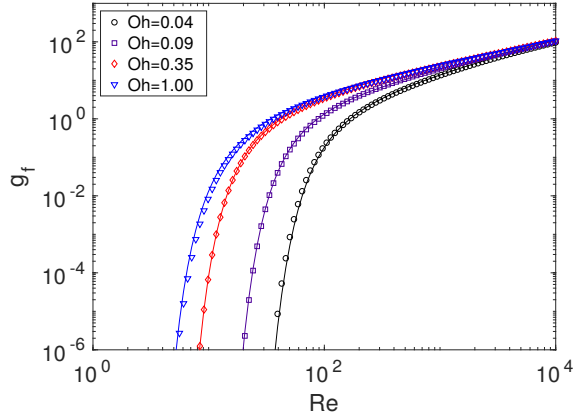}
\caption{The fit is represented by the dashed lines whereas the numerically computed integral are the various symbols for  $\Gamma=10.5$.}
\label{fig:fit_log_m}
\end{figure}

In this section we discuss the fit for the integral in the equation for the breakup frequency derived in ($\ref{eqn:br_freq_nd}$). 
\begin{equation}\label{eqn:fit_integral}
g_f(Re,Oh,\Gamma) = \int_0^1 r_e^{-11/3} (r_e+1)^2 \left(1+\left(\frac{r_e Re}{\gamma_2}\right)^{-2}\right)^{-1/3} \Omega(Oh,Re,\Gamma;r_e)dr_e.
\end{equation}
As it would be computationally intensive to evaluate an integral at every grid point for every timestep we develop an empirical form of the integral as a function of the two nondimensional parameters $Re$ and $Oh$, for discrete values of $\Gamma$. 
We begin by plotting the  integral $g_f$ for a wide range of Reynolds $Re$  and Ohnesorge $Oh$ numbers for a fixed value of $\Gamma=10.5$. This is shown in figure $\ref{fig:fit_log_m}$ as the symbols for different Oh numbers. The value of $\Gamma$ is chosen based on the physical properties of the oil in \citet{Murphy2016}.  We can see that we have a power law behavior for higher Re with a sharp cutoff for the small Re. The cut-off location
is a function of the Ohnesorge ($Oh$) number. This suggests that we could fit   $g_f$ using two power laws to capture the two extremes. The fit equation can be written as 
\begin{equation}\label{eqn:fit_l}
G(Re_i,Oh_i) = ax^b + cx^{d} -e,
\end{equation}
where $G=\log_{10}(g_{f})$, $x = \log_{10}(Re)$, and $a,b,c,d,e$ are functions of $Oh$. We use Matlab's curve fitting toolbox to carry out the fitting procedure. The toolbox uses a Levenberg-Marquardt algorithm to provide the best fit for the data. We find that the coefficient $b$ can be fixed at $b=0.45$. The other coefficients can be expressed as functions of the Ohnesorge number  using the following fits as a function of $y=Oh$:

\begin{equation}\label{eqn:fit_coeff}
\begin{gathered}
a(y) = a_1\exp(-a_2 y) + a_3\exp(-a_4 y) \\
b = 0.45\\
log_{10}[ - c(y)] = \frac{c_1 y^{-c_2}} {1+c_3 y^{c_4}}  \\
d(y) = -\frac{d_1 y^{-d_2}}{1+d_3y^{-d_4}} \\
 \log_{10}[e(y)] = e_1\exp(-e_2\log_{10}(y+1)) + e_3\exp(-e_4\log_{10}(y+1)) 
\end{gathered}
\end{equation}

\begin{figure}
\centering
\includegraphics[width=0.6\textwidth]{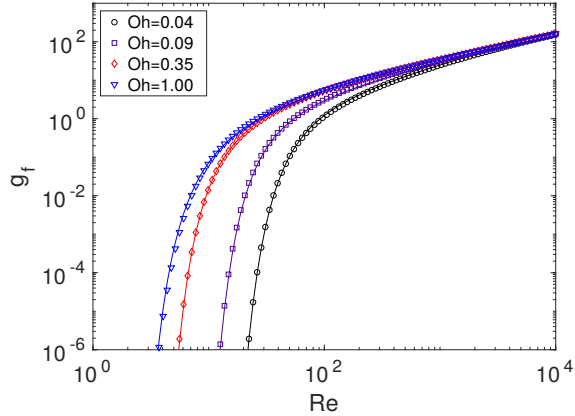}
\caption{The fit is represented by the dashed lines whereas the numerically computed integral are the various symbols for  $\Gamma=5.45$.}
\label{fig:fit_log_s}
\end{figure}
The coefficients in (\ref{eqn:fit_coeff}) for the particular value of $\Gamma$ chosen, are given in table \ref{tab:coeff} for $\Gamma=10.5$ and in table \ref{tab:coeff2} for $\Gamma=5.45$. The fit is valid for droplet Reynolds number less than $10^4$ and for $0.006\leq Oh \leq 2$.
We numerically evaluate the integral in ($\ref{eqn:fit_integral}$) and compare it with by evaluating the algebraic fit from ($\ref{eqn:fit_coeff}$) in figure $\ref{fig:fit_log_m}$ for four $Oh$ numbers. The fit is plotted using the dashed lines of different color, while the numerically evaluated integral is represented by the symbols. We see that we have good agreement in the parameter range considered.

We can use the same methodology for a different value of $\Gamma$. As an example, we show in figure $\ref{fig:fit_log_s}$ the fit for the breakup integral for $\Gamma =5.45$, based on the oil properties from \citet{Johansen2013}. We see that we can obtain a  good fit for different values of $\Gamma$. The coefficients for intermediate $\Gamma$ values can be obtained by interpolating the function $G(Oh_i,Re_i)$ between the two cases. For example, to obtain $g_f$ for $\Gamma=8$ we first linearly interpolate $G$ from $\Gamma_1=5.45$ and $\Gamma_2=10.5$ as,

\begin{table}
\centering
 \begin{tabular}{c c c c c}
 Variable &\multicolumn{4}{c}{Coefficients} \\ [0.5ex]
 $x_k$ & $k=1$ & $k=2$ & $k=3$ & $k=4$ \\
\midrule
a & 2.374 & 19.88 & 2.788 & 0.07416 \\ [0.4cm]
 c & 1.41 & 0.245 & 5.178 & 0.83 \\[0.4cm]
 d & 5.313 & 0.4541 & 0.4981 & 0.4219 \\[0.4cm]
 e & 0.415 & 41.09 & 0.5088 & 0.4604 \\
\end{tabular}
\caption{$\Gamma = 10.5$.}
\label{tab:coeff}
\end{table}

\begin{table}
\centering
 \begin{tabular}{c  rrrr} 
 Variable &\multicolumn{4}{c}{Coefficients} \\ [0.5ex] 
 $x_k$ & $k=1$ & $k=2$ & $k=3$ & $k=4$ \\
 \midrule
 a & 2.392 & 26.76 & 2.877 & 0.1244 \\ [0.4cm]
 c & 0.5446 & 0.3776 & 12.67 & 1.462 \\ [0.4cm]
 d & 4.172 & 0.5492 & 0.5079 & 0.4879 \\ [0.4cm]
 e & 0.4113 & 55.94 & 0.5125 & 0.7182 \\

\end{tabular}
\caption{$\Gamma = 5.45$.}
 \label{tab:coeff2}

\end{table}

\begin{figure}
\centering
\includegraphics[width =0.6\linewidth]{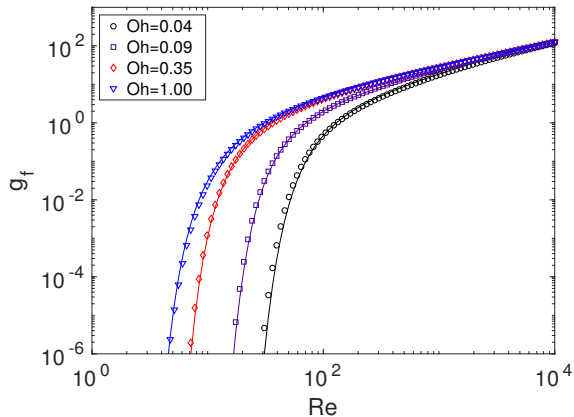}
\caption{The fit is represented by the dashed lines computed by interpolating between $\Gamma_1=5.45$ and $\Gamma_2=10.5$. The symbols represent the numerically computed integral for  $\Gamma=8$.}
\label{fig:fit_log_8}
\end{figure}

\begin{equation}
    G(Re_i,Oh_i;\Gamma) = G(Re_i,Oh_i;\Gamma_1) + \frac{G(Re_i,Oh_i;\Gamma_2) -G(Re_i,Oh_i;\Gamma_1)}{\Gamma_2-\Gamma_1}(\Gamma-\Gamma_1) 
\end{equation}
The integral can then be obtained as $g_f = 10^{G}$.
The results from the interpolation for $Oh=0.042,0.09, 0.35$ and $1$  are shown in figure $\ref{fig:fit_log_8}$. We see that even with a simple linear interpolation we can obtain satisfactory results. 

In order to quantify the speedup obtained by using the fits compared to the integral, we can calculate the CPU time per simulation time step for each case. We find that the LES with the fits is 60 times faster than an LES  with the breakup frequency calculated with the numerical integration of the integral at every grid point and time step. This speedup is more pronounced when the grid is refined. The fits are calculated using vectorized operations that are fast and efficient even on fine grids. The integral on the other hand has to be evaluated at every grid point as the integrand is a function of position.

\section{Specifying the jet injection velocity in coarse LES}

In this section we explain how the magnitude of the vertical jet velocity is determined so as to ensure that LES reproduces relevant quantities from the experiment.  As it would be computationally expensive to resolve the small scale structures within the nozzle in our LES, the injected jet is modeled using a locally acting body force. The jet in the experiment has a diameter of $d_{expt} = 4\;\mbox{mm}$ while the effective grid spacing of the LES is $5.2\;\mbox{mm}$. In order to properly resolve the turbulent inflow in the nozzle, much finer grids would be required in that region. In the LES model we take the view that we simulate a `coarser' jet injection process which matches the experiments further downstream where the jet has already grown to a specified length scale and the jet centerline velocity and turbulence fluctuations have decreased. We match the length-scales and centerline velocity at a particular downstream location where the jet has already grown sufficiently to match what the simulation can begin to resolve. 

\begin{figure} 
\hspace{-0.5cm}
\centering
\begin{subfigure}{0.45\textwidth}
        \centering
        \includegraphics[width=1.2\linewidth]{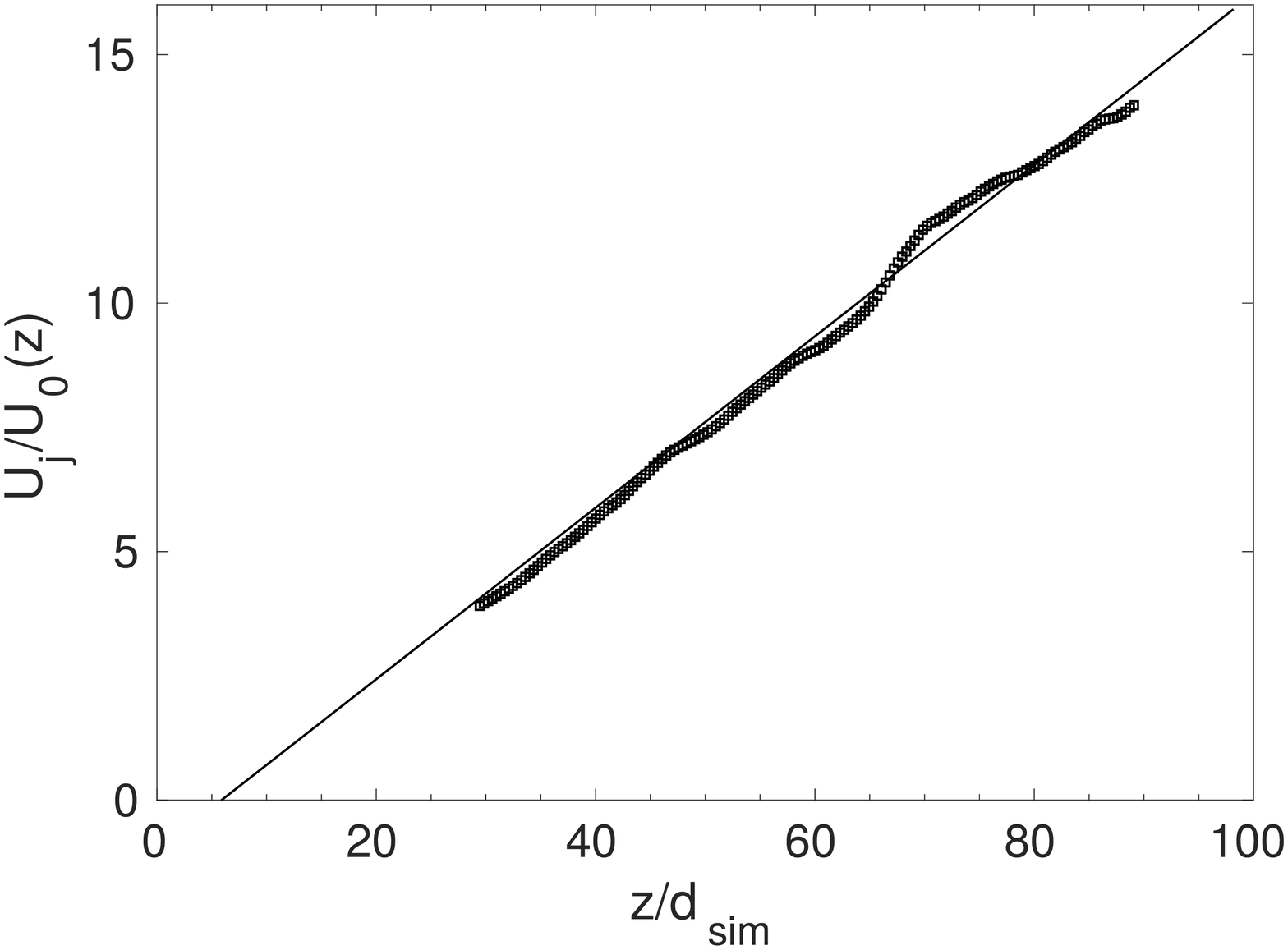}
    \end{subfigure}
    \hspace{0.5cm}
    \begin{subfigure}{0.45\textwidth}
 \centering
\includegraphics[width=1.2\linewidth]{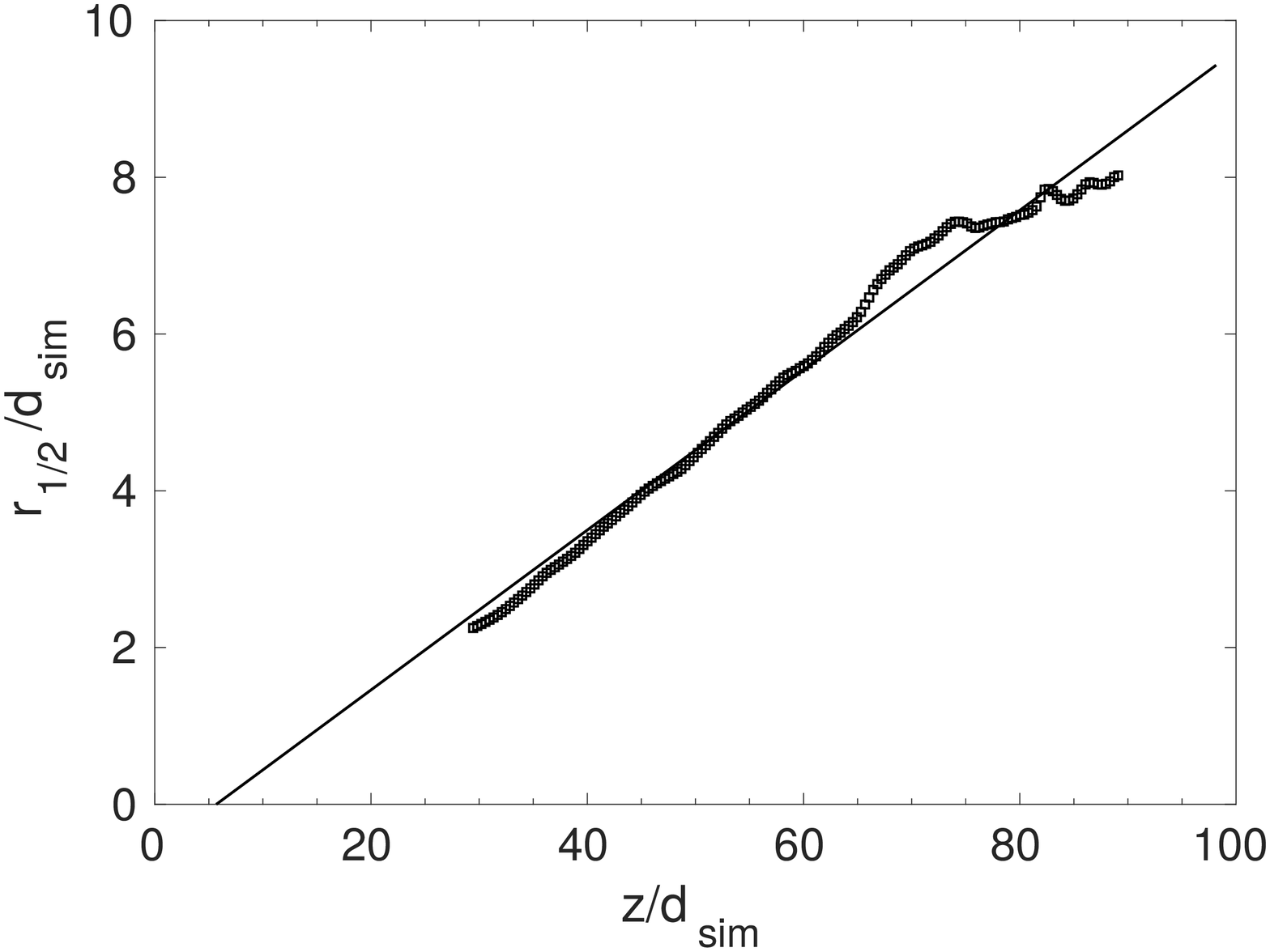}
\end{subfigure}
\caption{Left panel: Variation of the mean centerline velocity with axial distance in LES of jet without crossflow using imposed body force. Right panel: Half-width of jet plotted as a function of axial distance. In both cases the symbols represent the simulated data, and the solid line represents the fit. The length scales in both plots are normalized w.r.t the simulation diameter $d_{sim}=7.8\;\mbox{mm}$ }
\label{fig:jet_char}
\end{figure}

We calibrate the computational setup on a jet without crossflow and refer to well-established jet scaling laws and known correlations. First we verify that the simulated jet with an imposed smoothed vertical force reproduces well-known behaviors. 
We   examine the centerline velocity $U_0(z)$ and the half width of the jet ($r_{1/2}$) defined as, 
\begin{equation}
\langle U(z,r_{1/2}(z),0)\rangle = \frac{1}{2}U_0(z),    
\end{equation}
where $z$ is the distance downstream from the nozzle exit or simulated injection point.
A simulation is run using the same resolution and all other relevant parameters as in the crossflow simulation except the crossflow, with a vertical body force that results in a given maximum jet velocity $U_{j,sim}$ in the region where the force is applied.  
 We plot the centerline velocity and the half width from our simulation in figure $\ref{fig:jet_char}$. We can see that we recover the linear growth of the inverse centerline velocity as well as the linear scaling for the half-width.
 More specifically, the centerline velocity for a turbulent jet decays linearly with $z$ \citep{Hussein1994} and can be written as
 \begin{equation}\label{eqn:centerline}
        \frac{U_j}{U_0(z)} = \frac{(z-z_0)}{Bd},
    \end{equation}
    where $B$ is the velocity decay coefficient and $U_j$ is the jet injection velocity. We capture the decay of the jet centerline velocity, as shown in figure $\ref{fig:jet_char}$.
    The slope of this line is the inverse of the  product of the velocity decay constant $B$ times the nozzle diameter, which in the simulation is assumed to be the diameter, $d_{sim}$, of an ``effective nozzle''. Using a known value for $B$, namely $B\approx6$ \citep{Hussein1994}, we can deduce $d_{sim} \approx 7.8\;\mbox{mm}$. We note that the ``effective simulation diameter'' is almost twice the experimental jet diameter $d_{expt}$. 
 
 Regarding the linear growth of the jet, the spreading rate can be determined as the slope   $S$,
    \begin{equation}\label{eqn:rate}
        S=\frac{dr_{1/2}}{dz}
    \end{equation}
    By fitting the figure with a straight line, we obtain $S=0.102$ which matches the experimental values of \citet{Hussein1994} and \citet{Xu2002} very well.

Next, we choose to position the body force at a location where the experimental jet is expected to have achieved the half-width equal to the simulated jet's inflow radius, that is we require that $r_{1/2,expt} = d_{sim}/2$. The applied force is spatially smoothed in a region over three grid points in $x,y$ and $z$  using a super-Gaussian smoother (of order 5 and width $\sigma_G = \Delta x$) centered at $z_m$ as shown by the sketch. We recall that $r_{1/2,expt}(z_m) = S (z_m-z_0)$, where $z_0$ is the virtual origin of the experimental nozzle. Using the value $d_{sim} = 7.8\;\mbox{mm}$ found above, we solve for $z_m-z_0$ and find $z_m-z_0 = d_{sim}/(2 S) = 38.2\;\mbox{mm}$, i.e. we apply the force $38.2\;\mbox{mm}$ above the location of the real nozzle's virtual origin.   

The last parameter to determine is the jet centerline velocity, $U_{j,sim}$, at location $z=z_m$. The simulated injection jet velocity  will be set equal to the experimental centerline velocity at that location, thus reproducing the mean flow of the jet as the most basic condition to be met at that location, where the LES grid resolution is just sufficient to resolve a jet's mean velocity profile. Using the classical scaling, the centerline velocity at $z_m$ in the experiment may be obtained by 
\begin{equation} 
    U_{0,expt}(z_m) = \frac{ U_{j,expt}\, B \, d_{expt}}{z_m-z_0}.
\end{equation}
Setting $U_{j,sim} = U_{0,expt}(z_m)$ and replacing $z_m-z_0 =  r_{1/2,expt} / \,S =  d_{sim} / (2S)$ we obtain

\begin{equation} 
    U_{j,sim} = \frac{2 \,S \,U_{j,expt} B \, d_{expt}}{d_{sim}}.
\end{equation}

Substituting $d_{sim}=7.8\;\mbox{mm}$, $S=0.102$, $B=6$ and $U_{j,expt}=2.5\;\mbox{m}/\mbox{s}$ we can calculate $U_{j,sim} \approx 1.6\;\mbox{m}/\mbox{s}$.

The body force magnitude is adjusted in the LES so as to achieve this value of the maximum velocity in the region where the body force is applied. With this value enforced, we obtain the desired linear growth of the jet width and the inverse linear decay of centerline velocity matching those of a classical turbulent round jet.  

\begin{figure}
\centering
\includegraphics[width=0.5\linewidth]{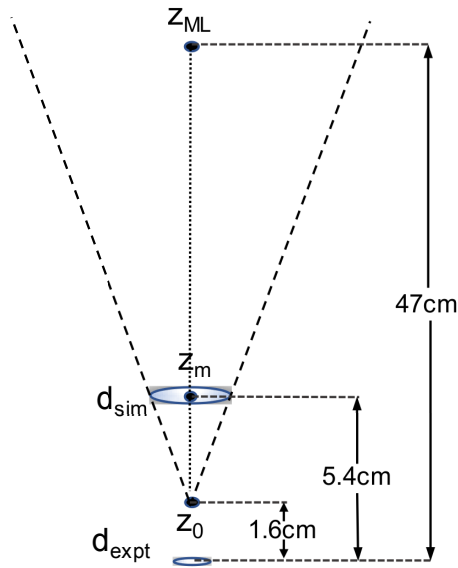}
\caption{Sketch depicting the nozzle placement in LES (at vertical position $z_m=54\;\mbox{mm}$ downstream of the experimental nozzle), the virtual origin of the experiment (assumed to be at $z_0=16\;\mbox{mm} = 4d_{expt}$ downstream of the nozzle), and the measurement location (at a height of $z_{ML}=470\;\mbox{mm}$ above the experimental nozzle position - in the experiment with cross-flow, there is additional displacement in the horizontal direction).}
\label{fig:jet_meas}
\end{figure}

To determine the corresponding location to measure the size distribution for the simulation, we note that for a classical round jet $z_0 \approx 4d$ \citep{Hussein1994,Xu2002}. We can then calculate  $z_m = d_{sim}/2S +4d \approx 54.2\;\mbox{mm}$. The measurement volume for the experiment was located at a height of $47$cm from the nozzle in the vertical direction. Therefore the measurement location for the simulation is at a vertical distance of $z_{ML}-z_m\approx 42\;\mbox{cm}$ from the applied force position (see figure $\ref{fig:jet_meas}$).

\clearpage
\bibliographystyle{jfm}
 \bibliography{Paper_1,PBE_paper_references}
\end{document}